\documentclass[preprint]{aastex}
\usepackage{amsmath}
\usepackage{color}
\usepackage{ulem}
\begin{document}

\title{Observations of Infalling and Rotational Motions on a 1,000-AU Scale around 17 Class 0 and 0/I Protostars: Hints of Disk Growth and Magnetic Braking?}

\author{Hsi-Wei Yen\altaffilmark{1}, Patrick M. Koch\altaffilmark{1}, Shigehisa Takakuwa\altaffilmark{1}, Paul T. P. Ho\altaffilmark{1,2}, Nagayoshi Ohashi\altaffilmark{1,3}, and Ya-Wen Tang\altaffilmark{1}}

\altaffiltext{1}{Academia Sinica Institute of Astronomy and Astrophysics, P.O. Box 23-141, Taipei 10617, Taiwan; hwyen@asiaa.sinica.edu.tw} 
\altaffiltext{2}{Harvard-Smithsonian Center for Astrophysics, 60 Garden Street, Cambridge, MA 02138, USA}
\altaffiltext{3}{Subaru Telescope, National Astronomical Observatory of Japan, 650 North A'ohoku Place, Hilo, HI 96720, USA}

\begin{abstract}
We perform imaging and analyses of SMA 1.3 mm continuum, C$^{18}$O (2--1) and $^{12}$CO (2--1) line data of 17 Class 0 and 0/I protostars to study their gas kinematics on a 1,000-AU scale. 
Continuum and C$^{18}$O (2--1) emission are detected toward all the sample sources and show central primary components with sizes of $\sim$600--1,500 AU associated with protostars. 
The velocity gradients in C$^{18}$O (2--1) have wide ranges of orientations from parallel to perpendicular to the outflows, 
with magnitudes from $\sim$1 to $\sim$530 km s$^{-1}$ pc$^{-1}$. 
We construct a simple kinematic model to reproduce the observed velocity gradients, estimate the infalling and rotational velocities, and infer the disk radii and the protostellar masses. 
The inferred disk radii range from $<$5 AU to $>$500 AU with estimated protostellar masses from $<$0.1 $M_\sun$ to $>$1 $M_\sun$. 
Our results hint that both large and small disks are possibly present around Class 0 protostars, which could be a sign of disk growth at the Class 0 stage. 
In addition, the directions of the overall velocity gradients in 7 out of the 17 sources are close to perpendicular to their outflow axes ($\Delta\theta > 65\degr$), 
which is a signature of significant rotational motions. 
From our model fitting, the specific angular momenta in these sources are estimated to be $>$2 $\times$ 10$^{-4}$ km s$^{-1}$ pc, 
suggesting that magnetic braking is unlikely efficient on a 1,000-AU scale in these Class 0 and 0/I sources.
In a sub-sample with observed magnetic field orientations, we find no source with large specific angular momenta together with closely aligned magnetic field and outflow axes. 
This possibly hints that the magnetic field, if originally aligned with the rotational axis, can play a role in removing angular momentum from infalling material at the Class 0 stage. 
We discuss our results in comparison with theoretical models of collapsing dense cores with and without  magnetic fields in the context of disk formation. 
\end{abstract}

\section{Introduction}
Circumstellar disks around young stellar objects are sites of planet formation (Williams \& Cieza 2011). 
Interferometric observations in the CO lines show that the motions of these disks are well explained by Keplerian rotation, 
and that the disk radii range from $\sim$100 AU to $\sim$800 AU (e.g., Guilloteau \& Dutrey 1994; Dutrey et al.~1998; Simon et al.~2000; Pi\'etu et al.~2007). 
The disk masses traced by dust continuum emission at millimeter and submillimeter wavelengths range from 10$^{-4}$ $M_\sun$ to 10$^{-1}$ $M_\sun$ (e.g., Qi et al.~2003, 2004; Andrews \& Williams 2007; Guilloteau et al.~2011; Andrews et al.~2012; P\'erez et al.~2012). 
Keplerian disks have also been observed around several Class I protostars (e.g., Brinch et al.~2007; Lommen et al.~2008; J{\o}rgensen et al.~2009; Lee 2010, 2011; Takakuwa et al.~2012; Yen et al.~2013; Brinch \& J{\o}gensen 2013; Harsono et al.~2014; Chou et al.~2014). 
The Keplerian disks around Class I protostars have outer radii ranging from $\sim$60 AU to $\sim$300 AU and masses from 10$^{-3}$ $M_\sun$ to $\sim$10$^{-1}$ $M_\sun$, comparable to those of disks around T Tauri stars. 
Therefore, Keplerian disks with a size of a 100-AU scale are likely already well developed at the Class I stage. 

Disks around protostars at earlier evolutionary stages are likely deeply embedded in protostellar envelopes, and are difficult to directly image (e.g., Looney et al.~2003; Chiang et al.~2008). 
Recent interferometric observations have successfully revealed Keplerian disks around a Class 0/I protostar, B59\#11\footnotemark (Hara et al.~2013), and several Class 0 protostars, L1527 IRS (Tobin et al.~2012a; Ohashi et al.~2014), VLA 1623 (Murillo et al.~2013) and HH 212 (Lee et al.~2014; Codella et al.~2014). 
The disks observed around these Class 0 and 0/I protostars have outer radii ranging from $\sim$50 AU to $\sim$350 AU, comparable to those of disks around more evolved protostars and T Tauri stars. 
On the other hand, around several Class 0 protostars, such as B335 (Yen et al.~2010, 2011, 2013), NGC 1333 IRAS 4B (Yen et al.~2013), and NGC 1333 IRAS 2A (Brinch et al.~2009; Maret et al.~2014), there is no clear sign of rotational motion in their protostellar envelopes within a scale of 1,000 AU,  
which suggests that the Keplerian disks in these sources (if present) likely have outer radii smaller than 10 AU. 
It is still unclear as to how and when Keplerian disks form around protostars and whether Keplerian disks with a size of 100 AU are common among Class 0 protostars.

\footnotetext{B59\#11 is located in a dense core with a size of $\sim$0.1 pc and a mass of $\sim$24 $M_\sun$, which contains several protostellar sources (Brooke et al.~2007). B59\#11 has a spectral index of 3.29 at wavelengths from 2.2 $\mu$m to 8 $\mu$m, $T_{\rm bol}$ of 70$\pm$10 K, and $L_{\rm submm}/L_{\rm bol}$ of 0.03$\pm$0.01 (Brooke et al.~2007). The spectral index is consistent with that of Class I or 0 protostellar sources. Its $L_{\rm submm}/L_{\rm bol}$ is consistent with Class 0 sources ($L_{\rm submm}/L_{\rm bol}$ $>$ 0.005; Andr\'e et al. 2000), while its $T_{\rm bol}$ is near the border between the definitions of Class 0 and I sources, i.e., $T_{\rm bol}$ of 70 K (Andr\'e et al. 2000). Hence, B59\#11 is classified as a Class 0/I protostar by Brooke et al.~(2007). Considering its low mass ratio between the surrounding envelope ($\sim$0.09 $M_\sun$) and the protostellar mass ($\sim$0.73 $M_\sun$; Hara et al.~2013), B59\#11 is likely in a stage close to Class I.}

As protostars form through gravitational collapse of dense cores ($n \sim 10^{4} - 10^{5}$ cm$^{-3}$) in molecular clouds (e.g., Andr\'{e} et al.~2000; Myers et al.~2000), a Keplerian disk is expected to form when collapsing material rotates fast enough to become centrifugally supported (e.g., Shu et al.~1987). 
Theoretical models of collapsing dense cores without magnetic fields suggest that the outer radius of the Keplerian disk increases as the collapse proceeds toward outer regions and material with a higher angular momentum falls in (Ulrich 1976; Cassen \& Moosman 1981; Terebey et al.~1984; Basu 1998; Bate 1998). 
On the other hand, 
previous magnetohydrodynamic (MHD) simulations show that the magnetic field can effectively remove the angular momentum of collapsing material by magnetic braking, 
and suppress the outer radii of Keplerian disks to within $\sim$10 AU (e.g., Allen et al.~2003; Mellon \& Li 2008, 2009; Machida et al.~2011; Li et al.~2011; Dapp et al.~2012; Tomida et al.~2013). 
Several mechanisms have been proposed and demonstrated with numerical simulations to reduce the efficiency of magnetic braking and enable formation of large-scale disks, such as dissipation of protostellar envelopes (e.g., Machida et al. 2011; Machida \& Hosokawa 2013), misalignment between the magnetic field and rotational axis of collapsing dense cores (e.g., Hennebelle \& Ciardi 2009; Joos et al. 2012; Li et al.~2013), and turbulence (e.g., Seifried et al.~2012, 2013; Joos et al.~2013). 
Recent theoretical simulations by Machida et al.~(2014), investigating the dependence of disk formation on the initial model settings such as the sink radius and the density profile, successfully form Keplerian disks with properties similar to those observed in protostellar sources. 
However, there is still a discrepancy in gas motions between their simulations and those by Li et al.~(2011) with similar initial density profiles. 
Recent single-dish and interferometric polarimetric surveys toward protostellar sources show that the magnetic field tends to be misaligned with outflows on inner-envelope scales of a few thousand AU, 
and that the magnetic field orientations might change from the dense-core scale of 0.1 pc to the inner-envelope scale (e.g., Hull et al.~2013, 2014). 
The results of this polarimetric survey might support the idea that the misalignment between magnetic field and outflow enables the formation of large-scale Keplerian disks.  These results further suggest the presence of an interplay between magnetic field and gas motions as seen in MHD simulations (e.g., Machida et al.~2005). 
Therefore, to investigate the role of the magnetic field in disk formation, a comparison between magnetic field structures and gas motions in a sample of protostellar sources is needed.

Formation and evolution of Keplerian disks in collapsing dense cores are likely closely related to angular momentum transfer from dense cores on a 10,000-AU scale, to inner infalling envelopes on a 1,000-AU scale and to central disks on a 100-AU scale.  
Single-dish observations in the NH$_3$ and N$_2$H$^+$ lines show that dense cores exhibit velocity gradients over a 10,000-AU scale with a mean magnitude of $\sim$1--2 km s$^{-1}$ pc $^{-1}$, suggestive of large-scale rotational motions (Goodman et al.~1993; Caselli et al.~2002; Tobin et al.~2011). 
Interferometric observations in the N$_2$H$^+$ lines have revealed a larger amount of velocity gradients with a mean magnitude of $\sim$7--8 km s$^{-1}$ pc $^{-1}$ in the inner protostellar envelopes on thousands of AU scale (Chen et al.~2007; Tobin et al.~2011), 
suggestive of faster rotational motions on smaller scales. 
We note that the overall velocity gradients seen in the dense cores and the protostellar envelopes should reflect combinations of rotational motion and other systematic gas motions, such as infalling and/or outflowing motions, as well as the envelope morphologies (Tobin et al.~2012b). 
Systematical studies and comparisons of gas motions from large to small scales in a large sample of Class 0 protostars are essential to understand disk formation at the early evolutionary stages. 

In order to investigate disk formation in infalling envelopes, we studied radial profiles of rotational motions in four Class 0 and two Class I protostellar sources. We find an evolutionary trend from slow to fast rotational motions and from rotation with a conserved angular momentum to Keplerian rotation (Yen et al.~2013). 
Such an evolutionary trend can be explained with the conventional picture of an inside-out collapse where the angular momentum is conserved (e.g., Terebey et al.~1984). 
In the present paper, we enlarge our sample and focus on protostars at the Class 0 stage and in transition from Class 0 to I stages, to (1) investigate whether the presence of 100-AU-scale Keplerian disks is common among Class 0 protostars -- as those in L1527 IRS, B59\#11, VLA 1623, and HH 212 -- or not, and to (2) study the relation between gas kinematics and magnetic field.  

On large scales from thousands of AU to 0.1 pc, the presence of non-axisymmetric structures, such as filaments, in protostellar envelopes have been seen in 8 $\mu$m extinction maps (e.g., Tobin et al.~2010) and millimeter observations (e.g., Tobin et al.~2011). 
For non-axisymmetric structures, it is not straightforward to extract kinematic information from observed images in molecular lines (Tobin et al.~2012b). 
On the other hand, on smaller scales of $\sim$1,000 AU, the envelope structures appear to be more or less symmetric (e.g., Tobin et al. 2011, 2012b). 
Assuming that molecular-line emission traces axisymmetric flattened envelopes around protostars and that outflow axes represent rotational axes, 
the velocity gradients perpendicular to the outflow directions can be interpreted as rotational motion, 
and those along the outflow directions can be due to infalling motion or contamination from the outflows (e.g., Arce \& Sargent 2006; Yen et al.~2010, 2013). 
Hence, the observed velocity structures can be decomposed into infalling and rotational motions.  
Such analyses have been applied to the observational data of several protostellar sources, such as B335 (Yen et al.~2010), HH 212 (Lee et al.~2006), L1527 IRS (Ohashi et al.~1997), L1551 IRS 5 (Momose et al.~1998), and IRAS 16293$-$2422 (Takakuwa et al.~2007).
Their observational results can, indeed, be explained by axisymmetric models of a combination of infalling and rotational motions. 

In the present paper, we use the data of the C$^{18}$O (2--1; 219.560358 GHz) line obtained with the Submillimeter Array (SMA) to trace envelope kinematics. 
C$^{18}$O can be abundant in inner protostellar envelopes as it is evaporated from dust grains when the inner envelopes are heated up to $\sim$20 K through a proceeding collapse.
On the other hand, the abundances of other dense-gas tracers, as e.g., N$_2$H$^+$ and NH$_3$, typically decrease
as they are directly or indirectly destroyed by CO (e.g., Lee et al.~2004; Aikawa et al.~2008). 
We apply axisymmetric models to our C$^{18}$O data,
measure velocities of infalling and rotational motions on a 1,000-AU scale, and estimate the ranges of possible protostellar masses and Keplerian disk sizes around 17 Class 0 and 0/I protostars. 
Some of our sample sources were observed in polarized dust emission, 
revealing their magnetic field orientations (Attard et al.~2009; Matthews et al.~2009; Dotson et al.~2010; Davidson et al.~2011; Chapman et al.~2013; Hull et al.~2013, 2014).
Our results are discussed in comparison with theoretical models without magnetic fields, with MHD simulations, and with observational results of magnetic field orientations. 

\section{Sample}
The 17 sources studied in this project are Class 0 and 0/I low-mass protostars selected from the source lists of Froebrich (2005), Tachihara et al.~(2007), and Enoch et al.~(2009). 
Froebrich (2005) searched the literatures and compiled a list of 95 confirmed or candidate Class 0 and 0/I protostars using the following criteria:
(1) the bolometric temperature is less than 80 K, (2) the ratio of submillimeter ($>$350 $\mu$m) to bolometric luminosities is larger than 0.5\%, and (3) no near infrared ($<$5 $\mu$m) counterpart is present\footnotemark. 
Sources that satisfy all the criteria are classified as Class 0 sources.
Sources that satisfy two out of the three criteria are classified as Class 0/I sources. 
Tachihara et al.~(2007) found a Class 0 protostar with a low bolometric temperature of 40 K and a low bolometric luminosity of 0.16 $L_\sun$ from the molecular line, millimeter continuum, and infrared surveys of the Lupus 3 cloud. 
Enoch et al.~(2009) identified protostars in Perseus, Serpens, and Ophiuchus using the data from the ``From Molecular Cores to Planet-forming Disks'' {\it Spitzer} Legacy program and the 1.1 mm Bolocam continuum surveys with the Caltech Submillimeter Observatory (Evans et al.~2003; Enoch et al.~2006, 2007; Young et al.~2006). 
They classified 39 sources having bolometric temperatures less than 70 K as Class 0 protostars. 
We selected low-mass protostars from these Class 0 and 0/I source lists, searched the SMA data archive, and conducted new SMA observations toward seven sources. 
Combining the archival data and our observations, 
we find 17 sources detected in the C$^{18}$O (2--1) line ($>$3$\sigma$) in more than three velocity channels. 
Our sample is composed of these 17 sources, which are all nearby (with a distance $d \lesssim 250$ pc except one source at $d = 400$ pc), have a range of bolometric temperatures from 25 to 90 K and a range of bolometric luminosities from 0.2 to 7.7 $L_\sun$. 
L1527 IRS, B59\#11, and HH 212, where the presence of 100-AU Keplerian disks is reported (Tobin et al.~2012a; Hara et al.~2013; Lee et al.~2014; Codella et al.~2014; Ohashi et al.~2014), are also included in our analyses for uniform comparison with other sources observed with the SMA.
Table \ref{sample} shows a summary of our sample sources.  Table \ref{oblog} lists all the data used for the work here.

\footnotetext{The classification by Froebrich (2005) was done before {\it Spitzer} data came out. In {\it Spitzer} data, Class 0 sources typically show nebulosity at wavelengths shorter than 5 $\mu$m (e.g., Tobin et al.~2007, 2008).}

\section{Observations}
We have conducted observations with the SMA$\footnotemark$ at 225 GHz toward seven sources in our sample: L1448-mm, NGC 1333 IRAS 4A and 4B, L1527 IRS, Lupus 3 MMS, B228 and IRAS 16253$-$2429.  
Details of our SMA observations with the subcompact and compact configurations toward L1448-mm, NGC 1333 IRAS 4B and L1527 IRS are described by Yen et al.~(2013). 
Our new SMA observations with the very extended configuration were conducted toward L1527 IRS in 2011, September 9 and toward L1448-mm and NGC 1333 IRAS 4A and 4B in 2011, September 11. 
In these observations, 3c111 and 3c84 were observed as gain calibrators, 3c454.3 as a bandpass calibrator, and Callisto as a flux calibrator. 
The typical system temperature during the observations was 80--150 K.
Our new SMA observations with the subcompact configuration toward B228 and IRAS 16253$-$2429 were conducted in 2013,  April 30, 
and those with the compact configuration toward Lupus 3 MMS in 2013, March 14. 
1517$-$243 and 1626$-$298 were observed as gain calibrators, and 3c279 as a bandpass calibrator. 
Neptune and Titan were observed as flux calibrators in the subcompact and compact observations, respectively. 
The typical system temperature during these observations was 100--220 K.
In our observations with the compact configuration, 512 channels were assigned to a chunk with a bandwidth of 104 MHz for the C$^{18}$O (2--1) line, 
and 1024 channels in the observation with the subcompact configuration, 
resulting in velocity resolutions of 0.28 km s$^{-1}$ and 0.14 km s$^{-1}$, respectively. 
1.3 mm continuum and $^{12}$CO (2--1) line emission were observed simultaneously in all the observations.
The rest of the data of our sample sources were obtained from the SMA data archive. 
The observing dates and PIs of those observations are listed in Table \ref{oblog}. 
All the data were calibrated using the MIR software package (Scoville et al.~1993). 
The calibrated visibility data were Fourier-transformed and CLEANed with MIRIAD (Sault et al.~1995) to produce images.
The resolutions and noise levels of the 1.3 mm and C$^{18}$O (2--1) images of all the sample sources are listed in Table \ref{obsum1}, 
and those of the $^{12}$CO (2--1) images of a subset of the sample are listed in Table \ref{obsum2}.
The $^{12}$CO (2--1) results are presented in Appendix \ref{12co}.
Inclination angles in this subset of sources are estimated based on morphologies and velocity structures of their $^{12}$CO outflows (Appendix \ref{i}).

\footnotetext{The details of the SMA are described by Ho et al.~(2004).}

\section{Results}
\subsection{1.3 mm Continuum Emission}
\begin{figure}
\epsscale{1}
\plotone{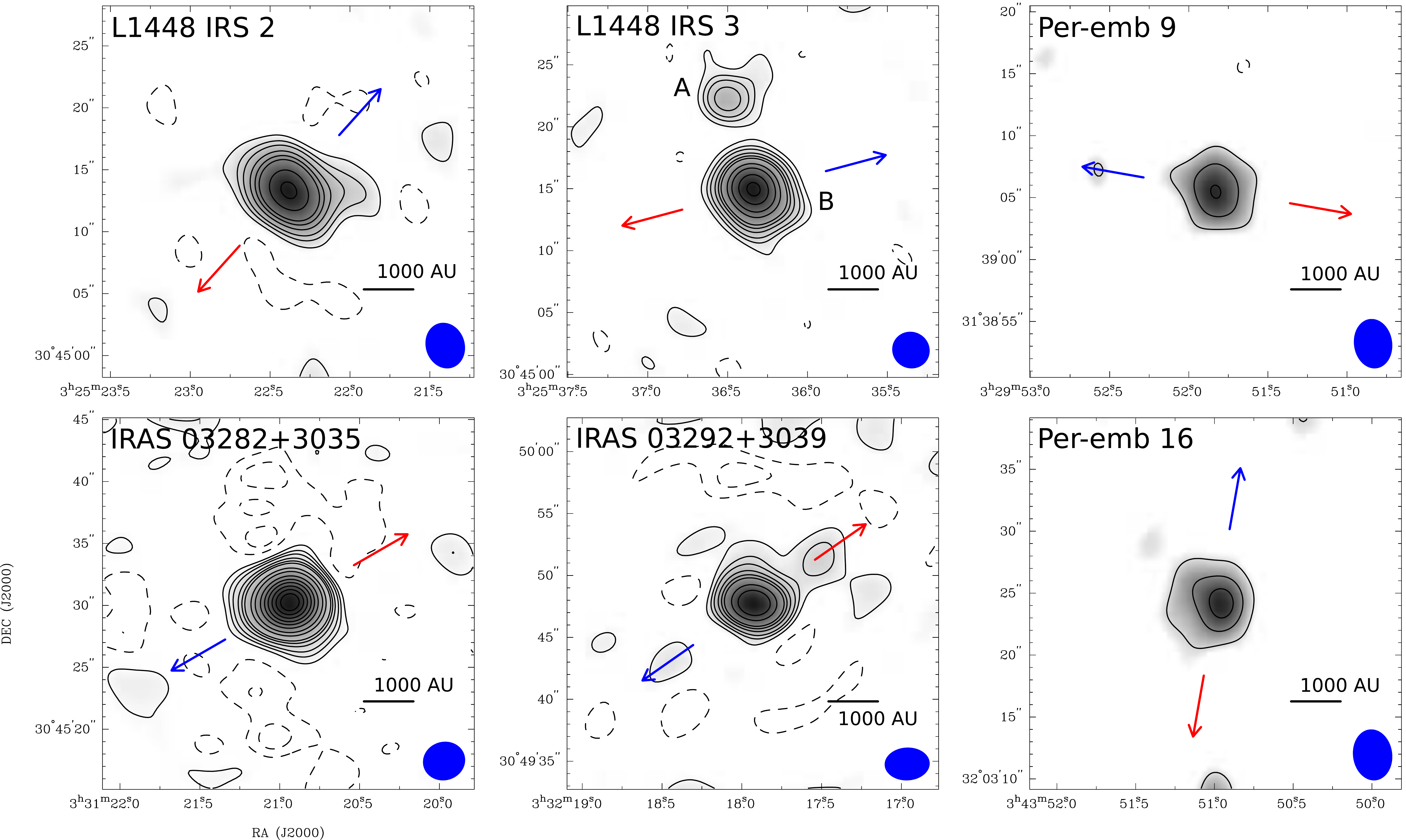}
\caption{1.3 mm continuum images of the sample sources. A filled ellipse at the bottom-right corner in each panel presents the beam size. Red and blue arrows denote the directions of the redshifted and blueshifted outflows, respectively. Contour levels are 3$\sigma$, 6$\sigma$, 9$\sigma$, 14$\sigma$, 19$\sigma$, 24$\sigma$, 39$\sigma$, 54$\sigma$, 79$\sigma$, 104$\sigma$, 129$\sigma$, and then in steps of 30$\sigma$. The beam sizes and the 1$\sigma$ noise levels are summarized in Table \ref{obsum1}.}
\label{config}
\end{figure}

\begin{figure}
\epsscale{1}
\plotone{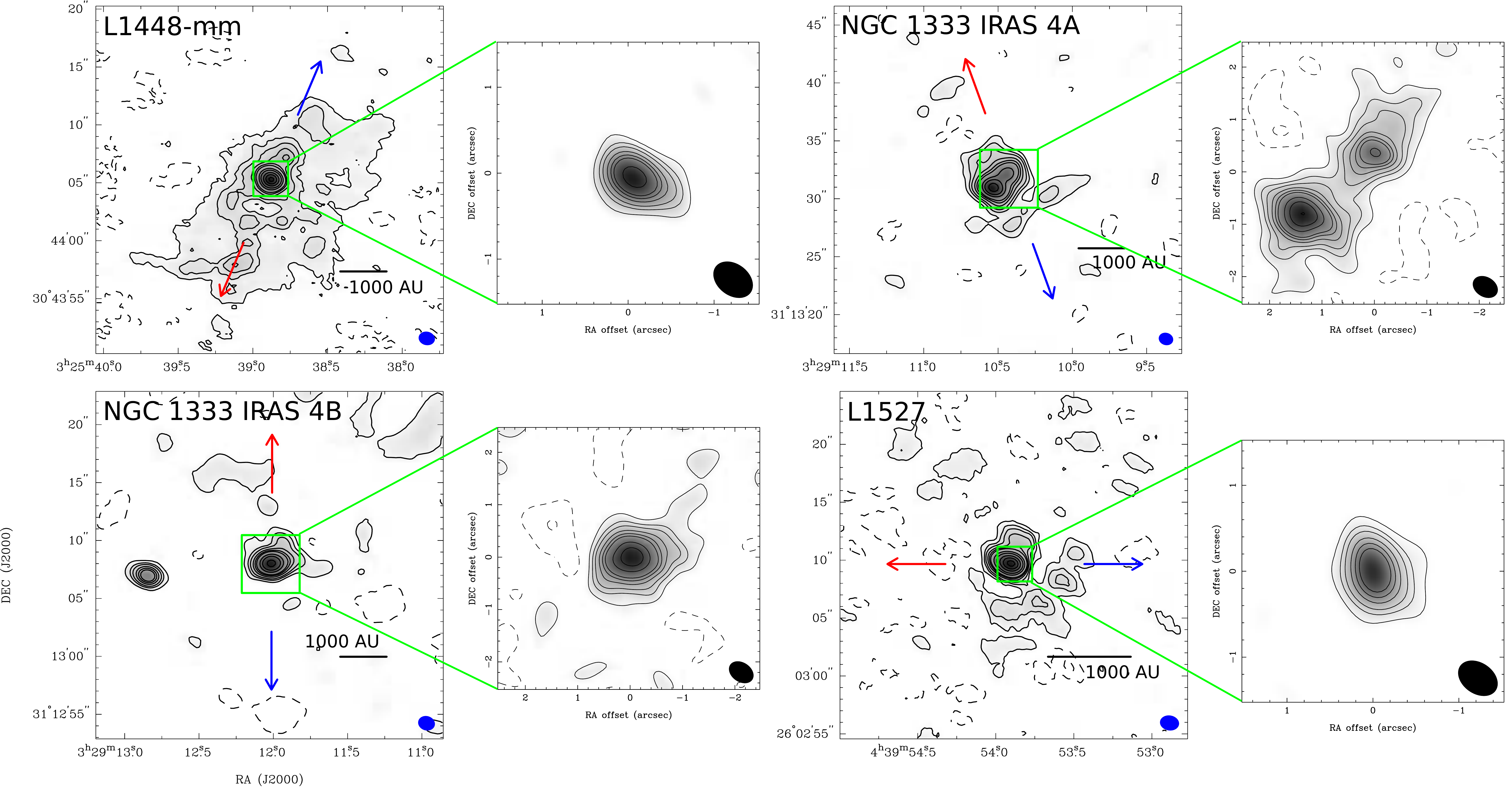}
\caption{1.3 mm continuum images of L1448-mm, NGC 1333 IRAS 4A and 4B, and L1527 IRS. The smaller panels present the high-resolution 1.3 mm continuum images using only the data of the very extended configuration. Green boxes in the larger panels 
show the areas of the high-resolution images. Filled ellipses at the bottom-right corner in each panel present the beam sizes. Red and blue arrows denote the directions of the redshifted and blueshifted outflows, respectively. Contour levels are 3$\sigma$, 6$\sigma$, 9$\sigma$, 14$\sigma$, 19$\sigma$, 24$\sigma$, 39$\sigma$, 54$\sigma$, 79$\sigma$, 104$\sigma$, 129$\sigma$, and then in steps of 30$\sigma$. The beam sizes and the 1$\sigma$ noise levels are summarized in Table \ref{obsum1}.}
\label{config2}
\end{figure}

Figure \ref{config} shows the 1.3 mm continuum images of the sample sources. 
L1448-mm, NGC 1333 IRAS 4A and 4B, and L1527 IRS have been additionally observed with the very extended configuration of the SMA at 1.3 mm, 
with an angular resolution of $\sim$0\farcs5 probing only the innermost ($\lesssim$3\arcsec) regions. 
The 1.3 mm continuum images of these four sources are shown in Figure \ref{config2}. 
Toward all the sample sources, central compact components with sizes of $\sim$100--1,000 AU are observed, 
and several sources additionally display extended structures.  
L1448-mm shows an extended component with a size of $\sim$20$\arcsec$ ($\sim$5,000 AU) elongated along the northwest--southeast direction
In L1527 IRS, several clumps with sizes of $\sim$2$\arcsec$ ($\sim$300 AU) are seen around the central component. 
B335 shows extensions with a length of $\sim$5$\arcsec$ ($\sim$650 AU) pointing toward the northwest and the southwest.  
L1157-mm displays an extension with a length of $\sim$3$\arcsec$ ($\sim$750 AU) pointing toward the southwest. 
Nevertheless, the presence of a dominant central component in all these sources is evident. 

Moreover, L1448 IRS 3, L1448-mm, NGC 1333  IRAS 4A and 4B, IRAS 03282$+$3035, and HH212 are known to be binary or multiple systems (e.g., Looney et al.~2000; Launhardt 2004; J{\o}rgensen et al.~2007; Lee et al.~2008; Chen et al.~2013). 
The reported secondary components are seen in our 1.3 mm continuum images of L1448 IRS 3, L1448-mm, and NGC 1333 IRAS 4A and 4B. 
L1448 IRS 3 exhibits three components: A to the north, B to the south, and NW to the northwest. 
NW is outside the plotting area of Figure \ref{config}.
In the present paper, we focus on L1448 IRS 3B, where the associated C$^{18}$O emission is detected, as will be shown below. 
In L1448-mm, there is a 9$\sigma$ peak located in the southeast, $\sim$7$\arcsec$ ($\sim$1,750 AU) away from the main component. 
This component embedded in the extended structure corresponds to the reported companion (Chen et al.~2013). 
In NGC 1333 IRAS 4A, the central condensation exhibits two peaks, 4A1 to the east and 4A2 to the west. 
In NGC 1333 IRAS 4B, the two components have a projected separation of $\sim$10\farcs3 ($\sim$2,600 AU). 
There is no previous estimate of the inclination angle of the secondary component located to the east. 
Its morphology and outflow velocity structures are not clearly detected with the present SMA data (J{\o}rgensen et al. 2007) and the CARMA+FCRAO data (Plunkett et al. 2013).  
More recent CARMA $^{12}$CO (2--1) observations show a clear detection (Hull et al.~2014), 
but details of kinematics and structures are not given. 
Thus, no sufficient information of the outflow is available to estimate its inclination angle.
Hence, in the present paper, we focus on the primary component in the west. 
Our continuum images of IRAS 03282$+$3035 and HH212 only show a single component. 
The projected binary separations of these two sources are reported to be 1\farcs6 (400 AU) and 0\farcs3 (120 AU), respectively (Chen et al.~2013), 
which are too small to be resolved by the SMA observations with the compact configuration (Table \ref{obsum1}).

A two-dimensional Gaussian distribution is fitted to the central components in all the sample sources.  
Depending on the intensity distributions in individual sources, pixels with values below 3$\sigma$--10$\sigma$ are excluded from the fitting to avoid contamination from the surrounding diffuse emission. 
The derived total flux, de-convolved size, and position angle of the major axis are shown in Table \ref{contable}. 
In the present paper, we adopt the Gaussian-fitted peak positions of the 1.3 mm continuum emission as the protostellar locations. 
Comparing the de-convolved position angles and the outflow directions, the central components of the 1.3 mm continuum emission in L1448 IRS2, L1448 IRS 3B, IRAS 03292$+$3039, B59\#11, and B335 are elongated perpendicularly to the outflow directions within $\lesssim$20$\degr$, suggesting that their protostellar envelopes are flattened and normal to the outflows or that the 1.3 mm continuum primarily arises from the circumstellar disks. 
Although the 1.3 mm continuum image of Lupus 3 MMS also shows a component elongated perpendicularly to the outflow direction, 
this component is not resolved with the SMA.
Hence, its apparent elongation is due to the convolution with the synthesized beam which is elongated perpendicularly to the outflow direction. 
On the other hand, 
those in B228 and L1157-mm are elongated along the outflow directions within $\lesssim$25\degr, 
which could be due to the contamination from the outflows.  
The 1.3 mm continuum emission in other sources, such as Per-emb 9, Per-emb 16, IRAS 03282$+$3035, HH 212, and IRAS 16253$-$2429, does not show a clear elongation or orientation preferentially parallel or perpendicular to the outflow directions. 

For the sources where we have both low- and high-resolution images, 
we can compare the structures of the continuum emission on scales of a few hundred AU with the innermost 100 AU.
In the low-resolution images of L1448-mm, NGC 1333 IRAS 4A2, and L1527 IRS,
the 1.3 mm continuum emission on a scale of 1$\arcsec$--2$\arcsec$ ($\sim$140--500 AU) does not show an elongation perpendicular to the outflow directions. 
However, interestingly, in the high-resolution images, the elongation of the continuum emission on smaller scales of $\sim$0\farcs5 ($\lesssim$125 AU) becomes more perpendicular to the outflow directions.
In addition, the aspect ratios between the major and minor axes of the 1.3 mm continuum emission also vary from the larger to smaller scales. 
These changes in orientations and aspect ratios could suggest that the 1.3 mm continuum emission on the 0\farcs5 scale in these sources traces a component that is different from the one on the 1$\arcsec$--2$\arcsec$ scale. 
On the 0\farcs5-scale, the 1.3 mm continuum emission in these sources, which is elongated perpendicularly to the outflow directions, could primarily trace the circumstellar disks with less contamination from the protostellar envelopes and the outflows, as compared to that on the 1$\arcsec$--2$\arcsec$ scale. 
On the other hand, in NGC 1333 IRAS 4B, both the low- and high-resolution images show similar orientations and aspect ratios,
suggesting that the 1.3 mm continuum emission from 1$\arcsec$--2$\arcsec$ to 0\farcs5 scales observed with the SMA most likely traces the same component.

The mass of the circumstellar material traced by the 1.3 mm continuum emission ($\tbond$$M_{\rm dust}$) can be estimated as 
\begin{equation}
M_{\rm dust} = \frac{F_{\rm 1.3mm}d^{2}} {\kappa_{\rm 1.3 mm} B(T_{\rm dust})},
\end{equation}
where $F_{\rm 1.3mm}$ is the total 1.3 mm flux, $d$ is the distance to the sources, $\kappa_{\rm 1.3 mm}$ is the dust mass opacity at 1.3 mm, $T_{\rm dust}$ is the dust temperature, and $B(T_{\rm dust})$ is the Planck function at a temperature of $T_{\rm dust}$. 
Under the assumption that the wavelength ($\tbond$$\lambda$) dependence of the dust mass opacity ($\equiv \kappa_{\lambda}$) is $\kappa_{\lambda} = 0.1 \times (0.3\ {\rm mm}/\lambda)^{\beta}$ cm$^{2}$ g$^{-1}$ (Beckwith et al.~1990), the mass opacity at 1.3 mm is 0.023 cm$^{2}$ g$^{-1}$ with $\beta = 1.0$ (e.g., J{\o}rgensen et al.~2007) and a gas-to-dust mass ratio of 100. 
$T_{\rm dust}$ is adopted to be the temperature at a radius of 500 AU of our best-fit kinematic model (Section \ref{fitting}). 
The 1.3 mm continuum components in L1448-mm, NGC 1333 IRAS 4A, and L1527 IRS observed with the SMA at the high angular resolution could trace the circumstellar disks in the inner 100 AU regions around the protostars. 
For typical circumstellar disks around T Tauri stars, the temperatures at a radius of 100 AU range from $\sim$20 to $\sim$50 K (e.g., Pi\'etu et al.~2007).  
Hence, $M_{\rm dust}$ traced by the high-resolution observations is estimated with $T_{\rm dust}$ = 50 K, leading to lower limits of $M_{\rm dust}$.
With the same $F_{\rm 1.3mm}$,  $T_{\rm dust}$ = 50 K gives three times smaller dust masses as compared to $T_{\rm dust}$ = 20 K.
$M_{\rm dust}$ is shown in Table \ref{contable}.

\subsection{C$^{18}$O (2--1) Emission}\label{c18oresult}
Figure \ref{c18ofig} shows moment-0 (i.e., integrated-intensity) maps overlaid on moment-1 (i.e., intensity-weighted mean velocity) maps of the C$^{18}$O (2--1) emission in our sample sources. 
Toward all the sources, the C$^{18}$O (2--1) emission shows compact components with sizes ranging from 1,000 AU to 3,000 AU associated with the protostellar positions. 
Elongated emission along the outflow directions is seen in NGC 1333 IRAS 4B, IRAS 03292$+$3039, Lupus 3 MMS, IRAS 16253$-$2429, B335, and L1157-mm. 
Directions orthogonal to the outflow axes are found in L1448 IRS 2, L1448 IRS 3B, IRAS 03282$+$3035, B228, and B59\#11. 
In addition to the central components, in L1527 IRS and L1448-mm, extended emission with sizes of $\sim$20$\arcsec$ (2,800 and 5,000 AU) is seen along the outflow cavity and the outflow direction, respectively. 

The moment-1 maps of the C$^{18}$O (2--1) emission in this sample show wide ranges of orientations and magnitudes of  velocity gradients. 
To measure these orientations and magnitudes, a linear relation, 
\begin{equation}\label{vgeq}
V_{\rm los} = M_{\rm vg} \cdot L + V_{\rm c},
\end{equation}
is fitted to the moment-1 maps, where $V_{\rm los}$ is the line-of-sight velocity, $M_{\rm vg}$ is the magnitude of the velocity gradient, $V_{\rm c}$ is a constant, and $L$ denotes the positional offset as
\begin{equation}
L = \Delta\alpha \cdot \sin \theta_{\rm vg} + \Delta\delta \cdot \cos \theta_{\rm vg}, 
\end{equation}
where $\Delta\alpha$ and $\Delta\delta$ are the RA and Dec offsets with respect to the protostellar position and $\theta_{\rm vg}$ is the position angle of the direction of the velocity gradient. 
$M_{\rm vg}$ and $\theta_{\rm vg}$ are determined by minimizing $\Sigma\ (V_{\rm ob}(\alpha, \delta) - V_{\rm los}(\alpha, \delta))$, where $V_{\rm ob}$ is the observed mean velocity in the moment-1 maps. 
The minimization is done using the IDL routine, $MPFIT$ (Markwardt 2009).
To focus on the velocity gradients in the central components with minimum contamination from surrounding extension or diffuse emission, 
we first fit a two-dimensional Gaussian distribution to the central component in the C$^{18}$O moment-0 map.  
Subsequently, the fitting of velocity gradients is only performed on the central region within a radius of twice the standard deviation ($=$FWHM$/2\sqrt{2\ln 2}$) along the major axis of the fitted Gaussian distribution.
The measured orientations and magnitudes of the overall velocity gradients are listed in Table \ref{c18ovg}, 
and their uncertainties are estimated based on the covariance matrixes. 
In L1448 IRS 3B, NGC 1333 IRAS 4A, Per-emb 9, IRAS 03292$+$3039, L1527 IRS, HH 212, and B59\#11, the directions of the velocity gradients of the central components are perpendicular to the outflow directions with $\Delta\theta \gtrsim 65\degr$, and their magnitudes range from $\sim$22 km s$^{-1}$ pc$^{-1}$ in HH 212 to $\sim$529 km s$^{-1}$ pc$^{-1}$ in B59\#11.  
On the other hand, the directions of the velocity gradients of the central components in NGC 1333 IRAS 4B, IRAS 03282$+$3035, B228, B335, and L1157-mm are along the outflow directions with $\Delta\theta \lesssim 20\degr$, and the magnitudes are from $\sim$9 km s$^{-1}$ pc$^{-1}$ in L1157-mm to $\sim$115 km s$^{-1}$ pc$^{-1}$ in B335. 
The directions of the velocity gradients of the central components in L1448 IRS 2, L1448-mm, Per-emb 16, Lupus 3 MMS, and IRAS 16253$-$2429 do not show a preference to be either parallel or perpendicular to the outflow directions. 
Their magnitudes range from $\sim$1 km s$^{-1}$ pc$^{-1}$ in Lupus 3 MMS to $\sim$78 km s$^{-1}$ pc$^{-1}$ in Per-emb 16.
Previous interferometric observations in N$_2$H$^+$ and NH$_3$ lines at lower angular resolutions of $\sim$3\arcsec--11\arcsec also found orientations of velocity gradients that vary from perpendicular to parallel to the outflow axes, 
with magnitudes that differ by an order of magnitude in a sample of protostellar sources (Tobin et al.~2011). 
The varying orientations of the {\it overall} velocity gradients could be a measure for the significance of the rotational motions relative to the infalling motions and  possible contaminations from the outflows. 
The sources with the velocity gradients perpendicular to the outflows likely exhibit more dominant rotational motions. (e.g., Arce \& Sargent 2006; Yen et al.~2013). 
Therefore, the wide ranges of orientations and magnitudes of velocity gradients suggest that low-mass Class 0 and 0/I protostars likely exhibit a variety of infalling and rotational motions in their envelopes, even though they are at similar evolutionary stages. 

In order to further assess the rotational motions in the protostellar envelopes in our sample sources, 
we isolate the velocity gradients {\it perpendicular} to the outflow axes and passing through the protostellar positions. 
This gradient primarily consists of rotational motions with a minimal contamination from infalling motions and outflows (e.g., Yen et al. 2013).  
We first extract the mean velocities along the axis perpendicular to the outflow axis 
from the moment-1 maps, 
and fit the mean velocities with Equation \ref{vgeq} using $MPFIT$. 
Here, $L$ is the positional offset along the axis perpendicular to the outflow axis.
The measured velocity gradients perpendicular to the outflow axes are shown in Table \ref{c18ovgpa}. 
In our sample, 5 out of 17 sources, L1448 IRS 3B, NGC 1333 IRAS 4A, IRAS 03292$+$3039, Per-emb 16, and B59\#11, show very significant velocity gradients with magnitudes larger than 90 km s$^{-1}$ pc$^{-1}$, suggesting the presence of fast rotational motions. 
On the other hand, NGC 1333 IRAS 4B, Lupus 3 MMS, and L1157-mm do not show significant velocity gradients perpendicular to their outflows, 
and the magnitudes are $\lesssim$15 km s$^{-1}$ pc$^{-1}$. 
Besides that, in B335, the direction of the velocity gradient perpendicular to the outflow is different from that of the rotational motion on scales of thousands of AU (Saito et al.~1999; Yen et al.~2011; Kurono et al.~2013). 
These results suggest that there is no clear signature of rotational motions on the inner 1,000-AU scale in these sources. 
Hence, even after isolating the velocity gradients perpendicular to the outflow axes, 
the results still suggest that the protostellar envelopes in our sample likely exhibit different rotational velocities, spanning over an order of magnitude.

\section{Kinematic Model}\label{fitting}
\subsection{Measuring Infalling and Rotational Velocities}
To investigate the infalling and rotational velocities of the protostellar envelopes and the sizes of the embedded Keplerian disks from the observed velocity gradients in the C$^{18}$O (2--1) line, 
we construct a simple model of infalling and rotational motions with the geometrically-thin approximation. 
This model is compared against our observational results. 
Theoretical models of collapsing dense cores without magnetic field show that more flattened inner regions (e.g., Ulrich 1976; Terebey et al.~1984). 
Including magnetic fields, the MHD simulations predict that the ratio of radius over height of the inner envelopes can grow to as large as five in regions with densities larger than 10$^{6}$ cm$^{-3}$ (e.g., Machida et al.~2014).
The C$^{18}$O (2--1) emission in protostellar sources is typically optically thin ($\tau \lesssim 0.3$) on a 1,000-AU scale as inferred from the observed C$^{18}$O (3--2) and (2--1) line ratio ($\sim$1--2) in several protostellar sources (Hogerheijde et al.~1998). 
Therefore, the geometrically-thin kinematic model can be applied to the C$^{18}$O (2--1) emission to study the velocity field on the equatorial plane. 
However, we note that if the inner envelope is highly flattened, the inner region could shield the outer region from being heated by the protostellar luminosity, 
and C$^{18}$O (2--1) may be frozen out on the equatorial plane, 
as in the case of protoplanetary disks (e.g., de Gregorio-Monsalvo et al.~2013; Rosenfeld et al.~2013). 
In addition, photons from the central region tend to escape along the outflow cavity, resulting in a warm region along the cavity (e.g., Spaans et al.~1995).   
The emission from the outflow cavity can then enhance and contaminate the velocity features of the inner envelope.

To minimize the influence of the outflow contamination on the velocity structures in C$^{18}$O (2--1), 
our models are only applied to the central compact C$^{18}$O (2--1) components, excluding outer 
extensions which could be associated with outflow activities. 
For typical molecular outflows, their outflowing velocities grow with radius $\propto R$ (e.g., Shu et al.~1991), 
while the free-falling velocity is $\propto R^{-0.5}$. 
Hence, in the central components at smaller radii, the velocity features of the outflows are expected to be minimal.
With our model, we estimate the masses of protostars and the specific angular momenta of their surrounding infalling envelopes based on the infalling and rotational motions on a 1,000-AU scale observed with the SMA.
Although the Keplerian disks (if present) in our sample sources may not be resolved with the present SMA observations at angular resolutions of $\sim$2$\arcsec$--8$\arcsec$, 
the possible radius of the Keplerian disk ($R_{\rm d}$) can be derived from the estimated protostellar mass ($M_*$) and the angular momentum of the infalling envelope ($j$) as
\begin{equation}\label{rd}
R_{\rm d} = \frac{j^2}{GM_*}, 
\end{equation}
under the assumption that the angular momentum of the infalling material is conserved. 
Infalling motions with conserved angular momenta are observed in several protostellar sources without large-scale ($>$100 AU) Keplerian disks
(e.g., Yen et al.~2013, 2014; Ohashi et al.~2014). 
We note that in these cases, the gas motions of the outer envelopes observed with the SMA are assumed to be smoothly connected to those of the inner disks.
Such smooth connections of gas motions are observed around a few protostars, such as L1527 IRS (Ohashi et al.~2014), TMC-1A (Aso et al.~2014), and L1551 IRS 5 (Chou et al.~2014). 
MHD simulations, however,  suggest that infalling and rotational motions could be dramatically braked in inner envelopes on a 100-AU scale due to an increasing magnetic pressure (e.g., Mellon \& Li 2008, 2009).  
Nevertheless, abruptly braking gas motions have not yet been observed. 
On the other hand, if our sample sources exhibit Keplerian disks with outer radii of several hundred AU, the disks can be resolved with the present SMA observations, and the protostellar mass can be estimated from the Keplerian rotation directly with our model.

The radial profiles of the C$^{18}$O surface density ($\Sigma$) and the gas kinetic temperature ($T$) of our model are described as 
\begin{equation}
\Sigma(r) = \Sigma_0 \cdot (\frac{r}{r_0})^p,
\end{equation}
\begin{equation}
T(r) = T_0 \cdot (\frac{r}{r_0})^q, 
\end{equation}
where $r_0$ is arbitrarily chosen to be 500 AU. 
$p$ is adopted to be $-1$, which is an approximated radial profile of the column density of a spherical envelope during an  inside-out collapse having a radial density profile $\propto$ $r^{-1.5}$ (Shu 1977). 
$q$ is adopted to be $-0.4$, which is the typical profile observed in protostellar sources (e.g., Shirley et al.~2000, 2001). 
The protostellar positions are adopted as the centres of mass, except for NGC 1333 IRAS 4A, where two protostars 
are located in a common envelope. 
Here, the peak position in the moment-0 map of the C$^{18}$O emission is adopted as the center of mass of the envelope.
Similarly to the fitting of the velocity gradients described in Section \ref{c18oresult}, 
the outer radius of our model ($\tbond$$R_{\rm out}$) is adopted to be twice the standard deviation along the major axis of the fitted Gaussian distribution of the central component in the C$^{18}$O moment-0 map. 
In our model, at $r > R_{\rm d}$, the gas is assumed to be infalling and rotating with a conserved angular momentum, and the infalling velocity is assumed to be a fraction ($\tbond$$f$) of the relevant free-fall velocity. 
Thus, the infalling and rotational velocities are described as  
\begin{equation}
V_{\rm rot}(r) = \frac{j}{r},
\end{equation}
\begin{equation}\label{vin}
V_{\rm in}(r) = f \cdot \sqrt{\frac{2GM_*}{r} - V_{\rm rot}^2(r)}. 
\end{equation}
At $r < R_{\rm d}$, the gas motion in our model is assumed to be a Keplerian rotation without any infalling component: 
\begin{equation}
V_{\rm rot}(r) = \sqrt{\frac{GM_*}{r}},
\end{equation}
\begin{equation}
V_{\rm in}(r) = 0.
\end{equation}
The line-of-sight velocity is given by
\begin{equation}\label{project}
V_{\rm los}(x,y) = V_{\rm rot}(r) \cdot \sin i \cdot \frac{x}{r} + V_{\rm in}(r) \cdot \sin i \cdot \frac{y}{r} + V_{\rm sys},
\end{equation}
where $x$ and $y$ are the coordinates along the major and minor axes with respect to the center of mass, respectively, and $i$ is the inclination angle, defined as the angle between the disk plane and the outflow axis. 
The line profile ($\phi_v$) of the C$^{18}$O line as a function of velocity ($v$) is assumed to be a Gaussian function as
\begin{equation}\label{vphi}
\phi_v \propto \exp(-\frac{(v-V_{\rm los})^2}{2{\sigma_v}^2}), 
\end{equation}
where $\sigma_v$ is the velocity dispersion. 
We assume $\sigma_v$ to be the thermal dispersion as 
\begin{equation}
\sigma_v = \sqrt{\frac{2kT(r)}{m}}, 
\end{equation}
where $m$ is the C$^{18}$O molecular mass and $k$ is the Boltzmann constant. 
In our models, the thermal dispersion at a radius of 500 AU ranges from $\sim$0.07 to $\sim$0.17 km s$^{-1}$.
The critical density of the C$^{18}$O (2--1) emission is $\sim$10$^4$ cm$^{-3}$, which is lower than the typical density of protostellar envelopes and disks on hundreds of AU scale ($\gtrsim$10$^5$ cm$^{-3}$).  
Hence, the C$^{18}$O emission is likely thermalized (e.g., Yen et al.~2011).
Under the assumption of LTE, 
the C$^{18}$O (2--1) line intensity ($I_\nu$) of our model is computed with the radiative transfer equation
\begin{equation}
I_{\nu} = B_{\nu}(T) \cdot (1 - \exp^{-\tau_\nu}), 
\end{equation}
with
\begin{equation}\label{rad}
\tau_\nu = \Sigma(r)\kappa_\nu\phi_v, 
\end{equation}
where $B_{\nu}(T)$ is the Planck function at a temperature of $T$, and $\tau_\nu$ and $\kappa_\nu$ are the optical depth and absorption coefficient of the C$^{18}$O (2--1) line, respectively. 
The population and absorption coefficient of the C$^{18}$O emission are computed with
\begin{equation}\label{Nj}
\frac{N_j}{N_{\rm C^{18}O}} = \frac{(2J+1)\exp^{-hB_{\rm e} J(J+1)/kT}}{kT/hB_{\rm e}},
\end{equation}
\begin{equation}
\kappa_\nu=-\frac{c^2}{8\pi {\nu_0}^2}\frac{g_{J+1}}{g_J}\frac{N_J}{N_{\rm C^{18}O}} A_{J+1,J} (1-\exp^{-h\nu_0/kT}), 
\end{equation}
where $J$ is the lower energy level, $N_J$ is the number density at the energy level $J$, $N_{\rm C^{18}O}$ is the total number density of C$^{18}$O molecules, $h$ is the Planck constant, $B_{\rm e}$ is the rotational constant of 54.89 GHz, $c$ is the speed of light, $\nu_0$ is the rest frequency, $g_J$ is the statistical weight of the energy level $J$, and $A_{J+1,J}$ is the Einstein coefficient.

There are five free parameters in our model: $\Sigma_0$, $T_0$, $M_*$, $j$, and $f$. 
The parameter $f$ in protostellar sources is not well understood. 
Without the effect of the magnetic field, infalling motions in protostellar sources are expected to be almost free fall, i.e., $f = 1$ (e.g., Ulrich 1976; Terebey et al.~1984). 
Theoretical studies incorporating magnetic fields show that the infalling velocity in protostellar envelopes is slower than the free-fall velocity (e.g., Krasnopolsky \& K\"onigl 2002; Li et al.~2011). 
As shown with Equation \ref{vin}, $M_*$ is proportional to ${V_{\rm in}}^2/f^2+{V_{\rm rot}}^2$. 
If infalling dominates over rotational motion, the estimated protostellar mass is approximately $\propto f^{-2}$. 
With the same amount of specific angular momenta, the inferred disk radius is $\propto f^2$ (Equation \ref{rd}).
Therefore, the uncertainty in $f$ can introduce significant errors in the estimated protostellar mass and disk radius. 
In recent observations revealing Keplerian disks embedded in infalling envelopes around L1527 IRS, L1551 NE, L1551 IRS 5, TMC-1A, and L1489 IRS, 
their protostellar masses can be estimated from Keplerian rotation. 
Hence, the infalling motions observed in these envelopes can be compared against expected free-fall motions. 
In the Class 0 protostar L1527 IRS, the infalling velocity is found to be half of the free-fall velocity ($f = 0.5$; Ohashi et al.~2014), 
while those in L1551 NE, L1551 IRS 5 and TMC-1A, which are Class I protostars, are about one third ($f = 0.3$; Takakuwa et al.~2013; Chou et al.~2014; Aso et al.~2014). 
On the other hand, in the Class I protostar L1489 IRS, the infalling motion can be explained with free-fall motion with a conserved angular momentum (Yen et al.~2014). 
Therefore, we perform two sets of model fitting, fixing $f$ to (1) $f = 0.5$ and (2) $f = 1$ which is often adopted to analyze the gas kinematics around protostars (e.g., Ohashi et al.~1997; Momose et al.~1998; Lee et al.~2006). 
In each model, $R_{\rm d}$ is derived with the fitted $M_*$ and $j$ (Equation \ref{rd}). 
Then, we compute model image cubes and generate model P--V diagrams along and perpendicular to the outflow axes.
The purpose of our model is to reproduce the observed velocity structures but not the entire intensity distributions which require more sophisticated models incorporating three-dimensional temperature and density structures and detailed radiative transfer calculations. 
Therefore, the fitting is performed on the P--V diagrams along and perpendicular to the outflow axes. To that purpose, 
we subtract the model P--V diagrams from the observed P--V diagram, sum the square of the residuals and search for the minimum of that to obtain the best fit. 

\begin{figure}
\epsscale{0.9}
\plotone{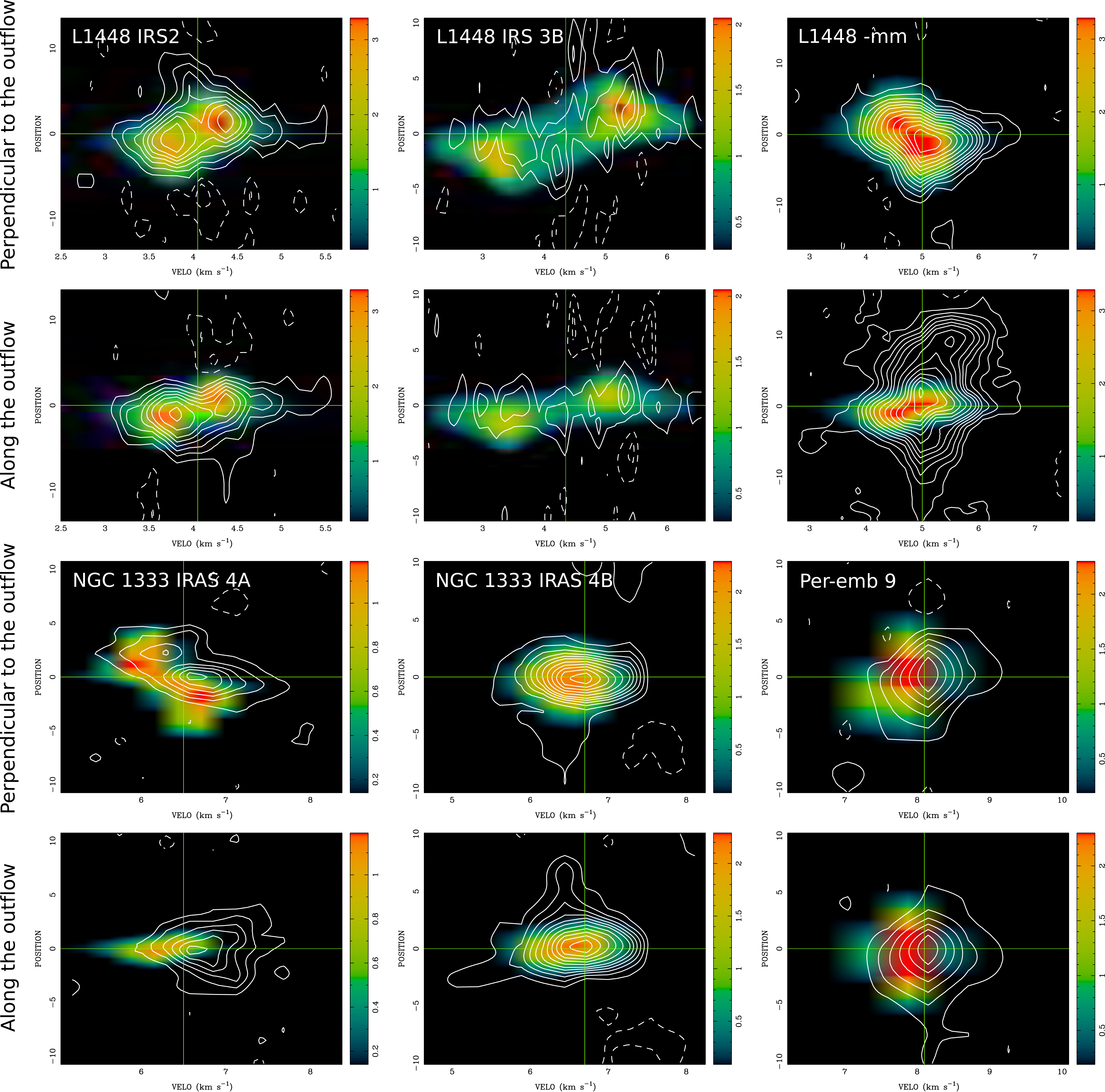}
\caption{P--V diagrams of the C$^{18}$O emission along and perpendicular to the outflow axes. Contours and color scales present the observed and model P--V diagrams. The model P--V diagrams are made from the best-fit kinematic models with $f = 0.5$. The panels in the first and third row show the P--V diagrams perpendicular to the outflow axes, and those in the second and fourth rows are along the outflow axes. Green vertical and horizontal lines denote the systemic velocity and the protostellar position, respectively. Contour levels are from 2$\sigma$ to 12$\sigma$ in steps of 2$\sigma$ and then in steps of 3$\sigma$, where the 1$\sigma$ noise levels are summarized in Table \ref{obsum1}.}
\label{fitfig}
\end{figure}

Two sets of best-fit parameters are obtained with $f=0.5$ and 1, listed as ranges of values in Table \ref{c18ofit}. 
Detailed results of individual sources are discussed in Appendix \ref{ind}.
In sources, such as Per-emb 9 and IRAS 03292$+$3039, where the rotational motion is dominant over the infalling motion, 
the fitting with the two different $f$-values results in almost the same best-fit $M_*$. 
Even though a different $f$ is adopted in the fitting, 
the best-fit models necessarily yield similar $V_{\rm rot}(r)$ and $V_{\rm in}(r)$ in order to explain the observed velocity structures.  
Hence, the fitting with different $f$ also results in similar model images, 
where the difference in peak intensity is typically less than 10\%.   
Figure \ref{fitfig} presents the comparison between the observed P--V diagrams and the best-fit model with $f$ = 0.5. 
The model images of the best-fit models with $f$ = 1 are almost identical to those with $f$ = 0.5 in a visual inspection. 
Hence, the images are not shown here. 

\begin{figure}
\epsscale{1}
\plotone{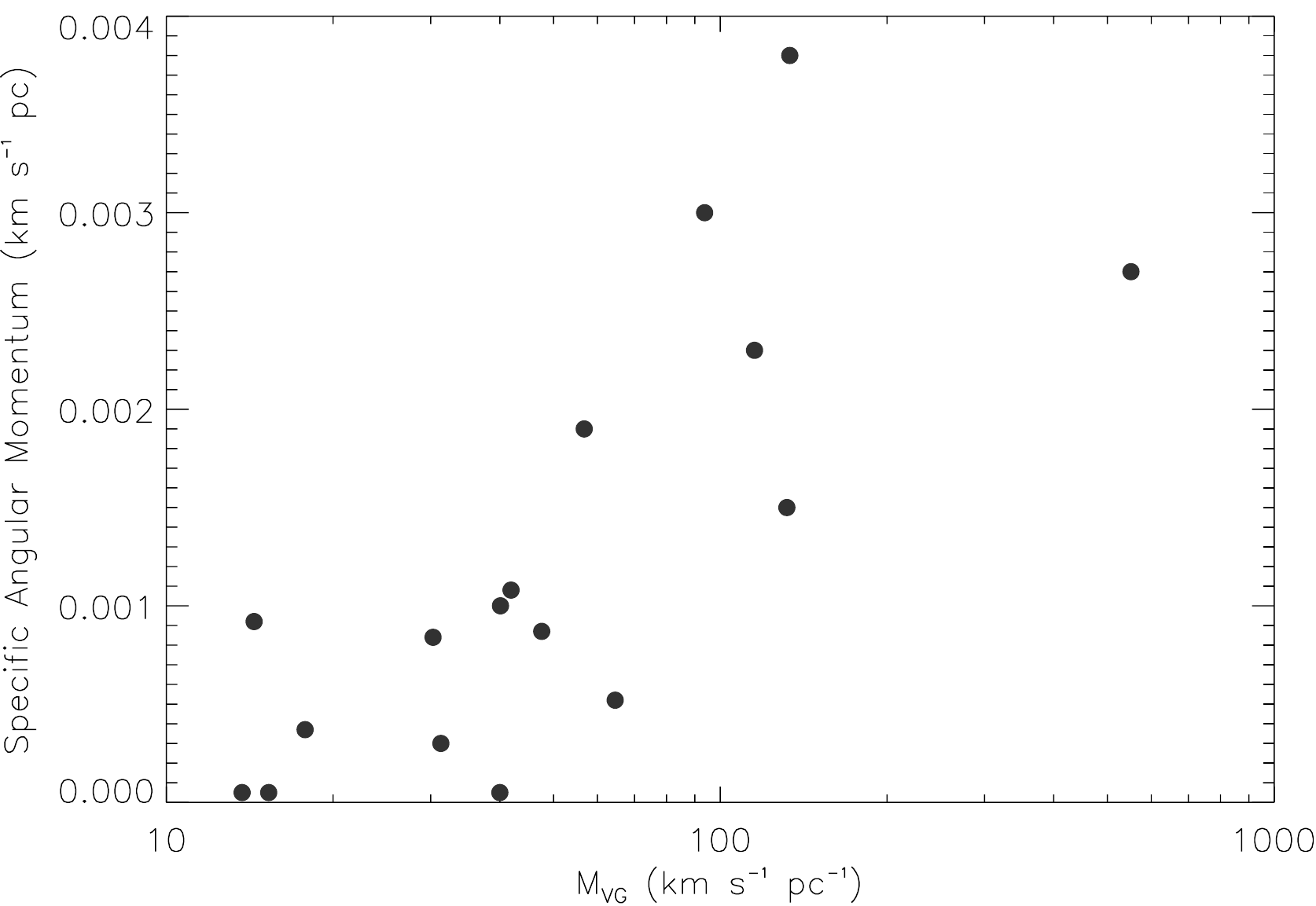}
\caption{Comparison between the observed magnitudes of the velocity gradients perpendicular to the outflow axes (Table \ref{c18ovgpa}) and the specific angular momenta estimated from the fitting with the kinematic models (Table \ref{c18ofit}).}
\label{mvgfig}
\end{figure}

The velocity gradients perpendicular to the outflow axes are typically considered as an indication of rotational motions in protostellar envelopes (e.g., Ohashi et al.~1997; Momose et al.~1998; Yen et al.~2013). 
In Figure \ref{mvgfig}, we compare the measured magnitude of the velocity gradients perpendicular to the outflow axes with the specific angular momenta estimated using our simple kinematic models.
Figure \ref{mvgfig} shows a clear correlation.
Note that the measured magnitudes originate from the projected gas motions, 
while in our kinematic models, we correct for the inclination angles to estimate the specific angular momenta.
In addition, we assume that the rotational motions are either Keplerian rotation or rotation with a conserved angular momentum in our kinematic models. 
Hence, it is expected that in individual sources,  the velocity gradients perpendicular to the outflow axes may not be fully attributed to rotational motions.
These two effects cause the data points to scatter around the correlation, as seen in Figure \ref{mvgfig}.  
Nevertheless, the obvious correlation shows that the velocity gradient perpendicular to the outflow axis can be a tracer for rotational motions on these scales.

\subsection{Uncertainty and Robustness of Fitting Results}
To assess possible uncertainties and shortcomings introduced by our simplified kinematic model and the low resolution of the data as compared to the observed disk sizes ($\sim$100--300 AU, e.g., Brinch et al.~2007; Lommen et al.~2008; Takakuwa et al.~2012), 
we compare our results of the kinematics in HH 212 and L1527 IRS with those by Lee et al.~(2014) and Ohashi et al.~(2014), respectively. 
They analyzed the kinematics of HH 212 and L1527 IRS observed with the Atacama Large Millimeter/Submillimeter Array (ALMA) at $\lesssim$1$\arcsec$ resolutions with three-dimensional models and more sophisticated radiative transfer calculations. 
In HH 212, the infalling and rotational velocities at a radius of 400 AU are estimated to be 0.9 km s$^{-1}$ and 0.35 km s$^{-1}$ with the ALMA data in HCO$^+$ (4--3), respectively. This corresponds to a protostellar mass of $\sim$0.2 $M_\sun$, a specific angular momentum of $\sim$6.8 $\times$ 10$^{-4}$ km s$^{-1}$ pc and a disk radius of 120 AU. 
Under the same assumption, namely the infalling motion being free fall as adopted by Lee et al.~(2014), 
our model fitting of the lower resolution SMA data shows a protostellar mass of  0.12 $M_\sun$, a specific angular momentum of 4.6 $\times$ 10$^{-4}$ km s$^{-1}$ pc and a disk radius of 80 AU.
Therefore, our results are consistent with the ALMA results within 50\%.
On the other hand, 
the protostellar mass and the specific angular momentum in the infalling envelope in L1527 IRS are estimated to be 0.33 $M_\sun$ and 6.1 $\times$ 10$^{-4}$ km s$^{-1}$ pc with the ALMA C$^{18}$O (2--1) line data. The disk radius is estimated to be 54 AU. 
The infalling velocity is found to be half of the free-fall velocity (Ohashi et al.~2014). 
With $f = 0.5$, our SMA model fitting shows a protostellar mass of 0.24 $M_\sun$, a specific angular momentum of 5.8 $\times$ 10$^{-4}$ km s$^{-1}$ pc and a disk radius of 70 AU in L1527 IR. 
Thus, our results are also consistent with the ALMA results within 50\%. 
We remark that the estimated $M_*$ and $j$ from the SMA data are systemically lower than those from the ALMA data.
This could be due to limited resolutions and sensitivity of the SMA to detect the inner components at higher velocities.   
However,  the comparison here suggests that our assumptions, such as the geometrically-thin approximation and the optically-thin and LTE conditions, are likely valid and do not introduce a significant error, as compared to the uncertainty due to the $f$ parameter, which can introduce a systematic error as large as a factor of four. 

We have additionally probed the dependence of our results on $p$, $q$, and $R_{\rm out}$.
Three representative cases, B335, L1527 IRS and B59\#11 with inferred disk radii $<$5 AU, $\sim$70--140 AU, and $\sim$230--340 AU, are analyzed with $p$ = $-0.5$ and $-1.5$, $q$ = $-0.2$ and $-0.8$, and varying $R_{\rm out}$ by 20\%. 
We find that the best-fit $M_*$ and $j$ with different $p$ and $q$ are consistent within 10\%. 
In the cases of B335 and B59\#11, where no extended emission surrounding the central main components and only weak extensions are found, 
the changes in the best-fit $M_*$ and $j$ due to the 20\% variation of $R_{\rm out}$ are less than 10\%.  
For L1527, the change is less than 20\%. 
The larger change in L1527 compared to those in the other two sources is likely due to the presence of the significant extended emission in this source.
We, thus, conclude that our fitting results likely do not significantly depend on the assumptions of $p$, $q$, and $R_{\rm out}$. 

Another uncertain variable is the inclination angle $i$
which is a fixed parameter in our model.
Both $V_{\rm rot}$ and $V_{\rm in}$ are approximately proportional to $1/\sin(i)$.
Hence, $M_*$ and $j$ are expected to approximately scale with $1/\sin^2(i)$ and $1/\sin(i)$, respectively. 
For the four sources, Per-emb 9, IRAS 03282$+$3035, IRAS 03292$+$3039, and Per-emb 16, whose inclination angles are estimated in this work (Appendix \ref{i}), 
we test the robustness of our results by increasing $i$ by 20$\degr$, i.e., source is closer to edge on than in our original assumptions. 
We find that $M_*$ and $j$ follow the above simple scalings within 30\%, 
except for $j$ in IRAS 03282$+$3035. This is likely because 
the velocity structures in the P--V diagram perpendicular to the outflow axis in IRAS 03282$+$3035 are only marginally resolved. 
With the larger inclination angle, the model fitting tends to interpret the velocity structures as no clear rotational motion ($j < 5 \times 10^{-5}$ km s$^{-1}$ pc) instead of the slow rotational motion ($j \sim 7 \times 10^{-4}$ km s$^{-1}$ pc) from the original model where IRAS 03282$+$3035 is closer to face on. 
In summary, our test shows that $M_*$ and $j$ 
follow the simple scaling relations with a small change in the inclination angle 
if the velocity structures are resolved. 
We finally note that these results can deviate from the simple scaling relations, 
if the uncertainty in the inclination angle is more significant due to a more complex change in the two-dimensional intensity distribution.

\section{Discussion}
The SMA observations in the C$^{18}$O (2--1) line of the 17 Class 0 or 0/I protostars show that the velocity gradients on a scale of 500 AU to 1,500 AU (4\arcsec--7\arcsec) have a wide range of magnitudes from no clear gradient ($\sim$1 km s$^{-1}$ pc$^{-1}$) to $\sim$529 km s$^{-1}$ pc$^{-1}$ together with different orientations from parallel to perpendicular to the outflow directions. 
These results suggest that the protostellar envelopes around the low-mass Class 0 and 0/I protostars likely exhibit a variety of infalling and rotational motions, even though they are at a similar evolutionary stage.
We construct simple models to reproduce the observed velocity gradients, estimate the infalling and rotational velocities, and infer the disk radii and the protostellar masses in the sample sources. 
Below, we discuss the possibility of disk growth at the Class 0 stage and the role of the magnetic field through comparison between our results, observational results of more evolved protostars, and theoretical models with and without the effect of the magnetic field. 

\subsection{Possible Sign of Disk Growth at the Class 0 Stage}\label{noB}
\begin{figure}
\epsscale{1}
\plotone{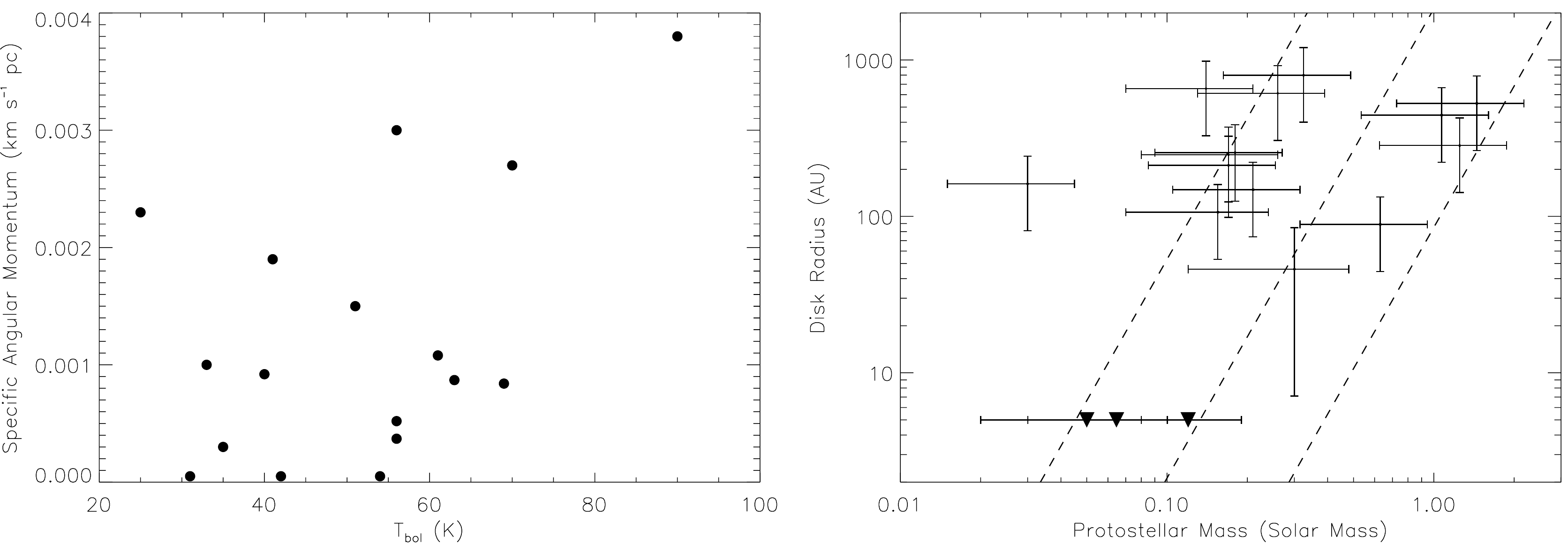}
\caption{Left panel: comparison between $T_{\rm bol}$ and the the specific angular momenta estimated from the fitting of kinematic models. Filled circles show the means of the ranges of the estimated specific angular momenta in Table \ref{poltab}. Right panel: possible disk radii as a function of estimated protostellar masses from the fitting of kinematic models. Error bars present the ranges of the best-fit values with $f = 0.5$ and $1$. If the ranges span less than 50\% of the best-fit values, the error bars present the 50\% uncertainty (see Section \ref{fitting}). Triangles denote the upper limits of the disk radii in NGC 1333 IRAS 4B, B335, and L1157-mm, where no sign of rotational motion is observed. The dashed lines show the expected trend in the theoretical model of collapsing dense cores without the effects of a magnetic field (Terebey et al.~1984) and with a sound speed of 0.2 km s$^{-1}$ and angular velocities for the core rotation of 1.5, 7.5, and 37.5 $\times$ 10$^{-14}$ s$^{-1}$ (from right to left).}
\label{rdfig}
\end{figure}

Figure \ref{rdfig} shows the estimated specific angular momenta as a function of $T_{\rm bol}$ and possible disk radii as a function of the estimated protostellar mass. 
The left panel in Figure \ref{rdfig} hints a trend that sources having higher $T_{\rm bol}$, which are expected to be more evolved, exhibit larger specific angular momenta on a 1,000-AU scale in their protostellar envelopes.
Such a trend is consistent with the theoretical model of inside-out collapsing dense cores where the angular momentum is conserved (e.g., Terebey et al.~1984). 
In this picture, the inner protostellar envelopes rotate faster as the outer material with a higher angular momentum starts to collapse. (e.g., Yen et al.~2013).
In our sample, the possible disk radii range from $<$5 AU to $>$700 AU under the assumption that the angular momenta of the infalling envelopes are conserved. 
There is no clear correlation between the possible disk radii and protostellar masses estimated from our kinematic models, as seen in the right panel in Figure \ref{rdfig}.  
Previous observations in CO emission show that disks around T Tauri stars have outer radii ranging from $\sim$100 AU to $\sim$800 AU (e.g., Guilloteau \& Dutrey 1994; Dutrey et a.~1998; Simon et al.~2000; Pi\'etu et al.~2007). 
Observations of several Class I protostars show that the outer radii of disks range from $\sim$50 AU to $\sim$300 AU (e.g., Brinch et al.~2007; Lommen et al.~2008; J{\o}rgensen et al.~2009; Takakuwa et al.~2012; Chou et al.~2014). 
In our sample of Class 0 and 0/I protostars, 11 out of the 17 protostars possibly exhibit large-scale disks with outer radii $>$100 AU up to $\sim$700 AU, comparable to the disk radii at the later evolutionary stages. 
Especially L1448 IRS 3B, Per-emb 9, IRAS 03292$+$3039, and B59\#11 show clear velocity gradients perpendicular to the outflow axes, and their rotational motions are most likely dominant over the infalling motions, suggesting the presence of Keplerian disks with outer radii of hundreds of AU (Table \ref{c18ovg} and \ref{c18ofit}). 
In addition, L1527 and HH 212 are known to exhibit Keplerian disks with an outer radius of $\sim$50--90 AU and $\sim$90--120 AU, respectively (Tobin et al.~2012a; Ohashi et al.~2014, Lee et al.~2014; Codella et al.~2014).
Therefore, our results could suggest that large-scale disks are already developed at the Class 0 stage. 
On the other hand, 
no observational signature of rotational motion is detected in NGC 1333 IRAS 4B, B335, and L1157-mm, suggesting that the disk radii (if present) in these sources are likely small ($<$5 AU).
Millimeter continuum observations at sub-arcsecond resolutions also suggest that the disk radii are smaller than 60 AU in B335 (Harvey et al.~2003) and smaller than 40 AU in L1157-mm (Chiang et al.~2012).
The presence of both large and small disks might suggest that the Class 0 stage could be the stage to build a large-scale disk. 
Note that the sample size in the present paper is still limited. 
A larger sample of protostellar sources with directly imaged disks at the Class 0 stage is required to estimate the time scale of disk growth. 

Theoretical models of disk formation in collapsing dense cores without the effect of magnetic fields, where the angular momentum is conserved, show that the size of Keplerian disks around protostars grows very fast as $R_{\rm d} \propto \omega^2c_{\rm s}t^3$, 
where $\omega$, $c_{\rm s}$, and $t$ are the angular velocity, the sound speed and the age of the collapsing cores (Terebey et al.~1984; Basu 1998). 
Here, $\omega$ describes the initial angular momentum of the dense cores, and $c_{\rm s}$ is related to the mass of the dense cores and their mass infalling rates. 
With the observed mean $\omega$ of 7.5 $\times$ 10$^{-14}$ s$^{-1}$ (e.g., Tobin et al.~2011) and typical $c_{\rm s}$ of 0.2 km s$^{-1}$ at 10 K, 
a Keplerian disk with a radius of 100 AU can form in $\sim$10$^5$ years, which is comparable to the time scale of the Class 0 stage (e.g., Enoch et al.~2009; Dunham et al.~2014a). 
The disk radii as a function of protostellar mass of this theoretical model are shown as dashed lines in Figure \ref{rdfig}b. 
Although the possible disk sizes in B335 and L1157-mm are small, 
their associated dense cores have rotational motions and masses similar within a factor of two to three to those of other sources which possibly exhibit larger disks with radii $\gtrsim$100 AU (Table \ref{comtable}).  
Their estimated protostellar masses and upper limits of disk radii are also within the expected relation between protostellar mass and disk radii in the theoretical model.  
The last source showing a slow rotational motion, NGC 1333 IRAS 4B, is in a cluster region, and its large-scale rotational motion in the associated dense core is less clear (Volgenau et al.~2006).
One possibility is that these three sources, where the disk radii are inferred to be small, are younger than the other sources,
and the disk sizes in these sources will grow as the collapse proceeds. 
Future high-resolution and high-sensitivity observations to directly image disks around Class 0 protostars are required to compare their properties with those around more evolved sources and to study disk evolution. 
Moreover, the comparison between the directly measured disk sizes, those inferred under the assumption of angular momentum conservation, and those in theoretical models incorporating magnetic fields (e.g., Mellon \& Li 2008, 2009; Machida et al.~2014) can reveal detailed physics in infalling envelopes. 

\subsection{Gas Motion vs Magnetic Braking and Magnetic Field Orientation}
MHD simulations show that the magnetic field can effectively remove the angular momenta of collapsing material and/or disks by magnetic braking and suppress the radii of Keplerian disks to within 10 AU (e.g., Mellon \& Li 2008, 2009; Machida et al.~2011; Li et al.~2011; Dapp et al.~2012). 
In our sample, L1448 IRS 3B, NGC 1333 IRAS 4A, IRAS 03292$+$3039, Per-emb 16, and B59\#11 show large velocity gradients ($M_{\rm vg} > 90$ km s$^{-1}$ pc$^{-1}$) perpendicular to their outflow axes and passing through their protostellar positions on a 1,000-AU scale.
Their specific angular momenta in their protostellar envelopes are estimated to be $>$10$^{-3}$ km s$^{-1}$ pc from our kinematic models. 
Furthermore, from the entire sample, the specific angular momenta on a 1,000-AU scale in 11 out of the 17 Class 0 or 0/I protostars are estimated to be $\gtrsim$5 $\times$ 10$^{-4}$ km s$^{-1}$ pc.
This all suggests that large-scale disks ($>$100 AU) possibly can form around these protostars.
Therefore, effective magnetic braking is unlikely common on a 1,000-AU scale among Class 0 protostellar sources. 
On the other hand, 
the disks around NGC 1333 IRAS 4B, B335, and L1157-mm likely have outer radii $<$5 AU, 
which could be the result of effective magnetic braking in these sources. 
Several mechanisms have been proposed to reduce the efficiency of magnetic braking and enable formation of large-scale disks, such as dissipation of protostellar envelopes (e.g., Machida et al. 2011), misalignment between the magnetic field and rotational axis of collapsing dense cores (e.g., Joos et al. 2012; Li et al.~2013), turbulence (e.g., Seifried et al.~2012, 2013; Joos et al.~2013), initial density profiles of dense cores and even simulation setups (e.g., Machida et al. 2014). 

The MHD simulations by Machida et al.~(2011) show that a 100-AU disk can form when its surrounding envelope is mostly dissipated and both the envelope and the disk have comparable masses, even though the parental dense core is strongly magnetized. 
In their more recent simulations with improved simulation setups (Machida et al.~2014), 
a 100-AU disk can form even when the surrounding envelope is more massive than the disk. 
In our observational results, the mass of the circumstellar material (protostellar envelope + disk) traced by the continuum emission can be considered as the upper limit of disk mass.   
L1527, surrounded by a Keplerian disk with a radius of $\sim$50--90 AU (Tobin et al.~2012a; Ohashi et al.~2014), 
and sources such as L1448 IRS 3B and L1448-mm with possible 100-AU disks, are
still embedded in their envelopes which are one order of magnitude more massive than the disks (Table \ref{comtable}). 
Therefore, envelope dissipation is unlikely an important effect to resolve the problem of magnetic braking in disk formation. 

\begin{figure}
\epsscale{1}
\plotone{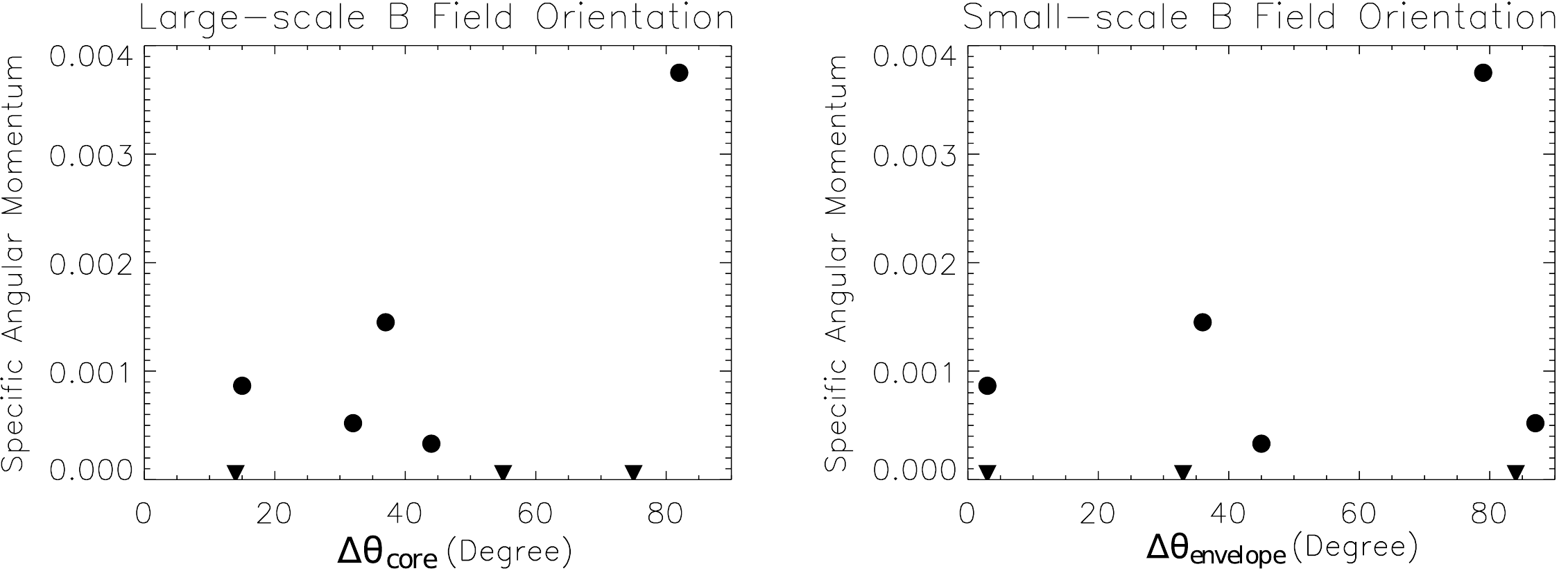}
\caption{Estimated specific angular momenta in units of km s$^{-1}$ pc compared with the misalignments between magnetic field and outflow axes at 10,000-AU (left panel) and 1,000-AU (right panel) scales. Filled circles show the means of the ranges of the estimated specific angular momenta shown in Table \ref{poltab}, and filled triangles present the upper limit of the specific angular momenta in the sources without clear signature of rotational motions. The position angles of the outflows are from the literature (Table \ref{sample}). The information of the mean magnetic field orientations is from Hull et al.~2014.}
\label{polfig}
\end{figure}

MHD simulations have also shown that if the magnetic field and rotational axis of a dense core are misaligned, 
the angular momentum can be more efficiently transported to the vicinity of protostars with infalling motions to form a large-scale Keplerian disk (e.g., Joos et al.~2012; Li et al.~2013). 
Some of our sample sources have been observed in polarized continuum emission from dust grains. Here, the polarization orientation is expected to be orthogonal to the orientation of the magnetic field (Attard et al.~2009; Matthews et al.~2009; Dotson et al.~2010; Davidson et al.~2011; Chapman et al.~2013; Stephens et al.~2013; Hull et al.~2013, 2014; Zhang et al.~2014).  
The axes of outflows are expected to trace the rotational axes in protostellar sources.  
The differences in position angles of their outflow axes and magnetic field mean orientations from Hull et al.~(2014) on 10,000- AU and 1,000-AU scales are listed in Table \ref{poltab}. 
Figure \ref{polfig} shows a comparison between the specific angular momenta estimated from our kinematic models and the outflow--magnetic field misalignments at 10,000-AU and 1,000-AU scales. 
Although there is not a very clear correlation between the angular momenta and the misalignments, 
all the data points tend to fall into a zone below a diagonal. 
The absence of sources with large specific angular momenta together with closely aligned magnetic field and outflow axes (zone above diagonal) can possibly hint that the magnetic field, if originally aligned with the rotational axis, can play a role in removing angular momentum from infalling material.
A larger sample for such a comparison is needed to reveal the genuine relation between the gas motion and the magnetic field.

Below, we compare the gas motions and magnetic field orientations in individual sources.
In L1157-mm, without a clear sign of rotational motions and with a possible disk radius 
$<$5 AU, the magnetic field orientation is well aligned with the outflow axis within 14$\degr$ from 10,000-AU to 1,000-AU scales.  
In L1448 IRS 3B, which shows a clear signature of fast rotation and possibly exhibits a large-scale disk, the magnetic field orientation and the outflow axis are clearly misaligned by more than 70\degr.  
These two sources could be examples illustrating that alignment suppresses and misalignment enables formation of large-scale disks. 
L1448-mm possibly exhibits a disk with an outer radius of $\sim$100 AU as estimated from our kinematic models, and has its magnetic field moderately misaligned with the outflow axes by $\sim$45$\degr$ at both 10,000 AU and 1,000 AU scales. 
In L1527 IRS, there is a Keplerian disk with a radius of $\sim$50--90 AU, as revealed with the ALMA observations (Tobin et al.~2012a; Ohashi et al.~2014), 
and the misalignment between the magnetic field orientation and the outflow axis increases from 32$\degr$ on a 10,000-AU scale to 87$\degr$ on a 1,000-AU scale.
The observational results of these two sources are also consistent with the theoretical expectation that the misalignment helps disk formation.  
On the other hand, in NGC 1333 IRAS 4B, there is no clear sign of rotational motion, and the magnetic field orientation is misaligned with the outflow axis by more than 60\degr. 
This result could be contradictory to the scenario where the misaligned magnetic field suppresses the efficiency of magnetic braking.
However, NGC 1333 IRAS 4B is located in a cluster region, where the large-scale rotational motion in the associated dense core and the environmental influence, such as impact by nearby outflows, are not clear. 
The other source showing no clear sign of rotational motion, B335, has complex magnetic field orientations from $>$10,000-AU to 1,000-AU scales. 
On a scale larger than 10,000 AU, the magnetic field orientation is inclined toward the outflow axis within 25$\degr$, as revealed with optical and near-infrared observations (Bertrang et al.~2014), 
while the single-dish observations at submillimeter wavelengths show that the magnetic field orientation on a scale of a few thousand AU is perpendicular to the outflow axis ($\Delta\theta_{\rm core}$ = 75\degr). 
On a 1,000-AU scale, its magnetic field is aligned with the outflow axis within 33$\degr$ as observed with CARMA at 1.3 mm. 
The effect of the magnetic field in these last two sources is not clear. 
As discussed in Section \ref{noB}, evolutionary effects might complicate the picture.
A counter-example to the scenario that disk formation requires the magnetic field to be misaligned with the rotation axis could be L1448 IRS 2.
The magnetic field orientation in L1448 IRS 2 is well aligned with the outflow axis from 10,000-AU to 1,000-AU scales within 15\degr, as in the case of L1157-mm. 
There is a clear velocity gradient perpendicular to the outflow ($M_{\rm vg} \sim 65$ km s$^{-1}$ pc$^{-1}$), which is the signature of significant rotational motion in its protostellar envelope.
In L1448 IRS 2, the specific angular momentum on a 1,000-AU scale is estimated to be $\sim 8 \times 10^{-3}$ km s$^{-1}$ pc, and the radius of the central Keplerian disk (if present) is possibly $\gtrsim$100 AU based on our kinematic model.
Nevertheless, as shown in MHD simulations, the influence of the magnetic field is also related to its strength and the ionization degree in the collapsing cores (e.g., Mellon \& Li 2008, 2009), which are both observationally difficult to measure. 
If the ionization degree in the inner envelopes is low, material can effectively decouple from the magnetic field (e.g., Padovani et al.~2013, 2014).  
In this case, efficient magnetic braking cannot occur, and large-scale disks may form.
To further observationally investigate the role of the magnetic field in disk formation, 
future observations revealing the distribution of angular momenta from 10,000-AU to 1,000-AU scales in Class 0 protostellar sources with different magnetic field orientations are essential. 

\section{Summary}
We perform imaging and analyses on SMA 1.3 mm continuum, C$^{18}$O (2--1) and $^{12}$CO (2--1) line data of 17 Class 0 and 0/I protostars to study their gas kinematics on a 1,000-AU scale in comparison with theoretical models of collapsing dense cores with and without magnetic fields. Our main results are summarized below. 

\begin{enumerate}
\item{Compact components with sizes of $\sim$100--1,400 AU are detected in the 1.3 mm continuum emission toward all the sample sources. The masses of the circumstellar material traced by the continuum emission range from 0.003 $M_\sun$ to 0.56 $M_\sun$, estimated from dust temperatures of 20--50 K. C$^{18}$O components with sizes of $\sim$500--1,500 AU associated with the continuum peaks are also seen toward all the sample sources. Our sources show a variety of velocity gradients in the C$^{18}$O emission. Their orientations range from parallel to perpendicular to the outflows, and the magnitudes are from $\sim$1 km s$^{-1}$ pc$^{-1}$ to $\sim$529 km s$^{-1}$ pc$^{-1}$. Clear velocity gradients perpendicular to the outflow axes with magnitudes larger than 90 km s$^{-1}$ pc$^{-1}$ are seen in L1448 IRS 3B, NGC 1333 IRAS 4A, IRAS 03292$+$3039, Per-emb 16, and B59\#11, suggestive of the presence of clear rotational motions.} 

\item{Assuming the C$^{18}$O (2--1) emission traces the axisymmetric flattened structures around the protostars, we construct a simple kinematic model of outer infalling and rotational motions and inner Keplerian rotation with the geometrically-thin assumption. We generate model images to explain the velocity gradients in the observed P--V diagrams along and perpendicular to the outflow axes in the C$^{18}$O (2--1) emission. With this method, we estimate the infalling and rotational velocities on a 1,000-AU scale and infer the disk radii and the protostellar masses under the assumption that the angular momentum of the infalling material is conserved. Our results show wide ranges of inferred disk radii from $<$5 AU to $>$500 AU and protostellar masses from $<$0.1 $M_\sun$ and $>$1 $M_\sun$. In particular, no sign of rotational motions is detected in NGC 1333 IRAS 4B, B335, and L1157-mm, which leads to inferred disk radii smaller than 5 AU. Rotational motions dominating over infalling motions are seen in L1448 IRS 3B, IRAS 03292$+$3039, and B59\#11, suggesting the presence of Keplerian disks with radii larger than 200 AU.}

\item{Our results suggest that both large ($>$100 AU) and small ($<$ 5 AU) disks are possibly present around Class 0 protostars, which hints a sign of disk growth at the Class 0 stage. Theoretical models without the effect of magnetic fields show that a Keplerian disk can grow to have an outer radius of 100 AU in a time scale of 10$^5$ years, which is comparable to the life time of the Class 0 stage. In this case, protostars with small disks can be younger than others, and their disk sizes will grow as the collapse proceeds. On the other hand, MHD simulations show that the magnetic field can suppress disk growth. Therefore, the small inferred disk radii in NGC 1333 IRAS 4B, B335, and L1157-mm may suggest that efficient magnetic braking is occurring in these sources. Nevertheless, 11 out of the 17 sources show signs of rotational motions with specific angular momenta larger than 5 $\times$ 10$^{-4}$ km s$^{-1}$ pc. They possibly exhibit Keplerian disks with radii larger than 100 AU. This suggests that efficient magnetic braking is unlikely common on a 1,000-AU scale among the Class 0 and 0/I sources.}

\item{The magnetic field orientations in a subset of our sample have been revealed with single-dish and interferometric polarimetric observations. By comparing the magnetic field orientations and our measured gas motions, we find no source with large specific angular momenta together with closely aligned magnetic field and outflow axes. This possibly hints that the magnetic field, if originally aligned with the rotational axis, can play a role in removing angular momentum from infalling material. L1157-mm, showing a slow rotational motion and a magnetic field aligned with the outflow, and L1448 IRS 3B, showing a fast rotational motion and a magnetic field misaligned with the outflow, are examples for a scenario where alignment suppresses and misalignment enables the formation of large-scale disks. A counter-example to this scenario could be L1448 IRS 2, which possibly exhibits a 100-AU Keplerian disk together with a magnetic field aligned with the outflow.}
\end{enumerate}

\acknowledgments
We thank all the SMA staff supporting this work. 
The SMA is a joint project between the Smithsonian Astrophysical Observatory and the Academia Sinica Institute of Astronomy and Astrophysics and is funded by the Smithsonian Institute and the Academia Sinica.
S.T. acknowledges a grant from the Ministry of Science and Technology (MOST) of Taiwan (MOST 102-2119-M-001-012-MY3) in support of this work. 
P.M.K. acknowledges support through grant MOST 103-2119-M-001-009.

\appendix
\section{$^{12}$CO (2--1) Emission}\label{12co}
\begin{figure}
\epsscale{0.9}
\plotone{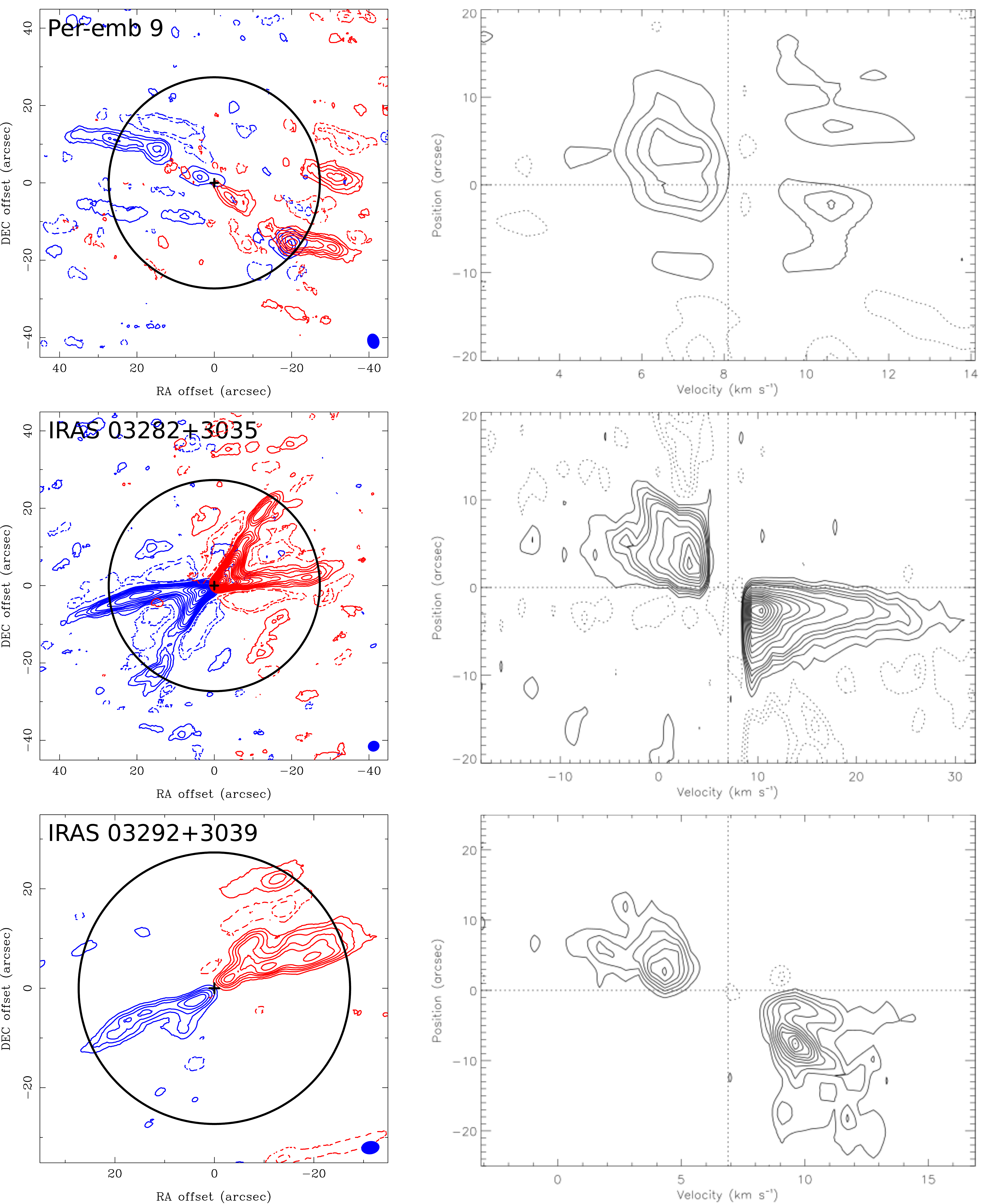}
\caption{{\em Left column.} Moment-0 maps of the blue- and red-shifted $^{12}$CO emission. A filled ellipse at the bottom-right corner in each panel presents the beam size. Crosses present the protostellar positions. Open circles show the size of the SMA primary beam at 230 GHz. Contour levels are from 3$\sigma$ to 15$\sigma$ in steps of 3$\sigma$, from 15$\sigma$ to 50$\sigma$ in steps of 5$\sigma$, and then in steps of 10$\sigma$. The beam sizes and the 1$\sigma$ noise levels are summarized in Table \ref{obsum2}. {\em Right column.} P--V diagrams of the $^{12}$CO emission along the outflow axes passing through the protostellar positions. Contour levels in the P--V diagrams of IRAS 03282$+$3035 and Lupus 3 MMS are from 2$\sigma$ to 10$\sigma$ in steps of 2$\sigma$ and then in steps of 5$\sigma$. Those of the other three sources are from 2$\sigma$ in steps of 2$\sigma$. The 1$\sigma$ noise levels are summarized in Table \ref{obsum2}.}
\label{12cofig}
\end{figure}

The outflows associated with the sample sources have been studied in details (see Table 1) except for Per-emb 9, IRAS 03282$+$3035, IRAS 03292$+$3039, Per-emb 16, and Lupus 3 MMS, where the orientation and/or inclination angles of the outflows have not been estimated before. 
The panels in the left column in Figure \ref{12cofig} show the moment-0 maps of the $^{12}$CO (2--1) emission at the red- and blue-shifted velocities in these sources. 
The red- and blue-shifted $^{12}$CO (2--1) emission in IRAS 03282$+$3035 and IRAS 03292$+$3039 show V-shaped structures with the apices approximately coincident with the protostellar positions, elongated along a northwest--southeast direction like the large-scale outflows observed with the JCMT in the $^{12}$CO (3--2) line (Hatchell et al.~2007, 2009). 
These V-shaped structures likely trace the walls of the outflow cavities and the material entrained in the outflows. 
The outflow in IRAS 03282$+$3035 has also been observed in the $^{12}$CO (1--0) emission with the OVRO (Arce \& Sargent 2006). 
Their $^{12}$CO (1--0) emission shows a structure consistent with the $^{12}$CO (2--1) emission from the SMA. 
The SMA $^{12}$CO (2--1) data of IRAS 03292$+$3039 were first published by Schnee et al.~(2012).
Our results show consistent structures with theirs.
From the V-shaped structures, the opening angles of the outflow cavities in IRAS 03282$+$3035 and IRAS 03292$+$3039 are estimated from our observations to be $\sim$45$\degr$ and $\sim$35$\degr$, respectively. 
In Per-emb 9, the $^{12}$CO (2--1) emission observed with the SMA appears to be clumpy and is distributed along the northeast--southwest direction. 
The weaker clumps located to the northwest and the southeast are not coincident with the side lobes of the stronger components, 
and are likely real emission. 
The distribution of these clumpy components could delineate V-shaped structures with an opening angle of $\sim$50$\degr$ at the blue- and red-shifted velocities. 
In Per-emb 16, the blue-shifted $^{12}$CO (2--1) emission also shows V-shape features with an opening angle of $\sim$40$\degr$, 
while it is less clear in the red-shifted emission. 
In Lupus 3 MMS, the red- and blue-shifted $^{12}$CO (2--1) emission is not aligned. 
The blue-shifted emission is elongated along the east--west direction with a position angle of $\sim$275$\degr$.
The red-shifted emission is primarily elongated along the northeast--southwest direction with a position angle of $\sim$65$\degr$.
As the distance to the protostar increases, the orientation of the eastern part becomes closer to the east--west direction. 
Part of the red-shifted emission is also seen to the west of the protostar. 
The blue-shifted emission together with the red-shifted emission in the west forms a V-shaped structure with an opening angle of $\sim$45$\degr$. 
The distribution of the $^{12}$CO (2--1) emission, with the red-shifted emission primarily to the east and the blue-shifted emission to the west, is consistent with the large-scale outflow observed with the ASTE in the $^{12}$CO (3--2) line (Dunham et al.~2014b). 
In the ASTE $^{12}$CO (3--2) map, the red- and blue-shifted emission are aligned along a position angle of $\sim$80$\degr$. 
Therefore, the $^{12}$CO (2--1) emission observed with the SMA likely traces only part of the wall of the outflow cavity in Lupus 3 MMS. 

\section{Inclination Angles}\label{i}
Except for Per-emb 9, IRAS 03282$+$3035, IRAS 03292$+$3039, Per-emb 16, and Lupus 3 MMS (Table \ref{sample}), the inclination angles of the sample sources have been estimated in the literature based on SEDs, morphology and velocity structures of outflows or proper motions of jets. 
For these sources, we investigate their inclination angles based on the morphologies and velocity structures of the outflows in the $^{12}$CO (2--1) emission. 
We compare the moment-0 maps and the P--V diagrams of the $^{12}$CO emission (Figure \ref{12cofig}) with those of model outflows having bipolar conical shapes computed by Cabrit et al.~(1986). 
For a bipolar conical outflow with an opening angle of $\theta_{\rm o.a.}$, 
if one side of the outflow only shows blue-shifted emission and the other red-shifted emission, 
its inclination angle ($\tbond$$i$) is $\theta_{\rm o.a.}/2 < i < \pi/2-\theta_{\rm o.a.}/2$. 
In addition, 
a bipolar conical outflow with a velocity structure of $V_{\rm outflow} \propto r$ shows two aligned fan-shaped structures with their apices pointing toward the protostellar position and a systemic velocity in its P--V diagram along the outflow axis. 
If the axis passing through the two fan-shaped structures is oriented closer to the axis of the systemic velocity in the P--V diagram, 
the inclination of the source is closer to edge-on. 
If the axis is oriented closer to the axis of the zero position in the P--V diagram, the inclination of the source is closer to face-on. 
The orientation of the fan-shaped structures in the P--V diagram is also related to the ratio $V_{\rm outflow}/r$. 
If the ratio is large, the axis passing through the two fan-shaped structures is oriented closer to the axis of the zero position, 
if small, closer to the axis of systemic velocity. 
Considering the projection into the line of sight, an observed $V_{\rm outflow}/r$ scales with inclination as $\tan i$ $\times$ intrinsic $V_{\rm outflow}/r$. 
The $^{12}$CO (2--1) emission in these five sources observed with the SMA most likely traces the walls of the outflow cavities. 
Typical $V_{\rm outflow}/r$ of the walls of outflow cavities in low-mass protostellar sources seen in the $^{12}$CO (2--1) emission is a few $\times$ 10$^{-3}$ km s$^{-1}$ AU$^{-1}$, such as those in B335 (Yen et al.~2010) and HH 211 (Gueth \& Guilloteau 1999). 
Based on these observational signatures, we estimate the inclination angles of these five sources to be closer to edge-on, intermediately inclined, or face-on,  
and we assume that the outflows in these five protostellar sources all have typical $V_{\rm outflow}/r$ of the order of 10$^{-3}$ km s$^{-1}$ AU$^{-1}$. 

{\em Per-emb 9.} The red- and blue-shifted $^{12}$CO (2--1) emission is seen in the west and the east, respectively.  Although it is less extended, this suggests that
the outflow lobes emit both red- and blue-shifted emission and the inclination angle is 
$\lesssim$25$\degr$ or $\gtrsim$65$\degr$, where the opening angle is $\sim$50$\degr$. 
In the P--V diagram of the $^{12}$CO (2--1) emission along the outflow axis, the emission primarily appears in the second and fourth quadrants, and the emission in the second quadrant extends to the third quadrants, which is the signature of an inclination angle of an outflow closer to 25$\degr$ as shown in Cabrit et al.~(1986). 
From the slope of the axis passing through the two strongest peaks and the center in the P--V diagram, the 
observed $V_{\rm outflow}/r$ is estimated to be $\sim$1 $\times$ 10$^{-3}$ km s$^{-1}$, which corresponds to $\sim$2 $\times$ 10$^{-3}$ km s$^{-1}$ with an inclination angle of 25\degr  which is consistent with the typical values.
Therefore, we assume the inclination angle of Per-emb 9 is 25$\degr$. 
In addition, assuming the 1.3 mm continuum emission in Per-emb 9 traces the flattened structures normal to the outflow axis around the protostar, 
from the aspect ratio of the major and minor axes, the inclination is estimated to be 29\degr, which is consistent with that from the outflow structures. 

{\em IRAS 03282$+$3035.} The eastern and western outflow lobes primarily show blue- and red-shifted emission, respectively, without a clear overlap, suggesting $23\degr \lesssim i \lesssim 68\degr$, where the opening angle is $\sim$45$\degr$.
In the P--V diagram, two fan-shaped structures point toward the upper left and the lower right along a direction with a position angle of $\sim$30\degr. 
From the slope of the axis passing through the two strongest peaks and the center in the P--V diagram, the 
observed $V_{\rm outflow}/r$ is estimated to be $\sim$6 $\times$ 10$^{-3}$ km s$^{-1}$.
Therefore, we assume that IRAS 03282$+$3035 has an intermediate inclination angle of $\sim$40\degr, which is an intermediate value between $23\degr$ and $68\degr$ in the sine function.
From the aspect ratio of the major and minor axes of the 1.3 mm conitnuum emission in IRAS 03282$+$3035, 
the inclination is estimated to be 40\degr which is consistent with the one from the outflow structures.

{\em IRAS 03292$+$3039.} Similar to IRAS 03282$+$3035, there is no clear overlap between the red- and blue-shifted emission in the outflow, suggesting $18\degr \lesssim i \lesssim 73\degr$, where the opening angle is $\sim$35$\degr$. 
The two fan-shaped structures seen in the P--V diagram are tilted along the axis with a position angle of almost 45$\degr$, suggesting IRAS 03292$+$3039 has an intermediate inclination angle between a face-on and an edge-on geometry. 
From the slope of the axis passing through the two strongest peaks and the center in the P--V diagram, the 
observed $V_{\rm outflow}/r$ is estimated to be $\sim$2 $\times$ 10$^{-3}$ km s$^{-1}$.
Therefore, the inclination angle of IRAS 03292$+$3039 is assumed to be 40$\degr$, an intermediate value between $73\degr$ and $18\degr$ in the sine function.
From the aspect ratio of the major and minor axes of the 1.3 mm conitnuum emission in IRAS 03292$+$3039, 
the inclination is estimated to be 67\degr, which is within the range we consider ($18\degr \lesssim i \lesssim 73\degr$). 

{\em Per-emb 16.} Both blue- and red-shifted emission is seen in the northern outflow lobe, suggesting its inclination angle is $\lesssim$20$\degr$ or $\gtrsim$70$\degr$, where the opening angle is $\sim$40$\degr$. 
In the P--V diagram, the emission is primarily in the first and third quadrants and extends to the forth and second quadrants, respectively, with some emission in the third quadrant also extending to the fourth quadrant. 
Overall, the emission components in the P--V diagram are tilted closer to the axis of the protostellar position than to the systemic velocity, suggesting that the inclination of Per-emb 16 is closer to face-on than edge-on.
From the slope of the axis passing through the two strongest peaks and the center in the P--V diagram, the 
observed $V_{\rm outflow}/r$ is estimated to be $\sim$2 $\times$ 10$^{-3}$ km s$^{-1}$, corresponding to $\sim$1 $\times$ 10$^{-2}$ km s$^{-1}$ with an inclination angle of 20\degr. 
Therefore, we assume that the inclination angle of Per-emb 16 is 20$\degr$.
From the aspect ratio of the major and minor axes of the 1.3 mm conitnuum emission in Per-emb 16, 
the inclination is estimated to be 33\degr, comparable to that from the outflow structures.

{\em Lupus 3 MMS.} The outflow observed with the SMA appears to be asymmetric. 
The eastern lobe is tilted away from the outflow axis as revealed with the single-dish observations (Dunham et al.~2014b) and could only trace part of the wall of the outflow cavity. 
On the other hand, 
the western lobe shows a V-shaped structure and is aligned with the outflow axis, which likely traces the conical wall of the outflow cavity. 
Hence, we discuss the inclination of Lupus 3 MMS based on the velocity structure of the western outflow lobe. 
The western outflow is seen in both red- and blue-shifted emission, suggesting $\lesssim$23$\degr$ or $\gtrsim$68$\degr$, where the opening angle is $\sim$45$\degr$. 
In the P--V diagram, the red-shifted emission of the western outflow lobe is located close to the systemic velocity, 
which is an observational signature for the inclination to be closer to edge-on than face-on, as shown by Cabrit et al.~(1986). 
From the slope of the axis passing through the two strongest peaks and the center in the P--V diagram, the 
observed $V_{\rm outflow}/r$ is estimated to be $\sim$2 $\times$ 10$^{-3}$ km s$^{-1}$, corresponding to $\sim$7 $\times$ 10$^{-4}$ km s$^{-1}$ with an inclination angle of 70\degr.
Therefore, we assume the inclination angle of Lupus 3 MMS is $\sim$70$\degr$, approximately $\sim$ $\pi/2-\theta_{\rm o.a.}/2$.

\section{Specific Comments on Individual Sources}\label{ind}
\begin{figure}
\epsscale{1}
\plotone{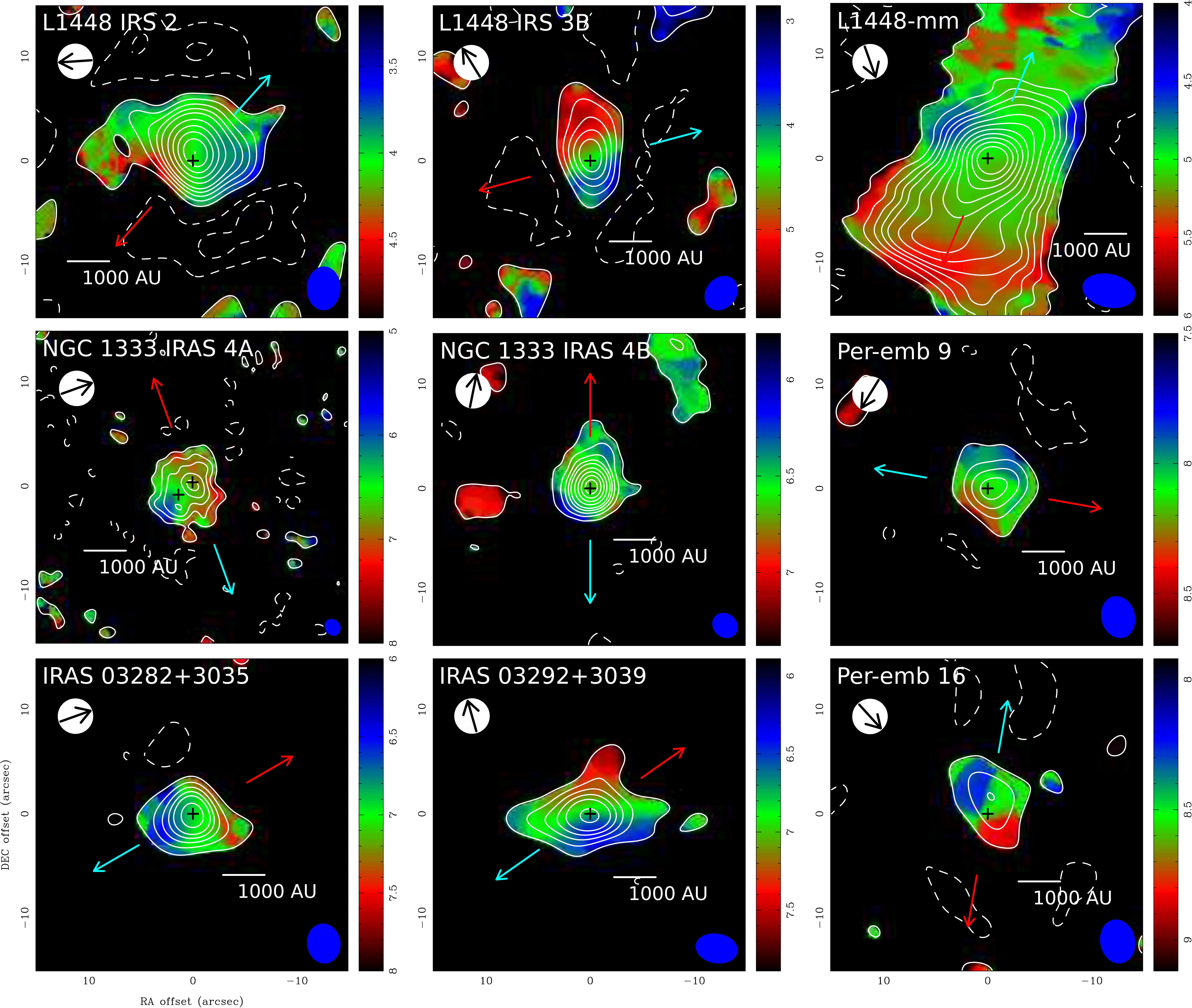}
\caption{Moment-0 maps (contours) overlaid on the moment-1 maps (color scale, units in km s$^{-1}$) of the C$^{18}$O emission. Black arrows on top of white filled circles
show the directions of the overall velocity gradients (Table \ref{c18ovg}). Filled ellipses at the bottom-right corners 
present the beam sizes. Crosses indicate the protostellar positions. Red and blue arrows denote the directions of the red- and blue-shifted outflows, respectively. Contour levels are from 3$\sigma$ to 15$\sigma$ in steps of 3$\sigma$, from 15$\sigma$ to 50$\sigma$ in steps of 5$\sigma$, and then in steps of 10$\sigma$. The beam sizes and the 1$\sigma$ noise levels are summarized in Table \ref{obsum1}. }
\label{c18ofig}
\end{figure}

The emission distributions in our model P--V diagrams tend to be less extended than those in the observed P--V diagrams because our models are only applied to the central components of the C$^{18}$O (2--1) emission. This is especially the case for L1448-mm, L1527 IRS, Lupus 3 MMS, and IRAS 16253$-$2429, where extended emission of 10$\arcsec$--20$\arcsec$ is clearly seen in addition to the central components (Figure \ref{c18ofig}). 
In L1448 IRS 2, NGC 1333 IRAS 4B, Per-emb 9, IRAS 03282$+$3035, IRAS 03292$+$3039, Per-emb 16, L1527 IRS, HH 212, B59\#11, B335, and L1157-mm, 
the morphologies, peak positions and velocity gradients of the C$^{18}$O (2--1) emission in their P--V diagrams both along and perpendicular to the outflow axes are well reproduced with our simple kinematic model. 
Below, we discuss the discrepancies between the model and observed P--V diagrams for the rest of the sources together 
with a comparison with previous results of gas motions in the individual sources.

{\em L1448 IRS 3B.} The peak positions and distributions of the red-shifted emission in the P--V diagrams both along and perpendicular to the outflow axis are reproduced with our model. Even though there is an apparent offset between
modelled and observed peak positions in the blue-shifted emission, 
the overall velocity gradients along and perpendicular to the outflow axis are reproduced with the model. A clear velocity gradient is present along the outflow axis, while no clear velocity gradient is seen perpendicular to the outflow axis. This suggests that in L1448 IRS 3B the rotational motion is dominant over the infalling motion. On large scales of $\sim$4$\arcmin$ ($\sim$60,000 AU), single-dish observations in the C$^{18}$O (1--0), H$^{13}$CO$^+$ (1--0), and N$_2$H$^+$ (1--0) lines show a velocity gradient from the north and the west to the south, and the southern region is more red-shifted than the northern and western regions (Volgenau et al.~2006). The direction of the large-scale velocity gradient is identical to the outflow direction.

{\em L1448-mm.} The observed P--V diagram perpendicular to the outflow axis is well reproduced with the model. The P--V diagram along the outflow axis shows an extended structure with a size of $\sim$25$\arcsec$, which is not included in our model fitting. The extended structure ($R > 5\arcsec$) exhibits a feature with a velocity $\propto R$, different from the central component where the velocities increase as radii decrease. The extended structure is likely related to the outflow activity. Nevertheless, the velocity feature of the central component along the outflow axis is reproduced with the model. We have measured the radial profile of the rotational velocities in L1448-mm, which is found to be $V_{\rm rot} \propto 1.3\cdot(R/100\ {\rm AU})^{-1.02}$ km s$^{-1}$, corresponding to an angular momentum of $j = 6.3 \times 10^{-4}$ km s$^{-1}$ pc, with an analytical method (Yen et al.~2013). The two independent methods show a consistent result of the rotational motion. On larger scales of thousands of AU, VLA observations in the NH$_3$ lines have found velocity gradients both along and perpendicular to the outflow axis, which are interpreted as rotational and infalling motions (Curiel et al.~1999). 

{\em NGC 1333 IRAS 4A.} The C$^{18}$O (2--1) emission is associated with the two protostellar sources (4A1 and 4A2). We adopt its emission 
peak as the center of mass in our model. 
The observed P--V diagram perpendicular to the outflow axis shows a clear velocity gradient, which can be reproduced with the model. However, the distribution of the C$^{18}$O (2--1) emission is asymmetric with respect to its emission peak (Figure \ref{c18ofig}), and hence our simple axisymmetric model shows an emission excess at offsets from $-3\arcsec$ to $-5\arcsec$ in the P--V diagram perpendicular to the outflow axis. The velocity feature in the P--V diagram along the outflow axis is not well explained with the model. This feature could be due to contamination from the outflow. The emission distribution is not symmetric around the systemic velocity. Thus, the emission excess at the blue-shifted velocities is seen in our model P--V diagram along the  outflow axis. Overall, the velocity gradient perpendicular to the outflow axis is clearer than the one along the outflow axis. This could suggest that the rotational motion is dominant over the infalling motion. In this case, the C$^{18}$O emission in NGC 1333 4A could trace a circumbinary Keplerian disk, as in the case of L1551 NE (Takakuwa et al.~2012, 2013), and the estimated protostellar mass could represent the total stellar mass of the binary system. A similar velocity gradient perpendicular to the outflow axis is also seen on larger scales of $\sim$20$\arcsec$ ($\sim$5,000 AU) in the PdBI observations in the N$_2$H$^+$ (1--0) line by Di Francesco et al.~(2001), which is interpreted as rotational motion. The observed inverse P Cygni profiles in several molecular lines suggest the presence of infalling motion which is estimated to be $\sim$0.68 km s$^{-1}$ at a radius of $\sim$270 AU (e.g., Di Francesco et al.~2001). Assuming that the infalling motion is free-fall, the infalling velocity estimated from the inverse P Cygni profiles corresponds to a protostellar mass of 0.14 $M_\sun$, which is consistent with our fitting results. The VLA observations in the NH$_3$ lines at an angular resolution of 0\farcs3--0\farcs4 (Choi et al.~2007, 2010) possibly reveal a circumstellar disk around each component. They further show that the NH$_3$ emission associated with the northern component 4A2 is red-shifted to the southeast and blue-shifted to the northwest, opposite to the direction of the velocity gradient observed with the SMA on the larger scale of 10$\arcsec$. The southern component 4A1 exhibits a velocity gradient and elongation along the northeast--southwest direction (Choi et al.~2011), different from those of 4A2 and the C$^{18}$O (2--1) emission observed with the SMA. If the velocity gradients on sub-arcsecond scales ($<$250 AU) observed with the VLA indeed trace the gas motions of the circumstellar disks around the individual binary components, the direction of the angular momentum axis likely varies from large to small scales. It is also possible that the velocity gradient seen in the P--V diagram perpendicular to the outflow axis does not reflect the gas motions but the systematic velocity offset between the two components in NGC 1333 IRAS 4A. Hence, the interpretation of the velocity gradient of the C$^{18}$O (2--1) emission observed with the SMA is still uncertain. Observations with a wide spatial dynamical range covering scales from 10$\arcsec$ to $0\farcs$5 are required to fully reveal the gas kinematics.

{\em NGC 1333 IRAS 4B.} The observed P--V diagrams both along and perpendicular to the outflow axis are well reproduced with the model. There is no clear velocity gradient seen in the P--V diagram perpendicular to the outflow axis, and hence there is no sign of rotational motion. Previous PdBI observations in the N$_2$H$^+$ (1--0) line (Di Francesco et al.~2001) and FCRAO+BIMA observations in the C$^{18}$O (1--0), H$^{13}$CO$^+$ (1--0), and N$_2$H$^+$ (1--0) lines (Volgenau et al.~2006) also show absence of detectable rotational motion, consistent with our results. 

{\em IRAS 03282$+$3035.} The single-dish observations 
in the N$_2$H$^+$ (1--0) line at an angular resolution of $\sim$27$\arcsec$ show that there is a large-scale velocity gradient with a magnitude of 1.3 km s$^{-1}$ pc$^{-1}$ and a position angle of 301$\degr$ over a 0.1-pc scale (Tobin et al.~2011). The VLA observations in the NH$_3$ lines at an angular resolution of $\sim$5$\arcsec$ show that on thousands of AU scale the velocity gradient is more significant with 8.7 km s$^{-1}$ pc$^{-1}$ with a similar position angle of 294$\degr$ (Tobin et al.~2011). A velocity gradient perpendicular to the outflow axis is also seen in these observations. The present SMA observations at the highest angular resolution of $\sim$3\farcs5 reveal an overall velocity gradient with a magnitude of 106.1 km s$^{-1}$ pc$^{-1}$ and a position angle of 291\degr on a 1,000-AU scale. In IRAS 03282$+$3035, the velocity gradients from 0.1-pc to 1,000-AU scales show similar orientations, almost parallel to the outflow, with increasing magnitudes.

{\em L1527 IRS.} The velocity features seen in the P--V diagram both  along and perpendicular to the outflow axis are well explained by our model. The infalling and rotational velocities at a radius of 2,000 AU are estimated to be 0.3 km s$^{-1}$ and 0.05 km s$^{-1}$, respectively, with the NMA observations in the C$^{18}$O (1--0) line (Ohashi et al.~1997). By extrapolating our fitting results of the gas motions at a radius $<$730 AU to the outer radii, the infalling and rotational velocities at a radius of 2,000 AU are estimated to be 0.23--0.24 km s$^{-1}$ and 0.05--0.06 km s$^{-1}$, respectively, which is approximately consistent with the NMA results. This suggests that the angular momentum of the infalling material is likely conserved from a 2,000-AU radius to an inner 730-AU radius. Indeed, the radial profile of the rotational velocities in L1527 IRS is measured to be $V_{\rm rot} \propto 1.0\cdot(R/100\ {\rm AU})^{-1.02}$ km s$^{-1}$ corresponding to an angular momentum of $j = 4.9 \times 10^{-4}$ km s$^{-1}$ pc derived with an analytical method (Yen et al.~2013).  These results are consistent with the present model fitting outcome. Sub-arcsecond-resolution observations with CARMA and ALMA reveal a Keplerian disk with an outer radius of 50--90 AU around a protostar with a mass of 0.2--0.3 $M_\sun$ (Tobin et al.~2012a; Ohashi et al.~2014). Our estimated disk radius and protostar mass with the low-resolution SMA data are comparable to those high-resolution results, if $f = 0.5$ is assumed (see also Section 5.2). 

{\em HH 212.} The SMA data were analyzed by Lee et al.~(2006) with three-dimensional models of an infalling and rotating envelope. Protostellar mass and specific angular momentum are estimated to be 0.15 $M_\sun$ and 6.7 $\times$ 10$^{-4}$ km s$^{-1}$ pc. Our results are approximately consistent with their results if $f = 1$ is adopted. Recent ALMA observations in the HCO$^+$ (4--3) and C$^{17}$O (3--2) lines reveal a Keplerian disk with a radius of $\sim$90--120 AU around a protostar with a mass of $\sim$0.2--0.3 $M_\sun$ (Lee et a.~2014; Codella et al.~2014), comparable to our estimates (see also Section 5.2). 

{\em B228.} The velocity gradient of the central C$^{18}$O component seen in the P--V diagram perpendicular to the outflow axis can be reproduced with our model, although our model P--V diagram shows two emission peaks in contrast with the observed one. In addition, there is a secondary emission peak at $V_{\rm LSR}$ of $\sim$4.4 km s$^{-1}$ in the P--V diagram along the outflow axis without any corresponding component from our model. Except that, the velocity features of the P--V diagram along the outflow axis can be explained by our model. Previous single-dish observations have found a velocity gradient along the northwest--southeast direction over a 2$\arcmin$ scale in the Lupus I cloud (Vilas-Boas et al.~2000; Tachihara et al.~2001), where the northwest region is more red-shifted and the southeastern region more blue-shifted. The direction of the large-scale velocity gradient is not the same as the one over the 10$\arcsec$ scale observed with the SMA. Tothill et al.~(2009) suggest that the large-scale velocity gradient does not reflect the gas motion but is due to a difference in systemic velocities of gas clumps located to the northwest and the southeast. 

{\em Lupus 3 MMS.} The C$^{18}$O emission is detected at relatively low signal-to-noise ratios ($\lesssim$5$\sigma$). Hence, the velocity structures are not clearly identified in the P--V diagrams. The overall distribution of the C$^{18}$O emission is elongated along the outflow direction. A possible velocity gradient, with blue-shifted emission at the positive offsets and with red-shifted emission at the negative offsets, appears in the P--V diagram perpendicular to the outflow axis. This can be explained with our model. On the other hand, no clear velocity feature is seen in the P--V diagram along the outflow axis. 

{\em IRAS 16253$-$2429.} 
Single-dish observations in the N$_2$H$^+$ (1--0) line reveal a velocity gradient with a magnitude of 1.2 km s$^{-1}$ pc$^{-1}$ over a 10,000-AU scale, where the blue-shifted emission is located to the southeast and the red-shifted emission to the northwest (Tobin et al.~2011). The N$_2$H$^+$ (1--0) emission observed with CARMA at an angular resolution of 9\farcs3 $\times$ 4\farcs9 shows a velocity gradient of 3.5 km s$^{-1}$ pc$^{-1}$ on thousands of AU scale with a direction identical to the one on the 10,000-AU scale (Tobin et al.~2011). Our SMA observations in the C$^{18}$O emission with the highest angular resolution of 7\farcs0 $\times$ 2\farcs8 reveal a velocity gradient with a magnitude of $\sim$33 km s$^{-1}$ pc$^{-1}$ in the central C$^{18}$O component on a 1,000-AU scale, where the blue-shifted emission is located to the southwest and the red-shifted emission to the northeast. 
The direction of the velocity gradient in IRAS 16253$-$2429 changes from 10,000-AU to 1,000-AU scales, and the magnitude increases.
In the P--V diagrams, the main peak positions and the velocity gradients can be explained with our model, except for the additional emission peak at $V_{\rm LSR}$ of $\sim$4.7 km s$^{-1}$, which is without any counterpart in our model, and except for the extension toward the offset of $\sim$ $-15\arcsec$ seen in the P--V diagram along the outflow axis, which is not included in our analysis.
The velocity structures of the N$_2$H$^+$ emission observed with CARMA suggest that the gas on thousands of AU scale is falling toward a protostar with a mass $\lesssim$0.1 $M_\sun$ (Tobin et al.~2011), which is consistent with our estimate. 

{\em B59\#11.} The SMA data were first presented by Hara et al.~(2013). They interpreted the entire C$^{18}$O component as a Keplerian disk and estimated the protostellar mass and the disk radius to be 0.73$^{+0.53}_{-0.39}$ $M_\sun$ and $<$350 AU, respectively. Hara et al.~(2013) measured the radial dependence of the rotational motion traced by the C$^{18}$O (2--1) emission and found $V_{\rm rot} \propto R^{-0.61}$, which is close to that of Keplerian rotation. On the other hand, that could also suggest that the rotational motion observed with the SMA is a mixture of an inner Keplerian rotation and an outer rotation with a conserved angular momentum.  Hence,  the radial dependence would be between $-0.5$ and $-1$, too. In our model, we regard the velocity gradient of the C$^{18}$O (2--1) emission as a combination of an outer infalling and rotational motion and an inner Keplerian rotation. Our results show a protostellar mass of 1.0--1.5 $M_\sun$ and a consistent disk radius of 230--340 AU. 

{\em B335.} The observed P--V diagram both along and perpendicular to the outflow axis can be reproduced with our model. No sign of rotational motion is detected with the SMA, as presented in our previous results (Yen et al.~2010, 2011, 2013). Single-dish observations of B335 in the C$^{18}$O (1--0) and (2--1) lines show a velocity gradient over a 10,000-AU scale perpendicular to the outflow axis with 0.3--0.8 km s$^{-1}$ pc$^{-1}$ (Saito et al.~1999; Yen et al.~2011). Previous NMA observations in the H$^{13}$CO$^+$ (1--0) line revealed both infalling and rotational motions on thousands of AU scale. The infalling velocity at a radius of 2,200 AU and the rotational velocity at a radius of 490 AU are both estimated to be 0.14 km s$^{-1}$ (Saito et al.~1999). By extrapolating our fitting results of the gas motions at a radius $<$750 AU to the outer radii, the infalling velocity at a radius of 2,200 AU is estimated to be 0.2 km s$^{-1}$, which is 40\% larger than the NMA results. With the more recent NMA and 45-m telescope observations in the H$^{13}$CO$^+$ (1--0) line, the magnitude of the velocity gradient over a 20,000 AU scale perpendicular to the outflow axis is measured to be 1.0 km s$^{-1}$ pc$^{-1}$. The protostellar mass is estimated to be 0.1 $M_\sun$ (Kurono et al.~2013) which is consistent with our estimate. Their model calculation to reproduce the observed NMA and 45-m images of the H$^{13}$CO$^+$ (1--0) emission suggests that the specific angular momentum within a radius of 1,000 AU is $\sim$7 $\times$ 10$^{-5}$ km s$^{-1}$ pc. This is comparable to our estimated upper limit, and hence, their results also suggest that the disk size is likely small ($<$5 AU).  

{\em L1157-mm.} No clear velocity gradient is seen in the central C$^{18}$O component in the P--V diagrams neither along nor perpendicular to the outflow axis. To reproduce the observed P--V diagrams, our model fitting results in a small protostellar mass (0.02--0.08 $M_\sun$) and a low specific angular momentum ($j$ $<$5 $\times$ 10$^{-5}$ km s$^{-1}$ pc) because the magnitude of the velocity gradient is small (Table \ref{c18ovg}). Single-dish observations of L1157-mm in the N$_2$H$^+$ (1--0) line show that there is a modest velocity gradient with 0.7 km s$^{-1}$ pc$^{-1}$ and a position angle of 303$\degr$ over a 10,000-AU scale (Tobin et al.~2011). This is about  a factor of three smaller than the typical value at the same scale ($\sim$2 km s$^{-1}$ pc$^{-1}$; Tobin et al.~2011). At a scale of thousands of AU, the CARMA and PdBI observations in the N$_2$H$^+$ (1--0) line and the VLA observations in the NH$_3$ lines at angular resolutions of $\sim$3$\arcsec$--7$\arcsec$ show velocity gradients around 4--6 km s$^{-1}$ pc$^{-1}$ and position angles of 179$\degr$--260$\degr$ (Tobin et al.~2011). The overall velocity gradient observed with the SMA in the C$^{18}$O (2--1) emission at a similar angular resolution ($\sim$3\arcsec) has a position angle of $\sim$347\degr. By comparing the velocity structures of the N$_2$H$^+$ (1--0) emission observed with CARMA with models of infalling and rotating spherical envelopes, Chiang et al.~(2010) suggest that the velocity gradient perpendicular to the outflow axis on a scale of thousands of AU can be interpreted as rigid-body rotation. On the other hand, Tobin et al.~(2012b) suggest that the velocity gradient perpendicular to the outflow axis on this scale can be due to infalling motions along the filamentary structures rather than rotational motions. On smaller scales, previous PdBI observations in the C$^{18}$O (1--0) line at an angualr resolution of $\sim$2$\arcsec$ have found a velocity gradient perpendicular to the outflow axis, shown as a red-shifted peak $\sim$2$\arcsec$ to the northeast and a blue-shifted peak $\sim$1$\arcsec$ to the west, suggesting a possible sign of the rotational motion (Gueth et al.~1997). More recent CARMA observation in the C$^{18}$O (2--1) line at a similar resolution show no clear sign of rotational motion on a 1,000-AU scale.

\begin{figure}
\epsscale{1}
\plotone{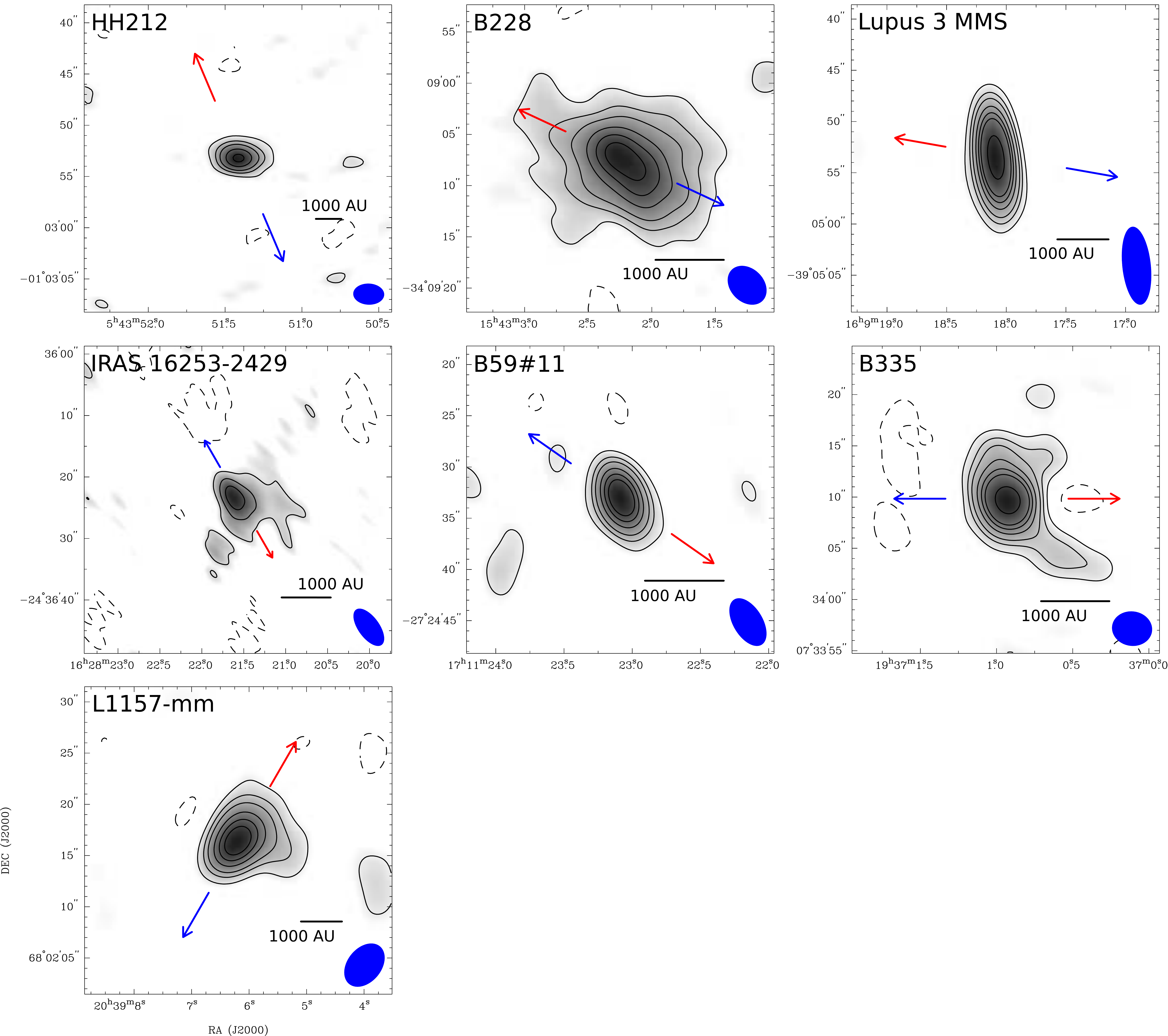}
Fig.~\ref{config}.--- Continued.
\end{figure}

\begin{figure}
\epsscale{1}
\plotone{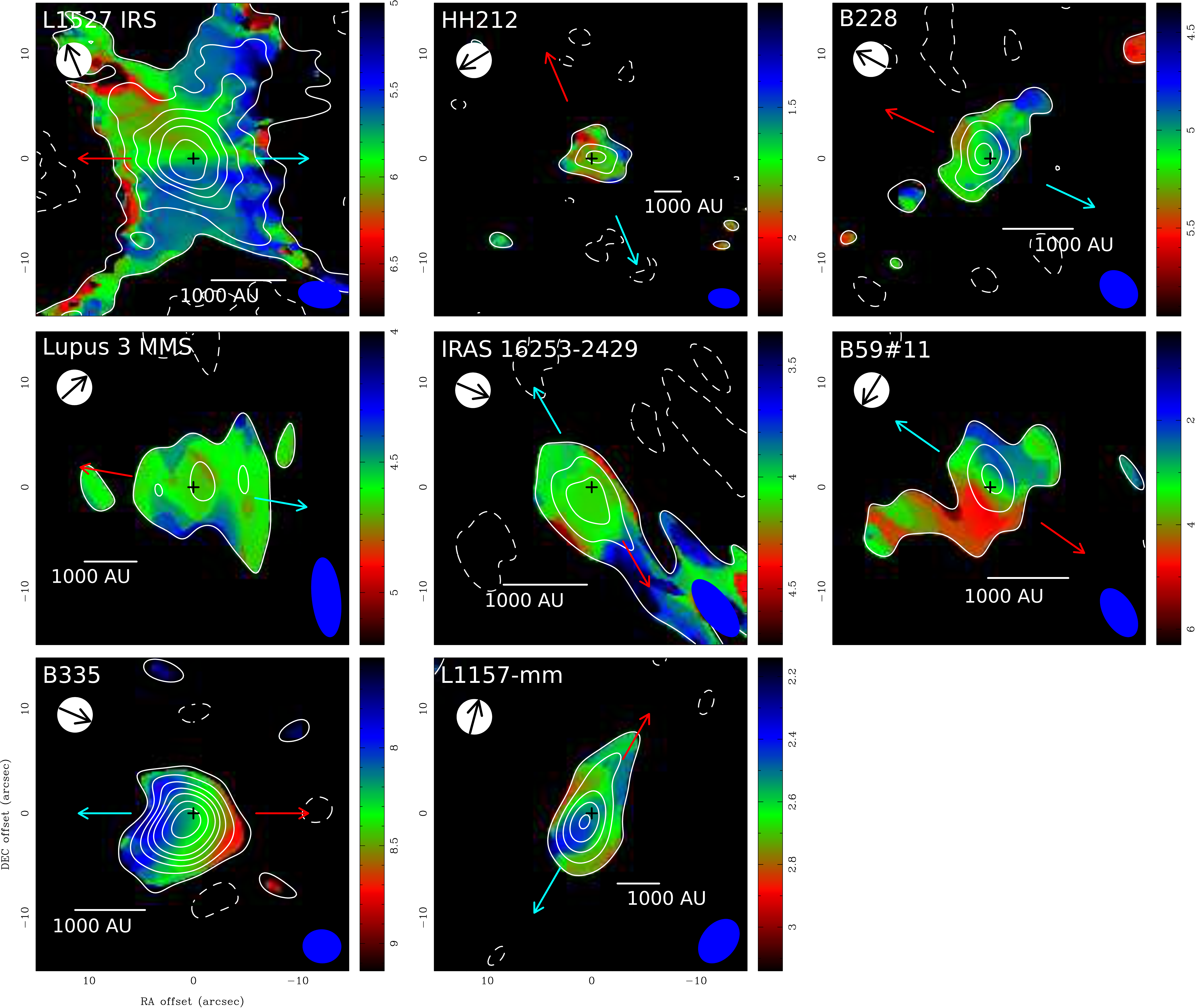}
Fig.~\ref{c18ofig}.--- Continued.
\end{figure}

\begin{figure}
\epsscale{0.9}
\plotone{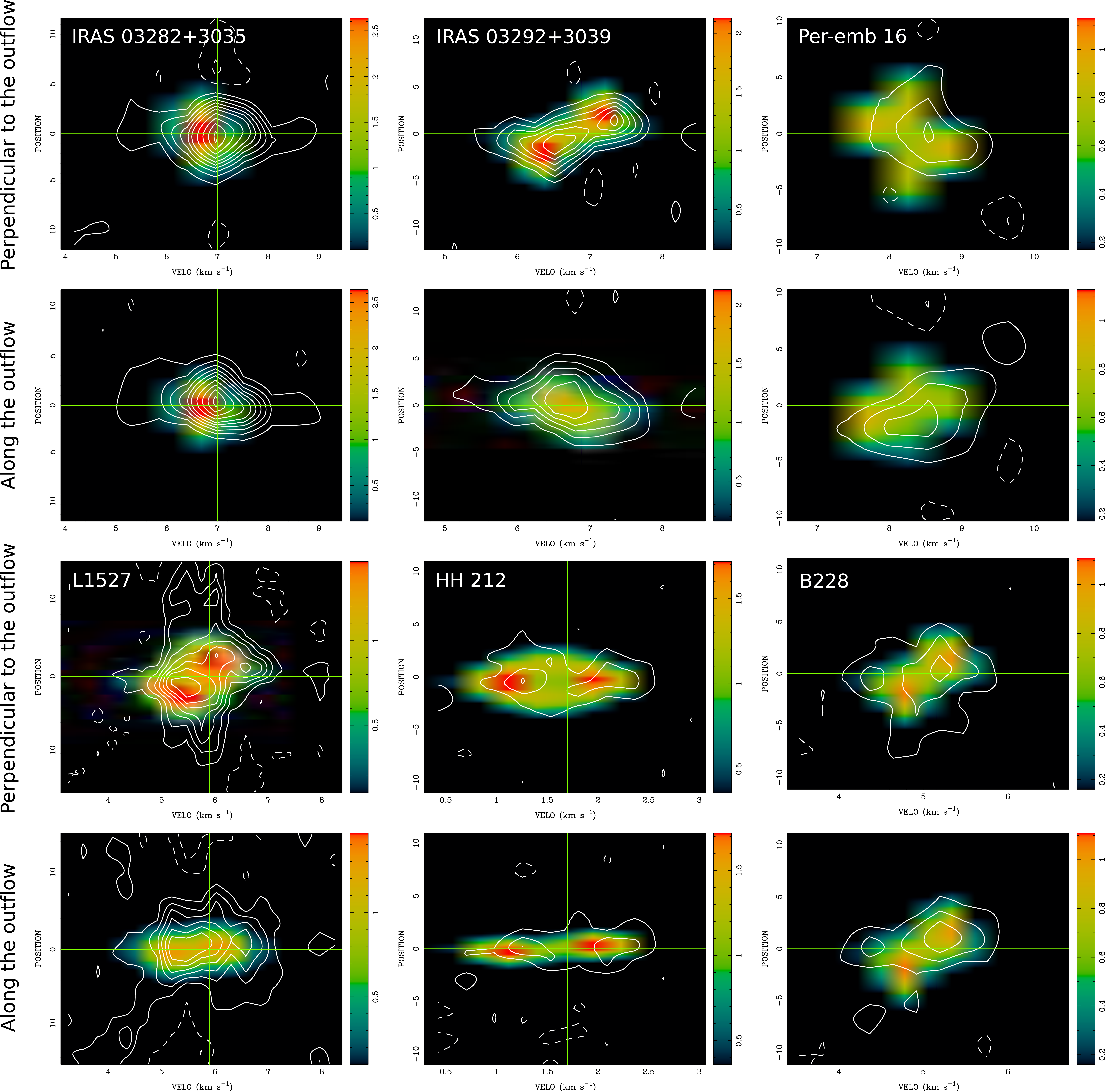}\\
Fig.~\ref{fitfig}.--- Continued.
\end{figure}

\begin{figure}
\epsscale{0.9}
\plotone{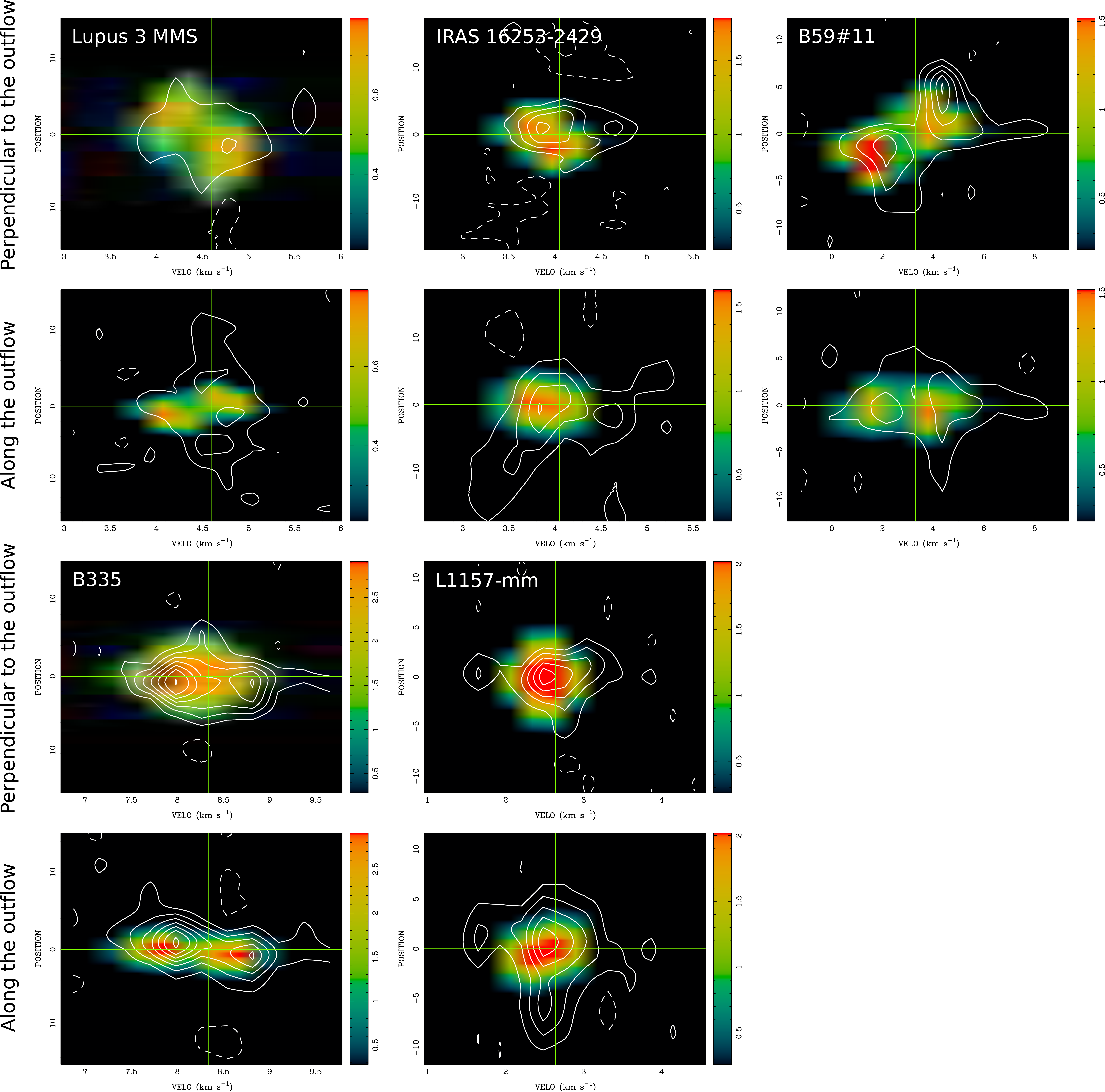}\\
Fig.~\ref{fitfig}.--- Continued.
\end{figure}

\begin{figure}
\epsscale{0.9}
\plotone{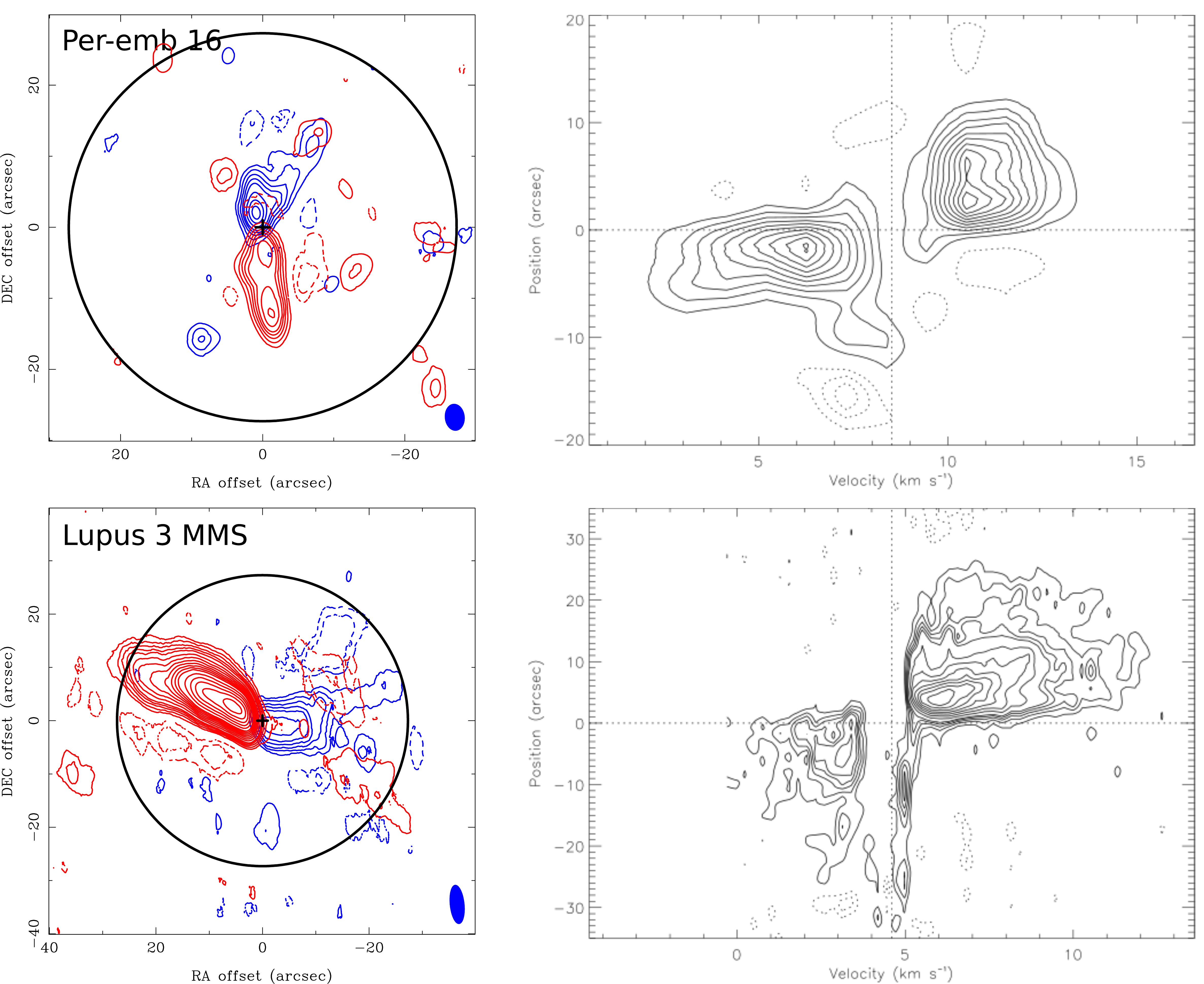}\\
Fig.~\ref{12cofig}.--- Continued.
\end{figure}

\clearpage

\begin{deluxetable}{lrrcccccccccc}
\rotate
\tablewidth{0pt}
\tablecaption{Source Sample}
\tablehead{\colhead{Source} & \multicolumn{2}{c}{Protostellar Position} & \colhead{Class} & \colhead{Distance} & \colhead{$L_{\rm bol}$} & \colhead{$T_{\rm bol}$} & $V_{\rm sys}$ & $i$\tablenotemark{a} & Outflow & Reference \\
\colhead{} & \multicolumn{2}{c}{(J2000)} & \colhead{} & \colhead{(pc)} & \colhead{($L_{\sun}$)} & \colhead{(K)} & \colhead{(km s$^{-1}$)} & \colhead{} & \colhead{(P.A.)} & \colhead{}
}
\startdata
L1448 IRS 2 & 03$^{h}$25$^{m}$22$\fs$38 & 30\arcdeg45\arcmin13$\farcs$3 & 0/I & 250 & 1.7 & 63 & 4.05 & 57\degr & 138\degr & 1,2,3,4,5\\
L1448 IRS 3B & 03$^{h}$25$^{m}$36$\fs$33 & 30\arcdeg45\arcmin14$\farcs$9 & 0/I & 250 & 4.3 & 90 & 4.35 & 63\degr & 105\degr & 1,2,3,4,6\\
L1448-mm & 03$^{h}$25$^{m}$38$\fs$87 & 30\arcdeg44\arcmin05$\farcs$4 & 0 & 250 & 4.4 & 69 & 5.0 & 70\degr & 157\degr & 1,2,7,8\\
NGC 1333 IRAS 4A & 03$^{h}$29$^{m}$10$\fs$43 & 31\arcdeg13\arcmin32$\farcs$5 & 0 & 250 & 4.2 & 51 & 6.5 & 79\degr & 20\degr & 1,2,9,10,11 \\
NGC 1333 IRAS 4B & 03$^{h}$29$^{m}$12$\fs$01 & 31\arcdeg13\arcmin08$\farcs$1 & 0 & 250 & 1.6 & 54 & 6.7 & 77\degr & 0\degr\tablenotemark{b} & 1,2,9,12\\
Per-emb 9 & 03$^{h}$29$^{m}$51$\fs$82 & 31\arcdeg39\arcmin05$\farcs$6 & 0 & 250 & 0.46 & 41 & 8.1 & 25\degr\tablenotemark{c} & 260\degr\tablenotemark{c} & 2,3 \\
IRAS 03282$+$3035 & 03$^{h}$31$^{m}$20$\fs$94 & 30\arcdeg45\arcmin30$\farcs$2 & 0 & 250 & 1.2 & 33 & 7.0 & 40\degr\tablenotemark{c} & 300\degr & 1,2,3,13 \\
IRAS 03292$+$3039 & 03$^{h}$32$^{m}$17$\fs$92 & 30\arcdeg49\arcmin47$\farcs$8 & 0 & 250 & 1.3 & 25 & 6.89 & 40\degr\tablenotemark{c} & 305\degr & 2,3,14 \\
Per-emb 16 & 03$^{h}$43$^{m}$50$\fs$99 & 32\arcdeg03\arcmin24$\farcs$3 & 0 & 250 & 0.38 & 56 & 8.52 & 20\degr\tablenotemark{c} & 170\degr\tablenotemark{c} & 2,3 \\
L1527 IRS & 04$^{h}$39$^{m}$53$\fs$91 & 26\arcdeg03\arcmin09$\farcs$8 & 0/I & 140 & 2.8 & 56 & 5.9 & 85\degr & 90\degr & 1,15,16 \\
HH 212 & 05$^{h}$43$^{m}$51$\fs$41 & $-01$\arcdeg02\arcmin53$\farcs$1 & 0 & 400 & 7.7 & $<$56 & 1.7 & 86$\degr$ & 23$\degr$ & 1,17,18,19 \\
B228 & 15$^{h}$43$^{m}$02$\fs$24 & $-34$\arcdeg09\arcmin07$\farcs$3 & 0/I & 150 & 2.3 & 61 & 5.15 & 15\degr & 65\degr & 1,20,21,22,23 \\
Lupus 3 MMS & 16$^{h}$09$^{m}$18$\fs$09 & $-39$\arcdeg04\arcmin53$\farcs$5 & 0 & 200 & 0.16 & 40 & 4.6 & 70\degr\tablenotemark{c} & 80\degr & 20,24,25,13\\
IRAS 16253$-$2429 & 16$^{h}$28$^{m}$21$\fs$59 & $-24$\arcdeg36\arcmin23$\farcs$6 & 0 & 125 & 0.25 & 35 & 4.05 & 75\degr & 210\degr & 1,2,16,26\\
B59\#11 & 17$^{h}$11$^{m}$23$\fs$08 & $-27$\arcdeg24\arcmin33$\farcs$1 & 0/I & 130 & 2.2 & 70 & 3.3 & 75$\degr$ & 235$\degr$ & 27,28,29\\
B335 & 19$^{h}$37$^{m}$00$\fs$93 & 07\arcdeg34\arcmin09$\farcs$8 & 0 & 150 & 1.5 & 31 & 8.34 & 80\degr & 270\degr & 1,30,31,32,33\\
L1157-mm & 20$^{h}$39$^{m}$06$\fs$17 & 68\arcdeg02\arcmin16$\farcs$6 & 0 & 250 & 2.7 & $<$42 & 2.64 & 80\degr & 330\degr & 1,34,16,35,36\\
\enddata
\tablenotetext{a}{The inclination angle ($i$) is defined as the angle between the disk plane and the plane of the sky, i.e., an inclination angle of 90$\degr$ corresponds to the edge-on geometry.}
\tablenotetext{b}{The direction of the jet, which could be precessing, traced by the water masers (van der Marel et al.~2013) is 29$\degr$ different from that of the outflow traced by the $^{12}$CO line (e.g., Hull et al.~2014). Here the outflow direction traced by the $^{12}$CO line is shown.}
\tablenotetext{c}{Estimated in this work.}
\tablerefs{(1) Froebrich 2005; (2) Enoch et al.~2009; (3) Kirk et al.~2007; (4) Tobin et al.~2007; (5) Wolf-Chase et al.~2000; (6) Kwon et al.~2006; (7) Curiel et al.~1999; (8) Girart \& Acord 2001; (9) Volgenau et al.~2006; (10) Choi et al.~2011; (11) Choi et al.~2006; (12) Marvel et al.~2008; (13) Dunham et al.~2014b; (14) Hatchell et al.~2007; (15) Tobin et al.~2008; (16) Tobin et al.~2011; (17) Lee et al.~2006; (18) Claussen et al.1998; (19) Lee et al.~2008; (20) Comer\'on 2008; (21) Vilas-Boas et al.~2000; (22) van Kempen et al.~2009; (23) Heyer \& Graham 1989; (24) Tachihara et al.~2007; (25) Nakajima et al.~2003; (26) van der Marel et al.~2013; (27) Brooke et al.~2007; (28) Lombardi et al.~2006; (29) Hara et al.~2014; (30) Stutz et al.~2008; (31) Chandler \& Sargent 1993; (32) Yen et al.~2011; (33) Hirano et al.~1988; (34) Chiang et al.~2010; (35) Gueth et al.~1996.}
\label{sample}
\end{deluxetable}

\begin{deluxetable}{lccc}
\tablewidth{0pt}
\tablecaption{SMA Observational Log}
\tablehead{Source & Observing Date & Configuration\tablenotemark{a} & PI}
\startdata
L1448 IRS 2 & 2007 Nov 04 & C & J. Foster \\ 
 & 2007 Nov 06 & C & J. Foster \\
L1448 IRS 3B & 2007 Nov 21 & C & J. Foster \\
L1448-mm & 2004 Nov 07 & C & J. J{\o}rgensen \\
 & 2011 Sep 12 & V & H.-W. Yen/This work \\
 & 2012 Jan 07 & S & H.-W. Yen/This work \\
NGC 1333 IRAS 4A & 2004 Nov 06 & C & J. J{\o}rgensen \\
 & 2004 Nov 22 & C & J. J{\o}rgensen \\
 & 2006 Jan 17 & E & J. J{\o}rgensen \\
 & 2011 Sep 12 & V & H.-W. Yen/This work \\
NGC 1333 IRAS 4B & 2004 Nov 06 & C & J. J{\o}rgensen \\
 & 2004 Nov 22 & C & J. J{\o}rgensen \\
 & 2006 Jan 17 & E & J. J{\o}rgensen \\
 & 2011 Sep 12 & V & H.-W. Yen/This work \\
 & 2012 Jan 07 & S & H.-W. Yen/This work \\ 
Per-emb 9 & 2009 Dec 21 & C & M. Hiramatsu \\
IRAS 03282$+$3035 & 2008 Dec 08 & C & X. Chen\\
 & 2009 Dec 25 & C & X. Chen \\
IRAS 03292$+$3039 & 2010 Oct 24 & C & S. Schnee\\
Per-emb 16 & 2009 Dec 21 & C & M. Hiramatsu \\
L1527 IRS & 2004 Nov 08 & C & J. J{\o}rgensen \\
 & 2011 Sep 09 & V & H.-W. Yen/This work \\
 & 2012 Jan 07 & S & H.-W. Yen/This work \\
HH 212 & 2004 Nov 29 & C & C.-F. Lee\\
 & 2005 Mar 19 & C & C.-F. Lee\\
B228 & 2009 Apr 29 & C & A. Hedden\\
 & 2013 Apr 30 & S & H.-W. Yen/This work \\
Lupus 3 MMS & 2013 Mar 14 & C & H.-W. Yen/This work \\
IRAS 16253$-$2429 & 2008 May 02 & C & K. Stapelfeldt\\
 & 2013 Apr 30 & S & H.-W. Yen/This work \\
B59\#11 & 2008 Mar 25 & C & T. Bourke\\
B335 & 2005 Jun 24 & C & J. J{\o}rgensen \\
L1157-mm & 2005 Jul 06 & C & J. J{\o}rgensen \\
\enddata
\tablenotetext{a}{S: Subcompact. C: Compact. E: Extended. V: Very extended.}
\label{oblog}
\end{deluxetable}

\begin{deluxetable}{llcclccc}
\rotate
\tablewidth{0pt}
\tablecaption{Summary of 1.3 mm Continuum and C$^{18}$O (2--1) Data}
\tablehead{ & \multicolumn{2}{c}{1.3 mm Continuum} & & \multicolumn{4}{c}{C$^{18}$O (2--1)} \\
\cline{2-3} \cline{5-8}
Source & Beam (P.A.) & Noise & & Beam (P.A.) & $\Delta V$\tablenotemark{b} & Noise\tablenotemark{c}  & Noise\tablenotemark{d}  \\ 
 & &  (mJy Beam$^{-1}$) & & & (km s$^{-1}$) & (Jy Beam$^{-1}$) & (Jy Beam$^{-1}$ km s$^{-1}$)
 }
\startdata
L1448 IRS 2 & 3\farcs7 $\times$ 3\farcs1 (18\degr) & 2 & & 3\farcs2 $\times$ 3\farcs2 (178\degr) & 0.14 & 0.2 & 0.1\\
L1448 IRS 3B & 3\farcs0 $\times$ 2\farcs9 (59\degr) & 5 & & 3\farcs5 $\times$ 2\farcs9 (139\degr) & 0.14 & 0.29 & 0.19\\
L1448-mm & 1\farcs4 $\times$ 1\farcs1 (78\degr) & 1 & & 5\farcs1 $\times$ 3\farcs2 (78\degr) & 0.28 & 0.11 & 0.09\\
 & 0\farcs5 $\times$ 0\farcs4 (53\degr)\tablenotemark{a} & 2 & \\ 
NGC 1333 IRAS 4A & 1\farcs2 $\times$ 1\farcs0 (72\degr) & 9 & & 1\farcs7 $\times$ 1\farcs4 (35\degr) & 0.28 & 0.14 & 0.11\\
 & 0\farcs5 $\times$ 0\farcs4 (59\degr)\tablenotemark{a} & 4 & \\ 
NGC 1333 IRAS 4B & 1\farcs4 $\times$ 1\farcs2 (73\degr) & 6 & & 2\farcs6 $\times$ 2\farcs2 (47\degr) & 0.28 & 0.09 & 0.07 \\
 & 0\farcs5 $\times$ 0\farcs4 (56\degr)\tablenotemark{a} & 3 & \\ 
Per-emb 9 & 4\farcs0 $\times$ 3\farcs0 (8\degr) & 2 & & 4\farcs0 $\times$ 3\farcs2 (15\degr) & 0.55 & 0.19 & 0.18\\
IRAS 03282$+$3035 & 3\farcs4 $\times$ 3\farcs1 (106\degr) & 1 & & 3\farcs8 $\times$ 3\farcs2 (8\degr) & 0.55 & 0.11 & 0.15\\
IRAS 03292$+$3039 & 3\farcs6 $\times$ 2\farcs6 (103\degr) & 6 & & 4\farcs0 $\times$ 2\farcs8 (80\degr) & 0.28 & 0.16 & 0.13\\
Per-emb 16 & 4\farcs1 $\times$ 3\farcs1 (7\degr) & 2 & &  4\farcs2 $\times$ 3\farcs3 (11\degr) & 0.55 & 0.17 & 0.17\\
L1527 IRS & 1\farcs6 $\times$ 1\farcs3 (81\degr) & 1 & & 4\farcs2 $\times$ 2\farcs5 (80\degr) & 0.28 & 0.1 & 0.1\\
 & 0\farcs5 $\times$ 0\farcs4 (55\degr)\tablenotemark{a} & 2 & \\ 
HH 212 & 3\farcs0 $\times$ 2\farcs0 (88\degr) & 4 & & 3\farcs0 $\times$ 1\farcs9 (82\degr) & 0.28 & 0.36 & 0.22\\
B228 & 4\farcs2 $\times$ 3\farcs2 (45\degr) & 2 & & 4\farcs1 $\times$ 3\farcs1 (46\degr) & 0.28 & 0.16 & 0.1 \\
Lupus 3 MMS & 7\farcs6 $\times$ 2\farcs7 (6\degr) & 4 & & 7\farcs6 $\times$ 2\farcs8 (6\degr) & 0.28 & 0.21 & 0.12\\
IRAS 16253$-$2429 & 7\farcs0 $\times$ 3\farcs4 (36\degr) & 2 & & 7\farcs0 $\times$ 2\farcs8 (37\degr) & 0.28 & 0.22 & 0.15\\
B59\#11 & 5\farcs1 $\times$ 2\farcs8 (31\degr) & 14 & & 5\farcs2 $\times$ 2\farcs9 (31\degr) & 1.11 & 0.21 & 0.5\\
B335 & 3\farcs9 $\times$ 3\farcs2 (82\degr) & 2 & & 3\farcs7 $\times$ 3\farcs2 (87\degr) & 0.28 & 0.28 & 0.14\\
L1157-mm & 4\farcs7 $\times$ 3\farcs3 (131\degr) & 5 & & 4\farcs7 $\times$ 3\farcs3 (131\degr) & 0.28 & 0.26 & 0.18
\enddata
\tablenotetext{a}{High-resolution maps made only with the data obtained with the very extended configuration. Other 1.3 mm continuum maps are made combining all the available data.}
\tablenotetext{b}{Velocity resolution.}
\tablenotetext{c}{1$\sigma$ noise level per channel.}
\tablenotetext{d}{1$\sigma$ noise level of moment 0 maps presented in Figure \ref{c18ofig}.}
\label{obsum1}
\end{deluxetable}

\begin{deluxetable}{llcccc}
\rotate
\tablewidth{0pt}
\tablecaption{Summary of $^{12}$CO (2--1) Data}
\tablehead{
Source & Beam (P.A.) & $\Delta V$ (km s$^{-1}$)\tablenotemark{a} & Noise\tablenotemark{b} & Noise at Blueshifted Velocity\tablenotemark{c} & Noise at Redshifted Velocity\tablenotemark{c} \\
 & & (km s$^{-1}$) & (Jy Beam$^{-1}$) & (Jy Beam$^{-1}$ km s$^{-1}$) & (Jy Beam$^{-1}$ km s$^{-1}$)}
\startdata
Per-emb 9 & 3\farcs9 $\times$ 3\farcs0 (16\degr) & 1.06 & 0.16 & 0.45 & 0.51 \\
IRAS 03282$+$3035 & 2\farcs9 $\times$ 2\farcs6 (111\degr) & 1.06 & 0.06 & 0.25 & 0.27\\
IRAS 03292$+$3039 & 3\farcs6 $\times$ 2\farcs6 (108\degr) & 0.53 & 0.17 & 0.3 & 0.32\\
Per-emb 16 & 3\farcs7 $\times$ 2\farcs7 (6\degr) & 1.06 & 0.21 & 0.59 & 0.5 \\
Lupus 3 MMS & 7\farcs3 $\times$ 2\farcs7 (6\degr) & 0.26 & 0.23 & 0.24 & 0.36
\enddata
\tablenotetext{a}{Velocity resolution.}
\tablenotetext{b}{1$\sigma$ noise level per channel.}
\tablenotetext{c}{1$\sigma$ noise level of moment 0 maps presented in Figure \ref{12cofig}.}
\label{obsum2}
\end{deluxetable}

\begin{deluxetable}{lclc}
\tablewidth{0pt}
\tablecaption{1.3 mm Continuum Emission}
\tablehead{Source & Flux (mJy) & Deconvolved Size (P.A.) & Mass ($M_\sun$)}
\startdata
L1448 IRS 2 & 189 & 3\farcs5 $\times$ 2\farcs0 (58\degr) & 0.046\\
L1448 IRS 3B & 798 & 2\farcs3 $\times$ 1\farcs7 (37\degr) & 0.2\\
L1448-mm & 190 & 0\farcs9 $\times$ 0\farcs8 (33\degr) & 0.11\\ 
& 118\tablenotemark{a} & 0\farcs3 $\times$ 0\farcs1 (72\degr) & 0.023\\
NGC 1333 IRAS 4A1 & 1625 & 1\farcs5 $\times$ 1\farcs0 (57\degr) & 0.56\\
& 891\tablenotemark{a}  & 0\farcs7 $\times$ 0\farcs5 (72\degr) & 0.17 \\
NGC 1333 IRAS 4A2 & 1446 & 1\farcs9 $\times$ 1\farcs7 (97\degr) & 0.35\\
& 447\tablenotemark{a}  & 0\farcs8 $\times$ 0\farcs5 (132\degr) & 0.085\\
NGC 1333 IRAS 4B & 893 & 1\farcs4 $\times$ 0\farcs8 (106\degr) & 0.22\\
& 588\tablenotemark{a}  & 0\farcs7 $\times$ 0\farcs5 (106\degr) & 0.14\\
Per-emb 9 & 41 & 4\farcs1 $\times$ 3\farcs6 (52\degr) & 0.008\\
IRAS 03282$+$3035 & 290 & 1\farcs6 $\times$ 1\farcs2 (70\degr) & 0.16\\
IRAS 03292$+$3039 & 546 & 1\farcs8 $\times$ 0\farcs7 (32\degr) & 0.1\\
Per-emb 16 & 45 & 4\farcs4 $\times$ 3\farcs7 (71\degr) & 0.026 \\
L1527 IRS & 204 & 1\farcs2 $\times$ 0\farcs8 (66\degr) & 0.036 \\
& 135\tablenotemark{a} & 0\farcs5 $\times$ 0\farcs2 (4\degr) & 0.008\\
HH 212 & 116 & 2\farcs2 $\times$ 1\farcs3 (84\degr) & 0.057\\
B228 & 252 & 9\farcs0 $\times$ 5\farcs5 (51\degr) & 0.022\\
Lupus 3 MMS & 200\tablenotemark{b} & \nodata & 0.024\\
IRAS 16253$-$2429 & 42 & 4\farcs7 $\times$ 4\farcs4 (6\degr) & 0.003\\
B59\#11 & 611 & 2\farcs6 $\times$ 1\farcs1 (159\degr) & 0.057\\
B335 & 170 & 4\farcs3 $\times$ 1\farcs9 (13\degr) & 0.012\\
L1157-mm & 229 & 3\farcs3 $\times$ 2\farcs6 (133\degr) &0.13\\
\enddata
\tablenotetext{a}{Fitting results of the high-resolution maps.}
\tablenotetext{b}{Failed to deconvolve.}
\label{contable}
\end{deluxetable}

\begin{deluxetable}{lccccc}
\rotate
\tablewidth{0pt}
\tablecaption{Overall Velocity Gradient of the C$^{18}$O (2--1) Emission}
\tablehead{
Source & $M_{\rm vg}$ (km s$^{-1}$ pc$^{-1}$) & $M_{\rm vg}$$\tablenotemark{a}$ (km s$^{-1}$ over 1,000 AU) & $V_{\rm C}$ (km s$^{-1}$) & Direction & $\Delta\theta\tablenotemark{b}$}
\startdata
L1448 IRS 2  & 65.3$\pm$0.2 & 0.317$\pm$0.001 & 4.07$\pm$0.01 & 92\fdg5$\pm$0\fdg2 & 46\degr \\ 
L1448 IRS 3B & 183.0$\pm$0.4 & 0.888$\pm$0.002 & 4.44$\pm$0.01  & 32\fdg3$\pm$0\fdg2 & 73\degr  \\
L1448-mm & 26.1$\pm$0.2 & 0.127$\pm$0.001 & 5.04$\pm$0.01 & 199\fdg1$\pm$0\fdg3 & 42\degr \\
NGC 1333 IRAS 4A & 116.1$\pm$0.6 & 0.564$\pm$0.003 & 6.80$\pm$0.01 & 287\fdg7$\pm$0\fdg3 & 88\degr \\
NGC 1333 IRAS 4B & 27.3$\pm$0.8 & 0.132$\pm$0.004 & 6.61$\pm$0.01 & 346\fdg6$\pm$1\fdg8 & 13\degr\\
Per-emb 9 & 58.3$\pm$0.5 & 0.283$\pm$0.003 & 8.15$\pm$0.01 & 147\fdg4$\pm$0\fdg6 & 67\degr\\
IRAS 03282$+$3035 & 106.1$\pm$1.0 & 0.515$\pm$0.005 & 6.96$\pm$0.01 & 290\fdg5$\pm$0\fdg5 & 10\degr\\
IRAS 03292$+$3039 & 110.6$\pm$0.3 & 0.537$\pm$0.002 & 6.82$\pm$0.01 & 15\fdg6$\pm$0\fdg2 & 71\degr\\
Per-emb 16 & 77.7$\pm$0.6 & 0.377$\pm$0.003 & 8.57$\pm$0.1 & 220\fdg1$\pm$0\fdg8 & 50\degr\\
L1527 IRS & 71.6$\pm$0.3 & 0.347$\pm$0.002 & 5.81$\pm$0.01 & 22\fdg1$\pm$0\fdg4 & 68\degr \\
HH 212 & 21.5$\pm$1.0 & 0.104$\pm$0.005 & 1.78$\pm$0.01 & 120\fdg8$\pm$2\fdg4 & 82\degr\\
B228 & 90.2$\pm$1.3 & 0.438$\pm$0.006 & 5.05$\pm$0.01 & 63\fdg9$\pm$0\fdg7 & 1\degr\\
Lupus 3 MMS & 1.2$\pm$0.2 & 0.006$\pm$0.001 & 4.60$\pm$0.01 & 310\fdg3$\pm$11\fdg9 & 50\degr\\
IRAS 16253$-$2429 & 32.9$\pm$0.6 & 0.160$\pm$0.003 & 4.08$\pm$0.01 & 246\fdg7$\pm$1\fdg3 & 37\degr\\
B59\#11 & 528.5$\pm$1.7 & 2.565$\pm$0.008 & 3.69$\pm$0.01 & 149\fdg1$\pm$0\fdg3 & 86\degr\\
B335 & 132.9$\pm$0.5 & 0.645$\pm$0.002 & 8.27$\pm$0.01 & 257\fdg9$\pm$0\fdg2 & 12\degr\\
L1157-mm & 9.2$\pm$0.4 & 0.045$\pm$0.002 & 2.61$\pm$0.01 & 346\fdg5$\pm$3\fdg5 & 17\degr\\
\enddata
\label{c18ovg}
\tablenotetext{a}{The magnitude of the velocity gradient scaled to be over 1,000 AU.}
\tablenotetext{b}{The difference in the position angles between the directions of the outflows and the velocity gradients.}
\end{deluxetable}

\begin{deluxetable}{lccc}
\tablewidth{0pt}
\tablecaption{Velocity Gradient of the C$^{18}$O (2--1) Emission Perpendicular to the Outflow Axes}
\tablehead{
Source & $M_{\rm vg \perp}$ (km s$^{-1}$ pc$^{-1}$) & $M_{\rm vg \perp}\tablenotemark{a}$ (km s$^{-1}$ over 1,000 AU) & $V_{\rm C}$ (km s$^{-1}$)}
\startdata
L1448 IRS2 & 47.6$\pm$0.8 & 0.231$\pm$0.002 & 4.08$\pm$0.01 \\
L1448 IRS 3B & 133.5$\pm$2.3 & 0.648$\pm$0.011 & 4.44$\pm$0.01 \\
L1448-mm & 30.3$\pm$0.8 & 0.147$\pm$0.003 & 5.05$\pm$0.01 \\
NGC 1333 IRAS 4A & 131.9$\pm$2.6 & 0.640$\pm$0.013 & 6.63$\pm$0.01 \\
NGC 1333 IRAS 4B & 15.3$\pm$3.5 & 0.074$\pm$0.017 & 6.61$\pm$0.01 \\
Per-emb 9 & 56.8$\pm$2.8 & 0.276$\pm$0.013 & 8.19$\pm$0.01 \\
IRAS 03282$+$3035 & 40.1$\pm$4.2 & 0.195$\pm$0.020 & 6.88$\pm$0.01 \\
IRAS 03292$+$3039 & 115.3$\pm$1.8 & 0.560$\pm$0.009 & 6.71$\pm$0.01 \\
Per-emb 16 & 93.7$\pm$6.5 & 0.455$\pm$0.032 & 8.57$\pm$0.01 \\
L1527 IRS & 64.6$\pm$1.7 & 0.313$\pm$0.008 & 5.81$\pm$0.01 \\
HH 212 & 17.8$\pm$3.3 & 0.086$\pm$0.016 & 1.75$\pm$0.01 \\
B228 & 41.9$\pm$5.5 & 0.204$\pm$0.027 & 5.04$\pm$0.01 \\
Lupus 3 MMS & 14.4$\pm$2.8 & 0.070$\pm$0.013 & 4.60$\pm$0.01 \\
IRAS 16253$-$2429 & 31.3$\pm$4.9 & 0.152$\pm$0.024 & 4.13$\pm$0.01 \\
B59\#11 & 551.4$\pm$8.0 & 2.677$\pm$0.039 & 3.89$\pm$0.01 \\
B335\tablenotemark{b} & 40.0$\pm$2.3 & 0.194$\pm$0.011 & 8.25$\pm$0.01 \\
L1157-mm & 13.7$\pm$2.8 & 0.067$\pm$0.013 & 2.57$\pm$0.01 \\ 
\enddata
\label{c18ovgpa}
\tablenotetext{a}{The magnitude of the velocity gradient scaled to be over 1,000 AU.}
\tablenotetext{b}{The direction of the velocity gradient is opposite to that of the rotational motion observed on thousands of AU scale.}
\end{deluxetable}

\begin{deluxetable}{lrrrrrr}
\rotate
\tablewidth{0pt}
\tablecaption{Fitting Results of the C$^{18}$O Emission}
\tablehead{Source & $M_*$ ($M_\sun$) & $j$ (km s$^{-1}$ pc) & $R_{\rm d}$ (AU) & $\Sigma_0$ (cm$^{-2}$) & $T_0$ (K) & $R_{\rm out}$\tablenotemark{a} (AU)} 
\startdata
L1448 IRS 2 & 0.1--0.26 & 8.3--9.0 $\times$ 10$^{-4}$ & 130--390 & 1 $\times$ 10$^{16}$ & 40--50 & 1080 \\
L1448 IRS 3B & 1.0--1.9 & 3.6--3.9 $\times$ 10$^{-3}$ & 330--720 & 2 $\times$ 10$^{16}$ & 30--40 & 1170 \\
L1448-mm & 0.11--0.31 & 5.6--10.2 $\times$ 10$^{-4}$ & 140--160  & 4--10 $\times$ 10$^{20}$ & 20 & 1230 \\ 
NGC 1333 IRAS 4A & 0.14--0.15 & 1.4--1.5 $\times$ 10$^{-3}$ & 630--720 & 5--7 $\times$ 10$^{16}$ & 30 & 1250\\
NGC 1333 IRAS 4B & 0.03--0.1 & $<$5 $\times$ 10$^{-5}$ & $<$5 & 1--2 $\times$ 10$^{17}$ & 40 & 660 \\
Per-emb 9 & 0.26 & 1.9 $\times$ 10$^{-3}$ & 690 & 5 $\times$ 10$^{15}$ & 50 & 1040 \\
IRAS 03282$+$3035 & 0.1--0.24 & 7.0--8.3 $\times$ 10$^{-4}$ & 100--330 & 7 $\times$ 10$^{15}$ & 50 & 740 \\
IRAS 03292$+$3039 & 0.32--0.33 & 2.2--2.4 $\times$ 10$^{-3}$  & 750--850 & 7 $\times$ 10$^{15}$ & 50 & 1040\\
Per-emb 16 & 0.74--1.4 & 2.7--3.2 $\times$ 10$^{-3}$ & 240--640 & 2 $\times$ 10$^{15}$ & 30 & 1380 \\
L1527 IRS & 0.07--0.24 & 4.6--5.8 $\times$ 10$^{-4}$ & 70--150 & 5--10 $\times$ 10$^{22}$ & 20 & 730 \\
HH 212 & 0.12--0.48 & 2.7--4.6 $\times$ 10$^{-4}$ & 10--80 & 1 $\times$ 10$^{17}$ & 50 & 1320 \\
B228 & 0.32--0.94 & 6.1--15.5 $\times$ 10$^{-4}$ & 60--120 & 1 $\times$ 10$^{15}$ & 30--40 & 590 \\
Lupus 3 MMS & 0.08--0.26 & 5.8--12.6 $\times$ 10$^{-4}$ & 200--290 & 1 $\times$ 10$^{15}$ & 20--30 & 1480 \\
IRAS 16253$-$2429 & 0.02--0.04 & 2.7--3.2 $\times$ 10$^{-4}$ & 90--240 & 4--5 $\times$ 10$^{15}$ & 40--50 & 670 \\
B59\#11 & 1.0--1.5 & 2.7 $\times$ 10$^{-3}$ & 230--340 & 4--9 $\times$ 10$^{16}$ & 30 & 640 \\
B335 & 0.05--0.19 & $<$5 $\times$ 10$^{-5}$ & $<$5 & 3 $\times$ 10$^{16}$ & 50 & 750 \\
L1157-mm & 0.02--0.08 & $<$5 $\times$ 10$^{-5}$ & $<$5 & 9 $\times$ 10$^{16}$ & 20 & 1060 \\
\enddata
\tablecomments{The ranges correspond to the fitting results with $f = 0.5$ and $f = 1$.}
\tablenotetext{a}{The outer boundary of our model, which is a fixed parameter in the fitting.}
\label{c18ofit}
\end{deluxetable}

\begin{deluxetable}{lccccc}
\tablewidth{0pt}
\tablecaption{Comparison with Rotation and Mass of Associated Dense Cores}
\tablehead{Source & $R_{\rm d}$ (AU) & $M_*$ ($M_\sun$) & $\omega$\tablenotemark{a} (km s$^{-1}$ pc$^{-1}$) & $M_{\rm core}$\tablenotemark{b} ($M_\sun$) & Reference} 
\startdata
L1448 IRS 3B & 330--720 & 1.0--1.9 & \nodata & 3.5 & 1\\
L1448-mm & 140--160 & 0.11--0.31 & \nodata & 1.5 & 1\\
NGC 1333 IRAS 4A & 630--720 & 0.14--0.15 & \nodata & 4.5 & 1\\
NGC 1333 IRAS 4B & $<$5 & 0.03--0.1 & \nodata & 1.7 & 1\\
IRAS 03282$+$3035 & 100--330 & 0.1--0.24 & 1.3 & 1.4 & 1,2\\
L1527 IRS & 70--150 & 0.07--0.24 & 2.2 & 0.8 & 1,2 \\
IRAS 16253$-$2429 & 90--240 & 0.02--0.04 & 1.2 & \nodata & 2 \\
B335 & $<$5 & 0.05--0.19 & 0.8 & 0.9 & 2,3 \\
L1157-mm & $<$5 & 0.02--0.08 & 0.9 & 2.0 & 1,2 \\
\enddata
\label{comtable}
\tablecomments{The ranges correspond to the fitting results with $f = 0.5$ and $f = 1$.}
\tablenotetext{a}{Velocity gradient perpendicular to the outflow in the associated cores observed on a 10,000 AU scale.}
\tablenotetext{b}{Mass of the associated core within a radius of 4,200 AU.}
\tablerefs{(1) Tobin et al.~2011; (2) Motte \& Andr\'e 2001; (3) Yen et al.~2011.}
\end{deluxetable}

\begin{deluxetable}{lccccc}
\tablewidth{0pt}
\tablecaption{Comparison with the Orientation of Magnetic Field and Outflow}
\tablehead{Source & $R_{\rm d}$ (AU) & $j$ (km s$^{-1}$ pc) & $\Delta\theta_{\rm core}$ & $\Delta\theta_{\rm envelope}$} 
\startdata
L1448 IRS 2 & 130--390 & 8.3--9.0 $\times$ 10$^{-4}$ & 15$\degr$ & 3$\degr$\\
L1448 IRS 3B & 330--720 & 3.6--3.9 $\times$ 10$^{-3}$ & 82\degr & 79\degr\\
L1448-mm & 140--160 & 5.6--10.2 $\times$ 10$^{-4}$ & 44$\degr$ & 45$\degr$\\
NGC 1333 IRAS 4A & 630--720 & 1.4--1.5 $\times$ 10$^{-3}$ & 37\degr & 36\degr\\
NGC 1333 IRAS 4B & $<$5 & $<$5.0 $\times$ 10$^{-5}$ & 55$\degr$ & 84$\degr$\\
L1527 IRS & 70--150 & 4.6--5.8 $\times$ 10$^{-4}$ & 32$\degr$ & 87$\degr$\\
B335 & $<$5 & $<$5 $\times$ 10$^{-5}$ & 75$\degr$ & 33$\degr$$\tablenotemark{a}$\\
L1157 & $<$5 & $<$5 $\times$ 10$^{-5}$ & 14$\degr$ & 3$\degr$\\ 
\enddata
\tablecomments{The ranges correspond to the fitting results with $f = 0.5$ and $f = 1$.}
\tablenotetext{a}{The polarization detections of the CARMA observations in B335 are all less then 3$\sigma$, and hence the magnetic field orientation is more uncertain.}
\tablerefs{Hull et al.~2014.}
\label{poltab}
\end{deluxetable}


\begin{thebibliography}{}
\bibitem[Aikawa et al.(2008)]{2008ApJ...674..984A} Aikawa, Y., Wakelam, V., Garrod, R.~T., \& Herbst, E.\ 2008, \apj, 674, 984 
\bibitem[Allen et al.(2003)]{2003ApJ...599..363A} Allen, A., Li, Z.-Y., \& Shu, F.~H.\ 2003, \apj, 599, 363 
\bibitem[Andrews et al.(2012)]{2012ApJ...744..162A} Andrews, S.~M., Wilner, D.~J., Hughes, A.~M., et al.\ 2012, \apj, 744, 162 
\bibitem[Andr\'{e} et al.(2000)]{And00} Andr\'{e}, P., Ward-Thompson, D., \& Barsony, M.\ 2000, in Protostars and Planets IV, ed. V.,
Mannings, A. P., Boss, \& S. S., Russel (Tucson, AZ: Univ. of Arizona Press), 59 
\bibitem[Andrews \& Williams(2007)]{2007ApJ...659..705A} Andrews, S.~M., \& Williams, J.~P.\ 2007, \apj, 659, 705 
\bibitem[Arce \& Sargent(2006)]{2006ApJ...646.1070A} Arce, H.~G., \& Sargent, A.~I.\ 2006, \apj, 646, 1070 
\bibitem[Aso et al.(2014)]{} Aso, Y., Ohashi, N., Saigo, K., et al.\ 2014, submitted to \apj
\bibitem[Attard et al.(2009)]{2009ApJ...702.1584A} Attard, M., Houde, M., Novak, G., et al.\ 2009, \apj, 702, 1584
\bibitem[Basu(1998)]{1998ApJ...509..229B} Basu, S.\ 1998, \apj, 509, 229 
\bibitem[Bate(1998)]{1998ApJ...508L..95B} Bate, M.~R.\ 1998, \apjl, 508, L95 
\bibitem[Beckwith et al.(1990)]{Bec90} Beckwith, S.~V.~W., Sargent, A.~I., Chini, R.~S., \& Guesten, R.\ 1990, \aj, 99, 924 
\bibitem[Bertrang et al.(2014)]{2014A&A...565A..94B} Bertrang, G., Wolf, S., \& Das, H.~S.\ 2014, \aap, 565, A94 
\bibitem[Brinch et al.(2009)]{2009A&A...502..199B} Brinch, C., J{\o}rgensen, J.~K., \& Hogerheijde, M.~R.\ 2009, \aap, 502, 199 
\bibitem[Brinch et al.(2007)]{2007A&A...475..915B} Brinch, C., Crapsi, A., J{\o}rgensen, J.~K., Hogerheijde, M.~R., \& Hill, T.\ 2007, \aap, 475, 915 
\bibitem[Brinch \& J{\o}rgensen(2013)]{2013A&A...559A..82B} Brinch, C., \& J{\o}rgensen, J.~K.\ 2013, \aap, 559, A82 
\bibitem[Brooke et al.(2007)]{2007ApJ...655..364B} Brooke, T.~Y., Huard, T.~L., Bourke, T.~L., et al.\ 2007, \apj, 655, 364
\bibitem[Cabrit \& Bertout(1986)]{1986ApJ...307..313C} Cabrit, S., \& Bertout, C.\ 1986, \apj, 307, 313 
\bibitem[Caselli et al.(2002)]{2002ApJ...572..238C} Caselli, P., Benson, P.~J., Myers, P.~C., \& Tafalla, M.\ 2002, \apj, 572, 238 
\bibitem[Cassen \& Moosman(1981)]{1981Icar...48..353C} Cassen, P., \& Moosman, A.\ 1981, \icarus, 48, 353 
\bibitem[Chandler \& Sargent(1993)]{Cha93} Chandler, C.~J., \& Sargent, A.~I.\ 1993, \apjl, 414, L29 
\bibitem[Chapman et al.(2013)]{2013ApJ...770..151C} Chapman, N.~L., Davidson, J.~A., Goldsmith, P.~F., et al.\ 2013, \apj, 770, 151 
\bibitem[Chen et al.(2013)]{2013ApJ...768..110C} Chen, X., Arce, H.~G., Zhang, Q., et al.\ 2013, \apj, 768, 110 
\bibitem[Chen et al.(2007)]{2007ApJ...669.1058C} Chen, X., Launhardt, R., \& Henning, T.\ 2007, \apj, 669, 1058 
\bibitem[Chiang et al.(2012)]{2012ApJ...756..168C} Chiang, H.-F., Looney, L.~W., \& Tobin, J.~J.\ 2012, \apj, 756, 168 
\bibitem[Chiang et al.(2010)]{2010ApJ...709..470C} Chiang, H.-F., Looney, L.~W., Tobin, J.~J., \& Hartmann, L.\ 2010, \apj, 709, 470 
\bibitem[Chiang et al.(2008)]{2008ApJ...680..474C} Chiang, H.-F., Looney, L.~W., Tassis, K., Mundy, L.~G., \& Mouschovias, T.~C.\ 2008, \apj, 680, 474 
\bibitem[Choi et al.(2011)]{2011PASJ...63.1281C} Choi, M., Kang, M., Tatematsu, K., Lee, J.-E., \& Park, G.\ 2011, \pasj, 63, 1281 
\bibitem[Choi et al.(2010)]{2010ApJ...723L..34C} Choi, M., Tatematsu, K., \& Kang, M.\ 2010, \apjl, 723, L34 
\bibitem[Choi et al.(2007)]{2007ApJ...667L.183C} Choi, M., Tatematsu, K., Park, G., \& Kang, M.\ 2007, \apjl, 667, L183 
\bibitem[Choi et al.(2006)]{2006ApJ...646.1050C} Choi, M., Hodapp, K.~W., Hayashi, M., et al.\ 2006, \apj, 646, 1050 
\bibitem[Chou et al.(2014)]{} Chou, T.-L., Takakuwa S., Yen, H.-W., Ohashi, N., \& Ho, P.~T.~P.\ 2014, accepted by \apj, arXiv:1410.3927 
\bibitem[Claussen et al.(1998)]{1998ApJ...507L..79C} Claussen, M.~J., Marvel, K.~B., Wootten, A., \& Wilking, B.~A.\ 1998, \apjl, 507, L79 
\bibitem[Codella et al.(2014)]{2014A&A...568L...5C} Codella, C., Cabrit, S., Gueth, F., et al.\ 2014, \aap, 568, L5
\bibitem[Comer\'on(1998)]{} Comer\'on, F. 2008, in Handbook of Star Forming Regions, Vol. II, ed. B. Reipurth, ASP Monograph Publ., 5, 295
\bibitem[Curiel et al.(1999)]{1999ApJ...527..310C} Curiel, S., Torrelles, J.~M., Rodr{\'{\i}}guez, L.~F., G{\'o}mez, J.~F., \& Anglada, G.\ 1999, \apj, 527, 310 
\bibitem[Dapp et al.(2012)]{2012A&A...541A..35D} Dapp, W.~B., Basu, S., \& Kunz, M.~W.\ 2012, \aap, 541, A35 
\bibitem[Davidson et al.(2011)]{2011ApJ...732...97D} Davidson, J.~A., Novak, G., Matthews, T.~G., et al.\ 2011, \apj, 732, 97 
\bibitem[de Gregorio-Monsalvo et al.(2013)]{2013A&A...557A.133D} de Gregorio-Monsalvo, I., M{\'e}nard, F., Dent, W., et al.\ 2013, \aap, 557, AA133 
\bibitem[Di Francesco et al.(2001)]{2001ApJ...562..770D} Di Francesco, J., Myers, P.~C., Wilner, D.~J., Ohashi, N., \& Mardones, D.\ 2001, \apj, 562, 770 
\bibitem[Dotson et al.(2010)]{2010ApJS..186..406D} Dotson, J.~L., Vaillancourt, J.~E., Kirby, L., et al.\ 2010, \apjs, 186, 406 
\bibitem[Dunham et al.(2014)]{2014arXiv1401.1809D} Dunham, M.~M., Stutz, A.~M., Allen, L.~E., et al.\ 2014a, in Protostars and Planets VI, ed. H., Beuther, R., Klessen, C., Dullemond, \& Th., Henning, (Tucson, AZ: Univ. of Arizona Press)
\bibitem[Dunham et al.(2014)]{2014ApJ...783...29D} Dunham, M.~M., Arce, H.~G., Mardones, D., et al.\ 2014b, \apj, 783, 29 
\bibitem[Dutrey et al.(1998)]{1998A&A...338L..63D} Dutrey, A., Guilloteau, S., Prato, L., et al.\ 1998, \aap, 338, L63 
\bibitem[Enoch et al.(2009)]{2009ApJ...692..973E} Enoch, M.~L., Evans, N.~J., II, Sargent, A.~I., \& Glenn, J.\ 2009, \apj, 692, 973 
\bibitem[Enoch et al.(2007)]{2007ApJ...666..982E} Enoch, M.~L., Glenn, J., Evans, N.~J., II, et al.\ 2007, \apj, 666, 982
\bibitem[Enoch et al.(2006)]{2006ApJ...638..293E} Enoch, M.~L., Young, K.~E., Glenn, J., et al.\ 2006, \apj, 638, 293
\bibitem[Evans et al.(2003)]{2003PASP..115..965E} Evans, N.~J., II, Allen, L.~E., Blake, G.~A., et al.\ 2003, \pasp, 115, 965
\bibitem[Froebrich(2005)]{2005ApJS..156..169F} Froebrich, D.\ 2005, \apjs, 156, 169 
\bibitem[Girart \& Acord(2001)]{2001ApJ...552L..63G} Girart, J.~M., \& Acord, J.~M.~P.\ 2001, \apjl, 552, L63 
\bibitem[Goodman et al.(1993)]{1993ApJ...406..528G} Goodman, A.~A., Benson, P.~J., Fuller, G.~A., \& Myers, P.~C.\ 1993, \apj, 406, 528 
\bibitem[Gueth \& Guilloteau(1999)]{1999A&A...343..571G} Gueth, F., \& Guilloteau, S.\ 1999, \aap, 343, 571 
\bibitem[Gueth et al.(1997)]{1997A&A...323..943G} Gueth, F., Guilloteau, S., Dutrey, A., \& Bachiller, R.\ 1997, \aap, 323, 943
\bibitem[Guilloteau et al.(2011)]{2011A&A...529A.105G} Guilloteau, S., Dutrey, A., Pi{\'e}tu, V., \& Boehler, Y.\ 2011, \aap, 529, A105 
\bibitem[Guilloteau \& Dutrey(1994)]{1994A&A...291L..23G} Guilloteau, S., \& Dutrey, A.\ 1994, \aap, 291, L23 
\bibitem[Hara et al.(2013)]{2013ApJ...771..128H} Hara, C., Shimajiri, Y., Tsukagoshi, T., et al.\ 2013, \apj, 771, 128 
\bibitem[Harsono et al.(2014)]{2014A&A...562A..77H} Harsono, D., J{\o}rgensen, J.~K., van Dishoeck, E.~F., et al.\ 2014, \aap, 562, A77
\bibitem[Harvey et al.(2003)]{2003ApJ...596..383H} Harvey, D.~W.~A., Wilner, D.~J., Myers, P.~C., \& Tafalla, M.\ 2003, \apj, 596, 383 
\bibitem[Hatchell \& Dunham(2009)]{2009A&A...502..139H} Hatchell, J., \& Dunham, M.~M.\ 2009, \aap, 502, 139 
\bibitem[Hatchell et al.(2007)]{2007A&A...472..187H} Hatchell, J., Fuller, G.~A., \& Richer, J.~S.\ 2007, \aap, 472, 187
\bibitem[Hennebelle \& Ciardi(2009)]{2009A&A...506L..29H} Hennebelle, P., \& Ciardi, A.\ 2009, \aap, 506, L29 
\bibitem[Heyer \& Graham(1989)]{1989PASP..101..816H} Heyer, M.~H., \& Graham, J.~A.\ 1989, \pasp, 101, 816
\bibitem[Hirano et al.(1988)]{Hir88} Hirano, N., Kameya, O., Nakayama, M., \& Takakubo, K.\ 1988, \apjl, 327, L69 
\bibitem[Ho et al.(2004)]{Ho04} Ho, P.~T.~P., Moran, J.~M., \& Lo, K.~Y.\ 2004, \apjl, 616, L1 
\bibitem[Hogerheijde et al.(1998)]{1998ApJ...502..315H} Hogerheijde, M.~R., van Dishoeck, E.~F., Blake, G.~A., \& van Langevelde, H.~J.\ 1998, \apj, 502, 315 
\bibitem[Hull et al.(2014)]{2014ApJS..213...13H} Hull, C.~L.~H., Plambeck, R.~L., Kwon, W., et al.\ 2014, \apjs, 213, 13 
\bibitem[Hull et al.(2013)]{2013ApJ...768..159H} Hull, C.~L.~H., Plambeck, R.~L., Bolatto, A.~D., et al.\ 2013, \apj, 768, 159 
\bibitem[Joos et al.(2013)]{2013A&A...554A..17J} Joos, M., Hennebelle, P., Ciardi, A., \& Fromang, S.\ 2013, \aap, 554, A17 
\bibitem[Joos et al.(2012)]{2012A&A...543A.128J} Joos, M., Hennebelle, P., \& Ciardi, A.\ 2012, \aap, 543, A128 
\bibitem[J{\o}rgensen et al.(2009)]{2009A&A...507..861J} J{\o}rgensen, J.~K., van Dishoeck, E.~F., Visser, R., et al.\ 2009, \aap, 507, 861 
\bibitem[J{\o}rgensen et al.(2007)]{2007ApJ...659..479J} J{\o}rgensen, J.~K., Bourke, T.~L., Myers, P.~C., et al.\ 2007, \apj, 659, 479 
\bibitem[Kirk et al.(2007)]{2007ApJ...668.1042K} Kirk, H., Johnstone, D., \& Tafalla, M.\ 2007, \apj, 668, 1042 
\bibitem[Krasnopolsky \& K\"onigl(2002)]{2002ApJ...580..987K} Krasnopolsky, R., \& K\"onigl, A.\ 2002, \apj, 580, 987 
\bibitem[Kurono et al.(2013)]{2013ApJ...765...85K} Kurono, Y., Saito, M., Kamazaki, T., Morita, K.-I., \& Kawabe, R.\ 2013, \apj, 765, 85
\bibitem[Kwon et al.(2006)]{2006ApJ...653.1358K} Kwon, W., Looney, L.~W., Crutcher, R.~M., \& Kirk, J.~M.\ 2006, \apj, 653, 1358 
\bibitem[Launhardt(2004)]{2004IAUS..221..213L} Launhardt, R.\ 2004, Star Formation at High Angular Resolution, 221, 213 
\bibitem[Lee et al.(2014)]{2014ApJ...786..114L} Lee, C.-F., Hirano, N., Zhang, Q., et al.\ 2014, \apj, 786, 114 
\bibitem[Lee(2011)]{2011ApJ...741...62L} Lee, C.-F.\ 2011, \apj, 741, 62 
\bibitem[Lee(2010)]{2010ApJ...725..712L} Lee, C.-F.\ 2010, \apj, 725, 712 
\bibitem[Lee et al.(2008)]{2008ApJ...685.1026L} Lee, C.-F., Ho, P.~T.~P., Bourke, T.~L., et al.\ 2008, \apj, 685, 1026 
\bibitem[Lee et al.(2006)]{Lee06} Lee, C.-F., Ho, P.~T.~P., Beuther, H., Bourke, T.~L., Zhang, Q., Hirano, N., \& Shang, H.\ 2006, \apj, 639, 292 
\bibitem[Lee et al.(2004)]{2004ApJ...617..360L} Lee, J.-E., Bergin, E.~A., \& Evans, N.~J., II 2004, \apj, 617, 360
\bibitem[Li et al.(2013)]{2013ApJ...774...82L} Li, Z.-Y., Krasnopolsky, R., \& Shang, H.\ 2013, \apj, 774, 82 
\bibitem[Li et al.(2011)]{2011ApJ...738..180L} Li, Z.-Y., Krasnopolsky, R., \& Shang, H.\ 2011, \apj, 738, 180 
\bibitem[Lombardi et al.(2006)]{2006A&A...454..781L} Lombardi, M., Alves, J., \& Lada, C.~J.\ 2006, \aap, 454, 781
\bibitem[Lommen et al.(2008)]{2008A&A...481..141L} Lommen, D., J{\o}rgensen, J.~K., van Dishoeck, E.~F., \& Crapsi, A.\ 2008, \aap, 481, 141 
\bibitem[Looney et al.(2003)]{2003ApJ...592..255L} Looney, L.~W., Mundy, L.~G., \& Welch, W.~J.\ 2003, \apj, 592, 255
\bibitem[Looney et al.(2000)]{2000ApJ...529..477L} Looney, L.~W., Mundy, L.~G., \& Welch, W.~J.\ 2000, \apj, 529, 477
\bibitem[Machida et al.(2014)]{2014MNRAS.438.2278M} Machida, M.~N., Inutsuka, S.-i., \& Matsumoto, T.\ 2014, \mnras, 438, 2278
\bibitem[Machida et al.(2011)]{2011PASJ...63..555M} Machida, M.~N., Inutsuka, S.-I., \& Matsumoto, T.\ 2011, \pasj, 63, 555 
\bibitem[Machida et al.(2005)]{2005prpl.conf.8280M} Machida, M.~N., Matsumoto, T., Hanawa, T., \& Tomisaka, K.\ 2005, Protostars and Planets V Posters, 8280 
\bibitem[Machida \& Matsumoto(2011)]{2011MNRAS.413.2767M} Machida, M.~N., \& Matsumoto, T.\ 2011, \mnras, 413, 2767 
\bibitem[Maret et al.(2014)]{2014A&A...563L...1M} Maret, S., Belloche, A., Maury, A.~J., et al.\ 2014, \aap, 563, L1 
\bibitem[Markwardt(2009)]{2009ASPC..411..251M} Markwardt, C.~B.\ 2009, Astronomical Data Analysis Software and Systems XVIII, 411, 251
\bibitem[Marvel et al.(2008)]{2008ApJ...685..285M} Marvel, K.~B., Wilking, B.~A., Claussen, M.~J., \& Wootten, A.\ 2008, \apj, 685, 285 
\bibitem[Matthews et al.(2009)]{2009ApJS..182..143M} Matthews, B.~C., McPhee, C.~A., Fissel, L.~M., \& Curran, R.~L.\ 2009, \apjs, 182, 143 
\bibitem[Mellon \& Li(2009)]{2009ApJ...698..922M} Mellon, R.~R., \& Li, Z.-Y.\ 2009, \apj, 698, 922 
\bibitem[Mellon \& Li(2008)]{2008ApJ...681.1356M} Mellon, R.~R., \& Li, Z.-Y.\ 2008, \apj, 681, 1356 
\bibitem[Momose et al.(1998)]{1998ApJ...504..314M} Momose, M., Ohashi, N., Kawabe, R., Nakano, T., \& Hayashi, M.\ 1998, \apj, 504, 314 
\bibitem[Motte \& Andr{\'e}(2001)]{2001A&A...365..440M} Motte, F., \& Andr{\'e}, P.\ 2001, \aap, 365, 440 
\bibitem[Murillo et al.(2013)]{2013A&A...560A.103M} Murillo, N.~M., Lai, S.-P., Bruderer, S., Harsono, D., \& van Dishoeck, E.~F.\ 2013, \aap, 560, A103
\bibitem[Myers(2000)]{Mye00} Myers, P. C., Evans, N. J., II, Ohashi, N. 2000, in Protostars and Planets IV, ed. V., Mannings, A. P., Boss, \& S. S., Russel (Tucson, AZ: Univ. of Arizona Press), 217
\bibitem[Nakajima et al.(2003)]{2003AJ....125.1407N} Nakajima, Y., Nagata, T., Sato, S., et al.\ 2003, \aj, 125, 1407 
\bibitem[Ohashi et al.(2014)]{} Ohashi, N., Saigo, K., Aso, Y., et al.\ 2014, accepted by \apj, arXiv:1410.0172 
\bibitem[Ohashi et al.(1997)]{1997ApJ...475..211O} Ohashi, N., Hayashi, M., Ho, P.~T.~P., \& Momose, M.\ 1997, \apj, 475, 211 
\bibitem[Padovani et al.(2014)]{2014arXiv1408.5901P} Padovani, M., Galli, D., Hennebelle, P., Commer{\c c}on, B., \& Joos, M.\ 2014, accepted by \aap
\bibitem[Padovani et al.(2013)]{2013A&A...560A.114P} Padovani, M., Hennebelle, P., \& Galli, D.\ 2013, \aap, 560, AA114 
\bibitem[P{\'e}rez et al.(2012)]{2012ApJ...760L..17P} P{\'e}rez, L.~M., Carpenter, J.~M., Chandler, C.~J., et al.\ 2012, \apjl, 760, L17 
\bibitem[Pi{\'e}tu et al.(2007)]{2007A&A...467..163P} Pi{\'e}tu, V., Dutrey, A., \& Guilloteau, S.\ 2007, \aap, 467, 163 
\bibitem[Plunkett et al.(2013)]{2013ApJ...774...22P} Plunkett, A.~L., Arce, H.~G., Corder, S.~A., et al.\ 2013, \apj, 774, 22 
\bibitem[Qi et al.(2004)]{2004ApJ...616L..11Q} Qi, C., Ho, P.~T.~P., Wilner, D.~J., et al.\ 2004, \apjl, 616, L11 
\bibitem[Qi et al.(2003)]{2003ApJ...597..986Q} Qi, C., Kessler, J.~E., Koerner, D.~W., Sargent, A.~I., \& Blake, G.~A.\ 2003, \apj, 597, 986 
\bibitem[Rosenfeld et al.(2013)]{2013ApJ...774...16R} Rosenfeld, K.~A., Andrews, S.~M., Hughes, A.~M., Wilner, D.~J., \& Qi, C.\ 2013, \apj, 774, 16 
\bibitem[Saito et al.(1999)]{1999ApJ...518..334S} Saito, M., Sunada, K., Kawabe, R., Kitamura, Y., \& Hirano, N.\ 1999, \apj, 518, 334 
\bibitem[Sault et al.(1995)]{1995ASPC...77..433S} Sault, R.~J., Teuben, P.~J., \& Wright, M.~C.~H.\ 1995, Astronomical Data Analysis Software and Systems IV, 77, 433 
\bibitem[Schnee et al.(2012)]{2012ApJ...755..178S} Schnee, S., Sadavoy, S., Di Francesco, J., Johnstone, D., \& Wei, L.\ 2012, \apj, 755, 178 
\bibitem[Scoville et al.(1993)]{1993PASP..105.1482S} Scoville, N.~Z., Carlstrom, J.~E., Chandler, C.~J., et al.\ 1993, \pasp, 105, 1482  
\bibitem[Shu et al.(1991)]{1991ApJ...370L..31S} Shu, F.~H., Ruden, S.~P., Lada, C.~J., \& Lizano, S.\ 1991, \apjl, 370, L31 
\bibitem[Shu et al.(1987)]{1987ARA&A..25...23S} Shu, F.~H., Adams, F.~C., \& Lizano, S.\ 1987, \araa, 25, 23 
\bibitem[Shu(1977)]{1977ApJ...214..488S} Shu, F.~H.\ 1977, \apj, 214, 488 
\bibitem[Seifried et al.(2013)]{2013MNRAS.432.3320S} Seifried, D., Banerjee, R., Pudritz, R.~E., \& Klessen, R.~S.\ 2013, \mnras, 432, 3320 
\bibitem[Seifried et al.(2012)]{2012MNRAS.423L..40S} Seifried, D., Banerjee, R., Pudritz, R.~E., \& Klessen, R.~S.\ 2012, \mnras, 423, L40
\bibitem[Simon et al.(2000)]{2000ApJ...545.1034S} Simon, M., Dutrey, A., \& Guilloteau, S.\ 2000, \apj, 545, 1034 
\bibitem[Spaans et al.(1995)]{1995ApJ...455L.167S} Spaans, M., Hogerheijde, M.~R., Mundy, L.~G., \& van Dishoeck, E.~F.\ 1995, \apjl, 455, L167
\bibitem[Stephens et al.(2013)]{2013ApJ...769L..15S} Stephens, I.~W., Looney, L.~W., Kwon, W., et al.\ 2013, \apjl, 769, LL15 
\bibitem[Stutz et al.(2008)]{Stu08} Stutz, A,~M., Rubin, M., Werner, M.~W., Rieke, G.~H., Bieging, J.~H., Keene, J., Kang, M., Shirley, Y.~L.; Su, K.~Y.~L., Velusamy, T., Wilner, D.~J.\ 2008, \apj, 687, 389 
\bibitem[Tachihara et al.(2007)]{2007ApJ...659.1382T} Tachihara, K., Rengel, M., Nakajima, Y., et al.\ 2007, \apj, 659, 1382 
\bibitem[Tachihara et al.(2001)]{2001PASJ...53.1081T} Tachihara, K., Toyoda, S., Onishi, T., et al.\ 2001, \pasj, 53, 1081
\bibitem[Takakuwa et al.(2013)]{2013ApJ...776...51T} Takakuwa, S., Saito, M., Lim, J., \& Saigo, K.\ 2013, \apj, 776, 51 
\bibitem[Takakuwa et al.(2012)]{2012ApJ...754...52T} Takakuwa, S., Saito, M., Lim, J., et al.\ 2012, \apj, 754, 52 
\bibitem[Takakuwa et al.(2007)]{2007ApJ...662..431T} Takakuwa, S., Ohashi, N., Bourke, T.~L., et al.\ 2007, \apj, 662, 431 
\bibitem[Terebey et al.(1984)]{1984ApJ...286..529T} Terebey, S., Shu, F.~H., \& Cassen, P.\ 1984, \apj, 286, 529 
\bibitem[Tomida et al.(2014)]{ }Tobin, J.~J., Looney, L.~W., Wilner, D.~J., et al.\ 2014, Protostars and Planets VI Posters, 1B092
\bibitem[Tobin et al.(2012)]{2012Natur.492...83T} Tobin, J.~J., Hartmann, L., Chiang, H.-F., et al.\ 2012a, \nat, 492, 83 
\bibitem[Tobin et al.(2012)]{2012ApJ...748...16T} Tobin, J.~J., Hartmann, L., Bergin, E., et al.\ 2012b, \apj, 748, 16 
\bibitem[Tobin et al.(2011)]{2011ApJ...740...45T} Tobin, J.~J., Hartmann, L., Chiang, H.-F., et al.\ 2011, \apj, 740, 45 
\bibitem[Tobin et al.(2010)]{2010ApJ...712.1010T} Tobin, J.~J., Hartmann, L., Looney, L.~W., \& Chiang, H.-F.\ 2010, \apj, 712, 1010 
\bibitem[Tobin et al.(2008)]{2008ApJ...679.1364T} Tobin, J.~J., Hartmann, L., Calvet, N., \& D'Alessio, P.\ 2008, \apj, 679, 1364 
\bibitem[Tobin et al.(2007)]{2007ApJ...659.1404T} Tobin, J.~J., Looney, L.~W., Mundy, L.~G., Kwon, W., \& Hamidouche, M.\ 2007, \apj, 659, 1404 
\bibitem[Tomida et al.(2013)]{2013ApJ...763....6T} Tomida, K., Tomisaka, K., Matsumoto, T., et al.\ 2013, \apj, 763, 6 
\bibitem[Tothill et al.(2009)]{2009ApJS..185...98T} Tothill, N.~F.~H., L{\"o}hr, A., Parshley, S.~C., et al.\ 2009, \apjs, 185, 98 
\bibitem[Ulrich(1976)]{1976ApJ...210..377U} Ulrich, R.~K.\ 1976, \apj, 210, 377 
\bibitem[van der Marel et al.(2013)]{2013A&A...556A..76V} van der Marel, N., Kristensen, L.~E., Visser, R., et al.\ 2013, \aap, 556, A76 
\bibitem[van Kempen et al.(2009)]{2009A&A...498..167V} van Kempen, T.~A., van Dishoeck, E.~F., Salter, D.~M., et al.\ 2009, \aap, 498, 167 
\bibitem[Vilas-Boas et al.(2000)]{2000ApJ...532.1038V} Vilas-Boas, J.~W.~S., Myers, P.~C., \& Fuller, G.~A.\ 2000, \apj, 532, 1038 
\bibitem[Volgenau et al.(2006)]{2006ApJ...651..301V} Volgenau, N.~H., Mundy, L.~G., Looney, L.~W., \& Welch, W.~J.\ 2006, \apj, 651, 301 
\bibitem[Williams \& Cieza(2011)]{2011ARA&A..49...67W} Williams, J.~P., \& Cieza, L.~A.\ 2011, \araa, 49, 67 
\bibitem[Wolf-Chase et al.(2000)]{2000AJ....120.1467W} Wolf-Chase, G.~A., Barsony, M., \& O'Linger, J.\ 2000, \aj, 120, 1467 
\bibitem[Yen et al.(2014)]{2014arXiv1407.2699Y} Yen, H.-W., Takakuwa, S., Ohashi, N., et al.\ 2014, arXiv:1407.2699 
\bibitem[Yen et al.(2013)]{2013ApJ...772...22Y} Yen, H.-W., Takakuwa, S., Ohashi, N., \& Ho, P.~T.~P.\ 2013, \apj, 772, 22
\bibitem[Yen et al.(2011)]{2011ApJ...742...57Y} Yen, H.-W., Takakuwa, S., \& Ohashi, N.\ 2011, \apj, 742, 57 
\bibitem[Yen et al.(2010)]{2010ApJ...710.1786Y} Yen, H.-W., Takakuwa, S., \& Ohashi, N.\ 2010, \apj, 710, 1786 
\bibitem[Young et al.(2006)]{2006AJ....132.1998Y} Young, C.~H., Bourke, T.~L., Young, K.~E., et al.\ 2006, \aj, 132, 1998 
\bibitem[Zhang et al.(2014)]{2014arXiv1407.3984Z} Zhang, Q., Qiu, K., Girart, J.~M., et al.\ 2014, arXiv:1407.3984 
\end{thebibliography}
\end{document}